\documentclass[12pt,twoside,a4paper]{book}
\usepackage{lmodern}
\usepackage[latin1]{inputenc} 
\usepackage[T1]{fontenc}
\usepackage[french,english,francais]{babel}
\usepackage{amsmath}
\usepackage{amsthm}
\usepackage{amssymb}
\usepackage{amsfonts}
\usepackage{amscd}
\usepackage{latexsym}
\usepackage{graphicx}
\usepackage{hyperref}
\usepackage{epsfig}
\usepackage{bm}
\usepackage{dsfont}
\usepackage{color}
\usepackage[resetlabels,labeled]{multibib}
\newcites{A}{Liste de mes articles cités dans cette Habilitation à diriger des Recherches}

\usepackage{fancyhdr}
\usepackage{titlesec} 
\titleformat{\chapter}{\normalfont\huge\bfseries}{\thechapter.}{20pt}{\Huge}
\titlespacing*{\chapter}{0pt}{-50pt}{40pt}

\hypersetup{colorlinks,urlcolor=magenta,citecolor=red,linkcolor=blue,filecolor=black}
\newtheorem{assumption}{Hypothèse}[section]

\newcommand\encadremath[1]{\vbox{\hrule\hbox{\vrule\kern8pt
\vbox{\kern8pt \hbox{$\displaystyle #1$}\kern8pt}
\kern8pt\vrule}\hrule}}
\def\enca#1{\vbox{\hrule\hbox{
\vrule\kern8pt\vbox{\kern8pt \hbox{$\displaystyle #1$}
\kern8pt} \kern8pt\vrule}\hrule}}

\newcommand{\theoremname}{Theorem}
\addto\captionsfrench{\renewcommand{\theoremname}{Th\'eor\`eme}}
\addto\captionsenglish{\renewcommand{\theoremname}{Theorem}}
\newtheorem{theorem}{\theoremname}[chapter]

\newcommand{\conjecturename}{Conjecture}
\addto\captionsfrench{\renewcommand{\conjecturename}{Conjecture}}
\addto\captionsenglish{\renewcommand{\conjecturename}{Conjecture}}
\newtheorem{conjecture}{\conjecturename}[chapter]

\newcommand{\remarkname}{Remark}
\addto\captionsfrench{\renewcommand{\remarkname}{Remarque}}
\addto\captionsenglish{\renewcommand{\remarkname}{Remark}}
\newtheorem{remark}{\remarkname}[chapter]

\newcommand{\propositionname}{Proposition}
\addto\captionsfrench{\renewcommand{\propositionname}{Proposition}}
\addto\captionsenglish{\renewcommand{\propositionname}{Proposition}}
\newtheorem{proposition}{\propositionname}[chapter]

\newcommand{\lemmename}{Lemma}
\addto\captionsfrench{\renewcommand{\lemmename}{Lemme}}
\addto\captionsenglish{\renewcommand{\lemmename}{Lemma}}
\newtheorem{lemma}{\lemmename}[chapter]

\newcommand{\corollaryname}{Corollary}
\addto\captionsfrench{\renewcommand{\corollaryname}{Corollaire}}
\addto\captionsenglish{\renewcommand{\corollaryname}{Corollary}}
\newtheorem{corollary}{\corollaryname}[chapter]

\newcommand{\definitionname}{Theorem}
\addto\captionsfrench{\renewcommand{\definitionname}{D\'efinition}}
\addto\captionsenglish{\renewcommand{\definitionname}{Definition}}
\newtheorem{definition}{\definitionname}[chapter]

\def\br{\begin{remark}\rm\small}
\def\er{\end{remark}}
\def\bt{\begin{theorem}}
\def\et{\end{theorem}}
\def\bd{\begin{definition}}
\def\ed{\end{definition}}
\def\bp{\begin{proposition}}
\def\ep{\end{proposition}}
\def\bl{\begin{lemma}}
\def\el{\end{lemma}}
\def\bc{\begin{corollary}}
\def\ec{\end{corollary}}
\newcommand{\beaa}{\begin{eqnarray*}}
\newcommand{\eeaa}{\end{eqnarray*}}
\newcommand{\bea}{\begin{eqnarray}}
\newcommand{\eea}{\end{eqnarray}}
\newcommand{\beq}{\begin{equation}}
\newcommand{\eeq}{\end{equation}}
\newcommand{\beqq}{\begin{equation*}}
\newcommand{\eeqq}{\end{equation*}}

\def\PI{(P_{\rm I})}
\def\PII{(P_{\rm II})}

\def\PIII{(P_{{\rm III}})}

\def\PIV{(P_{\rm IV})}
\def\PV{(P_{\rm V})}
\def\PVI{(P_{\rm VI})}

\def\DD{\mathcal D}
\def\RR{\mathcal R}

\newcommand{\td}{\tilde}

\newcommand{\genus}{{\normalfont g}}

\newcommand{\Tr}{{\,\rm Tr}\:}

\newcommand{\Res}{\mathop{\,\rm Res\,}}

\newcommand{\Pint}{{\int\kern -1.em -\kern-.25em}}

\newcommand{\virg}{{\qquad , \qquad}}

\makeatletter
\@addtoreset{equation}{section}

\def\Tr{\mathrm {Tr}}

\def\det{\mathrm {det}}
\def\ln{\mathrm {ln}}

\def\&{&{\hskip -20pt}}

\def\bea{\begin{eqnarray}}
\def\eea{\end{eqnarray}}

\def\Tr{\mathrm {Tr}}

\def\det{\mathrm {det}}
\def\ln{\mathrm {ln}}

\def\&{&{\hskip -20pt}}

\begin{document}

\setcounter{page}{1}
\thispagestyle{empty}
\selectlanguage{french}
\begin{center}
\large{Université Jean Monnet - Université de Lyon}\\ 
\large{Ecole doctorale \textbf{SIS}}\\
Spécialité: \textbf{Mathématiques Appliquées}
\end{center}

\medskip
\medskip

\begin{center}
\huge{\textbf{Application de la récurrence topologique aux intégrales de matrices aléatoires et aux systèmes intégrables}}
\end{center}

\medskip
\medskip
\medskip

\begin{center}
\large{\textbf{Habilitation à diriger des recherches}}
\end{center}

\medskip
\medskip
\begin{center}Soutenue publiquement le 16 Octobre 2017 par\end{center}

\medskip
\medskip

\begin{center}
\huge{\textbf{Olivier Marchal}}
\end{center}

\medskip
\medskip

\begin{center}
devant le jury composé de:
\end{center}

\begin{center}
\begin{tabular}{lll}
\text{M. Borot Gaëtan}, & Institut Max Planck de Mathématiques, &\text{Rapporteur}\\
\text{M. Cafasso Mattia}, & Université d'Angers, &\text{Coordinateur}\\
\text{M. Garban Christophe}, & Université Claude Bernard, &\text{Examinateur}\\
\text{M. Gaussent Stéphane}, & Université Jean Monnet, &\text{Examinateur}\\
\text{Mme Guionnet Alice}, & Ecole normale supérieure de Lyon, &\text{Examinatrice}\\
\text{M. Lisovyi Oleg}, & Université François Rabelais, &\text{Rapporteur}\\
\text{M. Ribault Sylvain}, & Institut de Physique théorique, CEA Saclay, &\text{Examinateur}\\
\text{M. Roubtsov Volodya}, & Université d'Angers, &\text{Rapporteur}\\
\end{tabular}
\end{center}

\medskip
\medskip
\medskip
\medskip
\medskip
\medskip

\begin{center}
 \copyright \, Olivier Marchal, 2017.
\end{center}
\newpage
\strut
\thispagestyle{empty}
\newpage

\setcounter{page}{3}
\thispagestyle{fancy}
\selectlanguage{french}
\huge{Résumé} \normalsize{\,}

\medskip
\medskip

Cette Habilitation à Diriger des Recherches présente les liens entre deux applications de la récurrence topologique développée par B. Eynard et N. Orantin en 2007 et dont les domaines d'application n'ont cessé de croître depuis. Elle présente les principaux résultats obtenus depuis mon doctorat en 2010, sans pour autant en couvrir l'intégralité. Les résultats sélectionnés ont été choisis afin de souligner l'efficacité de la récurrence topologique à relier deux domaines: celui des probabilités et celui des systèmes intégrables, a priori très éloignés l'un de l'autre.

\medskip

Le premier aspect développé dans cette habilitation concerne le domaine historique d'apparition de la récurrence topologique: les intégrales de matrices aléatoires. Je présente ainsi la manière dont la récurrence topologique peut être utilisée pour calculer le développement en $\frac{1}{N}$ de diverses intégrales matricielles de taille $N\times N$. En effet, à partir de l'ordre dominant de ces asymptotiques (appelé ``courbe spectrale'') dont diverses techniques de calcul existent, la récurrence topologique permet d'obtenir récursivement tous les ordres sous-dominants avec des calculs élémentaires d'analyse complexe. Si l'étude des intégrales de matrices hermitiennes par cette méthode est maintenant bien avancée, je montre que cette technique peut être utilisée dans le cas d'intégrales de matrices aléatoires unitaires et aussi sur des exemples simples d'intégrales connues, qui permettent ainsi de tester de façon concrète la méthode générale. L'extension au cas des ``ensembles $\beta$'' est également discutée brièvement.

\medskip

Le second point développé dans cette habilitation concerne l'utilisation de la récurrence topologique dans l'étude des systèmes intégrables ayant une représentation de Lax. Cette partie est illustrée par le cas des six équations de Painlevé. Notons que le comportement local des valeurs propres (dans le centre de la distribution limite ou aux bords) des matrices hermitiennes de grande taille est précisément donné par certains cas particuliers des équations de Painlevé montrant ainsi un lien direct avec la première partie. En ce sens, le formalisme de la récurrence topologique permet de relier de façon claire ces deux domaines apparemment très éloignés. Enfin, remarquons que l'application de la récurrence topologique dans le domaine des systèmes intégrables est actuellement en plein développement avec en particulier la recherche et la compréhension d'une version ``quantique'' de la courbe spectrale de manière systématique. Il ne fait aucun doute que des résultats importants sont à attendre dans les prochaines années sur ce sujet.

\medskip

Ainsi, à la frontière entre les probabilités et les systèmes intégrables, la récurrence topologique, couplée à la méthode des équations de boucle, constitue un outil intéressant en physique théorique et présentant plus particulièrement du potentiel dans la théorie des cordes topologiques et les modèles de théorie conforme en dimension deux. Compte tenu des possibilités de généralisation dues à un formalisme souple, elle représente un outil important dans tous ces domaines.

\medskip
\medskip
\medskip
\medskip

\textbf{\underline{Mots-clés}: Récurrence topologique, Intégrales de matrices aléatoires, Probabilités intégrables, Courbe spectrale, Systèmes intégrables, Equations de Painlevé.}

\selectlanguage{english}
\thispagestyle{fancy}
\huge{Abstract}\normalsize{~}

\medskip
\medskip

The goal of this ``Habilitation à diriger des recherches'' is to present two different applications, namely computations of certain partition functions in probability and applications to integrable systems, of the topological recursion developed by B. Eynard and N. Orantin in 2007. Since its creation, the range of applications of the topological recursion has been growing and many results in different fields have been obtained. In relation with these applications, the formalism of the topological recursion has been precised and stabilized. Thus it appears that it is the right time to present an up to date review of the possibilities offered by the topological recursion in both probability and integrable systems. Since the scope of the applications of the topological recursion is quite large, I have decided to focus only on two important subjects in relation to my research activities even though it does not cover the full extent of my research. 

\medskip

The first aspect that I will develop in this document deals with the historical domain of the topological recursion: random matrix integrals. I will review the formalism of the topological recursion as well as how it can be used to obtain asymptotic $\frac{1}{N}$ series expansion of various matrix integrals. In particular, a key feature of the topological recursion is that it can recover from the leading order of the asymptotic (known in the field as a ``spectral curve'') all sub-leading orders with elementary computations. This method is particularly well known and fruitful in the case of hermitian matrix integrals, but I will also show that the general method can be used to cover integrals with hard edges, integrals over unitary matrices and much more. In the end, I will also briefly mention the generalization to $\beta$-ensembles and its challenges.

\medskip

In a second chapter, I will review the connection between the topological recursion and the study of integrable systems having a Lax pair representation. Most of the results presented there will be illustrated by the case of the famous six Painlevé equations. Though the formalism used in this chapter may look completely disconnected from the previous one, it is well known that the local statistics of eigenvalues in random matrix theory exhibit a universality phenomenon and that the encountered universal systems are precisely driven by some solutions of the Painlevé equations. As I will show, the connection can be made very explicit with the topological recursion formalism. The use of the topological recursion in integrable systems is developing rapidly at the moment and many results are expected in the next few years. In particular, the construction of a ``quantum'' version of the spectral curve in relation with integrable systems is presently an open and hot topic.

\medskip

At the border between probability and integrable systems, the topological recursion with the loop equation method is a very effective tool in theoretical physics, topological string theory and conformal field theory in dimension $2$. With a wide generalization potential and a quite customizable formalism, it certainly provides an excellent tool in all those fields. 

\medskip
\medskip
\medskip
\medskip

\textbf{\underline{Keywords}: Topological recursion, Random matrix integrals, Integrable probabilities, Spectral curve, Integrable systems, Painlevé equations.}\selectlanguage{french}
\thispagestyle{fancy}
\selectlanguage{french}
\huge{Remerciements} \normalsize{\,}

\medskip
\medskip

Ecrire une habilitation à diriger les recherches n'est pas une chose facile car cela consomme une chose précieuse lorsque l'on est en poste: du temps. C'est la raison pour laquelle j'ai décidé d'écrire la plus grande partie de la mienne durant la période estivale de 2016 afin de pouvoir bénéficier de conditions de travail les plus idéales possibles. Le plus difficile dans ce type d'entreprise est bien souvent de se lancer véritablement puisque les raisons sont nombreuses pour repousser au lendemain le début d'un travail que l'on sait d'avance long et laborieux. En ce qui me concerne, c'est surtout la question de l'intérêt de la rédaction de l'habilitation en elle-même qui n'était pas évidente: écrire un document de synthèse présente un intérêt scientifique mais le fait de devoir l'écrire en français en limite considérablement la visibilité dans le contexte international actuel de la recherche en mathématiques. Certes, il reste la perspective d'obtenir le titre nécessaire pour pouvoir diriger seul une thèse, mais cela est fortement limité par la possibilité de pouvoir trouver de bons étudiants qui se font rares à l'heure actuelle surtout lorsque l'on enseigne dans une université de taille modeste comme celle de Saint-Etienne. Bien entendu, l'habilitation donne enfin la possibilité de pouvoir (après obtention de la qualification correspondante) candidater sur des postes de Professeurs d'Université. Mais pour beaucoup de maîtres de conférences le faible nombre de postes proposés actuellement combiné à des règles de non-localité présentes dans les sections de mathématiques, laissent apparaitre comme très hypothétique la perspective de pouvoir postuler concrètement sur des postes en respectant leurs contraintes familiales, qui peuvent être fortes pour certains, lorsque le conjoint n'est pas forcément mobile professionnellement ou que les enfants sont scolarisés.

Compte tenu de ce contexte, je pense que l'écriture de ce document aurait été fortement différée voire impossible sans le soutien des différentes personnes que je tiens à présent à remercier. Mes premiers et plus sincères remerciements vont tout d'abord à Pauline Chappuis qui a su faire de ma vie le bonheur qu'elle est actuellement. Elle a su aussi trouver les arguments pour me décider à me lancer dans la rédaction de cette habilitation et m'a apporté un soutien et un amour précieux agrémentés de nombreuses patisseries dont elle a le secret pour me donner des forces. J'espère que notre belle aventure se poursuivra pour toujours et que notre petit nid familial s'agrandisse prochainement. J'en profite également pour remercier ses parents Evelyne et Didier, ses soeurs Emilie et Andrea, ses grands-parents et ses très nombreux oncles, tantes, cousins, cousines et amis qui m'ont toujours accueilli de façon chaleureuse et fait vivre de très bons moments. Mes prochains remerciements vont à ma famille et plus précisément à mon frère jumeau (le doublon) Julien et son amie Claire, mon grand-frère David et bien sûr mes parents. Que de bons moments partagés depuis ces nombreuses années! Même si vous ne pouvez pas forcément m'aider dans le domaine mathématique, il est clair qu'une part importante de ma réussite académique vous est due. Les passages à Beaumotte, les diverses courses, trails, raids, sorties VTT, etc. sont autant de petits bonheurs et souvenirs qui me permettent de surmonter les épreuves de la vie. J'espère qu'il y en aura encore beaucoup d'autres à venir.

Je remercie maintenant tous les amis de longue date qui m'ont entouré depuis de nombreuses années. Même avec le temps qui passe, mon séjour au Canada pendant plusieurs années et les contraintes familiales et professionnelles qui s'accumulent, il est toujours bon de pouvoir compter sur une soirée ou un week-end entre amis. Chronologiquement ou par paires je remercie plus particulièrement pour les beaux souvenirs du passé, du présent et j'espère du futur: Jérôme Charles (la jej), Guillaume Tobler et Rebecca Toth (et William), Pascal Striker, Julyan Arbel, Elodie Emonoz, Mathilde Gacek, Cécilie Possot, Sophie Redoutey, Anne-Luce Zahm et JP Sixdenier, Emmanuel Jacob et Mikael de la Salle que j'ai retrouvé par hasard à l'ENS Lyon.

Passons maintenant aux mathématiciens et physiciens et aux amitiés Stéphanoises et Lyonnaises plus récentes. Ma première pensée va à mon ancien directeur de thèse Bertrand Eynard qui a été et reste pour moi un directeur de recherche exemplaire et qui a mis en lumière l'objet principal de cette habilitation. J'ai toujours pris un grand plaisir à travailler en sa compagnie. Ses nombreuses idées, sa façon d'expliquer clairement font de lui un des meilleurs chercheurs que j'ai rencontré et j'espère que notre collaboration se poursuivra dans le futur. Ma seconde pensée va à Nicolas Orantin et Gaëtan Borot qui sont toujours disponibles pour me débloquer ou répondre à mes interrogations. J'espère que prochainement nous pourrons produire un article tous ensemble. Je remercie également tous mes autres collaborateurs: Michel, Leonid, Vincent, Kohei, Piotr, Raphaël, Elba, Axel et tous les autres. Je tiens également à remercier l'ensemble du département de mathématiques de Saint-Etienne pour leur accueil et leur joie de vivre. J'inclus bien évidemment tous ceux de l'IAE, de Télécom, des mines, etc. que je cotoie régulièrement. En vrac cela donne: Michaël Bulois (le cobureau parfait), Nolwenn et la petite famille, Frédéric Chardard (toujours de bonne humeur), Julian Tugaut (le fan de game of thrones), Mathieu Sart (le seul plus jeune que moi), Filippo Nuccio (et son style italien), Laurence Grammont, Anne Pichereau, Driss Essouabri, Stéphane Gaussent, les deux autres Olivier (Gipouloux et Robert), la famille Canon (Marie-Claude et Eric), Sylvie Champier, Laëtitia Paoli, Mahdi Boukrouche, Valentina Busuioc, Federico Pellarin, Florence Millet, Mario Ahues et François Hennecart. Côté Lyonnais, la liste étant trop longue, je tiens à remercier spécialement Christophe Sabot et Elisabeth Mironescu pour leur aide à mon retour en France. Plus généralement, je remercie l'ensemble de l'équipe de probabilités, statistique et physique mathématique de Lyon 1 et de l'ENS que je vois régulièrement pour les séminaires ainsi que le secrétariat de l'Université Jean Monnet et en particulier Pascale Villet pour sa disponibilité dans de nombreuses démarches administratives.

Voilà, je suis sûr d'avoir oublié plein de monde mais les remerciements restent toujours un exercice d'exhaustivité imparfaite. Merci encore à toutes celles et ceux qui m'entourent, sans vous tout cela ne serait pas possible.

\thispagestyle{fancy}
\selectlanguage{french}
\huge{Liste de mes articles de recherche publiés ou soumis à des journaux internationaux et en cours d'évaluation}\normalsize{\,}
\medskip
\medskip
\medskip
\medskip
\medskip
\medskip

\begin{enumerate}
\item O.~Marchal, J.~Arbel, ``On the sub-Gaussianity of the Beta and Dirichlet distributions'', \textit{Electronic Communications in Probability}, Vol. \textbf{22}, No. 54, pp 1-14, 2017.
\newline
\item O.~Marchal, ``WKB solutions of difference equations and reconstruction by the topological recursion'', \textit{Nonlinearity}, 2017.
\newline
\item O. Marchal, ``Asymptotic expansions of some Toeplitz determinants via the topological recursion'', \href{https://arxiv.org/abs/1611.05627}{\textit{arXiv:1611.05627}}, Soumis à \textit{Random Matrices: Theory and Applications}, 2016.
\newline
\item B. Eynard, O. Marchal, R. Belliard, ``Integrable differential systems of topological type and reconstruction by the topological recursion'', \textit{Annales Henri Poincar\'{e}}, Vol. \textbf{18}, No. 10, pp 3193-3248, 2017.
\newline
\item B. Eynard, O. Marchal, R. Belliard, ``Loop equations from differential systems'', \textit{Annales Henri Poincar\'{e}}, 2016.
\newline
\item K. Iwaki, O. Marchal, A. Saenz, ``Painlev\'{e} equations, topological type property and reconstruction by the topological recursion'', \textit{Journal of Geometry and Physics}, 2017.
\newline
\item O. Marchal, ``Matrix models, Toeplitz determinants and recurrence times for powers of random unitary matrices'', \textit{Random Matrices: Theory and Applications}, Vol. \textbf{4}, No. 3, 2014. 
\newline
\item K.~Iwaki, O.~Marchal, ``Painlev\'{e} 2 equation with arbitrary monodromy parameter, topological recursion and determinantal formulas'', \textit{Annales Henri Poincaré}, Vol. \textbf{18}, No. 8, pp 2581-2620, 2017.
\newline
\item O. Marchal, ``Elements of proof for conjectures of Witte and Forrester about the combinatorial structure of Gaussian Beta Ensembles'', \textit{Journal of High Energy Physics}, Vol. \textbf{3}, 2014.
\newline 
\item O. Marchal, B. Eynard, M. Bergère, ``The sine-law gap probability, Painlev\'{e} 5, and asymptotic expansion by the topological recursion'', \textit{Random Matrices: Theory and Applications}, Vol. \textbf{3}, 2014.
\newline
\item V. Bouchard, A. Catuneanu, O. Marchal, P. Sulkowski, ``The remodeling conjecture and the Faber-Pandharipande formula'', \textit{Letters in Mathematical Physics}, Vol. \textbf{103}, 2013.
\newline
\item M. Bergère, B. Eynard, O. Marchal, A. Prats-Ferrer, ``Loop equations and topological recursion for the arbitrary $\beta$-two-matrix model'', \textit{Journal of High Energy Physics}, Vol. \textbf{3}, 2012.
\newline
\item O. Marchal, ``One cut solution of the $\beta$-ensembles in the Zhukovsky variable'', \textit{Journal of Statistical Mechanics}, P01011, 2012.
\newline
\item O. Marchal, ``Aspects géométriques et intégrables des modèles de matrices aléatoires'', Thèse de doctorat, disponible à \href{http://arxiv.org/abs/1012.4513}{\textit{arXiv:1012.4513}}, 2010.
\newline
\item L.O. Chekhov, B. Eynard, O. Marchal, ``Topological expansion of the $\beta$-ensemble model and quantum algebraic geometry in the sectorwise approach'', \textit{Theoretical and Mathematical Physics}, Vol. \textbf{166}, No. 2, pp. 141-185, 2011.
\newline
\item B. Eynard, A.K. Kashani-Poor, O. Marchal, ``A Matrix Model for the Topological String II: The Spectral Curve and Mirror Geometry'', \textit{Annales Henri Poincaré}, Vol. \textbf{14}, No. 1, pp. 119-158, 2013.
\newline
\item B. Eynard, A.K. Kashani-Poor, O. Marchal, ``A Matrix Model for the Topological String I: Deriving the Matrix Model'', \textit{Annales Henri Poincaré}, Vol. \textbf{15}, No. 10, pp. 1867-1901, 2014.
\newline
\item O Marchal, M. Cafasso, ``Double-scaling limits of random matrices and minimal $(2m,1)$ models: the merging of two cuts in a degenerate case'', \textit{Journal of Statistical Mechanics}, Vol. \textbf{2011}, 2011.
\newline
\item L.O. Chekhov, B. Eynard, O. Marchal, ``Topological expansion of the Bethe ansatz, and quantum algebraic geometry'', \href{http://arxiv.org/abs/0911.1664}{\textit{arXiv:0911.1664}}, 2009.
\newline
\item B. Eynard, O. Marchal, ``Topological expansion of the Bethe ansatz, and non-commutative algebraic geometry'', \textit{Journal of High Energy Physics}, Vol. \textbf{3}, 2009.
\newline
\item M. Bertola, O. Marchal, ``The partition function of the two-matrix model as an isomonodromic tau-function'', \textit{Journal of Mathematical Physics}, Vol. \textbf{50}, 013529, 2009. 
\end{enumerate}

\pagebreak

\huge{Liste de mes autres publications scientifiques} \normalsize{\,}

\medskip
\medskip
\medskip

\begin{enumerate}
\item O. Marchal, ``Fondements des probabilités avec exercices corrigés'', \textit{Ellipses Marketing}, 216 pages, 2017.
\newline
\item O. Marchal, ``Statistiques appliquées avec introduction au logiciel R'', \textit{Ellipses Marketing}, 264 pages, 2017.
\newline
\item F. Magne, E. Gomez, O. Marchal, P. Malvestio, A. Chaouat, F. Chabot, ``Evolution et facteurs prédictifs d'amélioration du SAHOS après chirurgie bariatrique dans une population d'obèses grades II et plus'', \textit{Revue des Maladies Respiratoires}, Vol. \textbf{34}, 2017.
\newline
\item P. Chappuis, G. Duru, O. Marchal, P. Girier, S. Dalle, L. Thomas, ``Dermoscopy: A useful tool for general practitioners in melanoma screening: a nationwide survey'', \textit{British Journal of Dermatology}, DOI: 10.1111/bjd.14495 2015.
\newline
\item O. Marchal, ``Locks and keys: How fast can you open several locks with too many keys?'', \href{http://arxiv.org/abs/1509.00844}{\textit{arXiv:1509.00844}}, 2015.

\end{enumerate}

\medskip
\medskip
\medskip
\medskip

\huge{Encadrement scientifique} \normalsize{\,}

\medskip
\medskip

\begin{enumerate}
\item \underline{Encadrement de stage de M2}: 2016: Encadrement du stage de Thomas Gérard de l'ENS Lyon. Intitulé du mémoire: ``Matrices de transition aléatoires''.
Stage de cinq mois effectué d'Avril 2016 à Septembre 2016.
\newline
\item \underline{Encadrement de TER} (niveau L3) en probabilités. En 2015: ``Jeux de Parrondo''. En 2014 : ``Valeurs propres des matrices aléatoires hermitiennes''.
\newline
\item \underline{Encadrement de projets transdisciplinaires M2 Enseignement}: Projets transdisciplinaires à l'interface biologie/mathématiques/physique. Thème 2014: Dérèglement climatique. Thème 2015: ``Dynamique interne de la Terre''. Thème 2016: ``Risques associés au changement climatique''.
\newline
\item \underline{Encadrement de stages ``Hippocampe''}: Initiation à la recherche pour des étudiants de CPGE pendant 3 jours sur des thèmes de recherche proposés par des enseignants-chercheurs. Projets encadrés en 2015, 2016 et 2017. Le stage de 2016 sur le paradoxe de Parrondo a donné lieu à une note de vulgarisation écrite par H. Davaux et publiée dans la gazette des mathématiciens.  
\newline
\item \underline{Analyses statistiques pour des thèses de médecine}:\newline
\begin{itemize}
\item Pauline Chappuis (dirigée par les professeurs P. Girier et L. Thomas). \newline
\textit{Titre de la thèse}: ``Evaluation des méthodes de dépistage du mélanome en médecine générale, apport de l'utilisation de la dermoscopie. Enquête quantitative auprès de médecins généralistes français''.
\newline
\item Fanny Magne-Vigneaud (dirigée par le professeur A. Chaouat). \newline
\textit{Titre de la thèse}: ``Evolution et facteurs prédictifs de persistance du syndrome d'apnées-hypopnées du sommeil (SAHOS) chez les patients obèses, après une chirurgie bariatrique''. Résultats présentés également au Congrès de Pneumologie de Langue Française.
\newline
\item Cornélia Ciudin (dirigée par les Professeurs Creuzot-Garcher et Muselier). \newline 
\sloppy{\textit{Titre de la thèse}: ``Apport d'un e-learning sur les pratiques des MG de Bourgogne Franche-Comté et Auvergne Rhône-Alpes concernant les délais d'orientation des patients en ophtalmologie''.}

\end{itemize}
\end{enumerate}

\tableofcontents \thispagestyle{fancy}
\selectlanguage{french}
\chapter{Introduction}
\thispagestyle{fancy}

La récurrence topologique est apparue dans sa forme actuelle en 2007 dans l'article \cite{EO} après des travaux initiés par de nombreux chercheurs. Néanmoins, il n'est pas rare dans certains articles ou certaines présentations d'entendre parler de la récurrence topologique sous le nom de récurrence d'Eynard et Orantin auquel on ajoute parfois le nom de L. Chekhov qui a également contribué à l'émergence du formalisme actuel. Cette récurrence a depuis connue certaines reformulations et certains liens avec la théorie conforme de Liouville, les systèmes intégrables ainsi que des versions locales ou modifiées ont été élaborés pour en faire aujourd'hui un élément important en physique théorique ainsi que dans l'étude des intégrales de matrices aléatoires. Certains détails techniques concernant directement la récurrence topologique \cite{F1,XY,FreeEnergy} ont également été précisés par la suite mais il apparait évident que le contenu de \cite{EO} reste aujourd'hui comme l'un des articles fondateurs utilisé couramment dans le domaine. 

A l'origine, la récurrence topologique a été créée pour calculer de façon efficace le développement perturbatif (i.e. en $\frac{1}{N}$) des fonctions de corrélations issues d'intégrales de matrices aléatoires hermitiennes. Si ce domaine d'application reste évidemment encore d'actualité, la récurrence a été rapidement généralisée à n'importe quelle ``courbe spectrale'' qu'elle soit issue d'une intégrale de matrices ou non. Cette généralisation a permis son application dans des domaines de géométrie énumérative a priori très éloignés du domaine des probabilités. On citera par exemple la gravité $2D$ \cite{BergereEynard,DGZ}, les volumes de Weil-Petersson \cite{BookEynard,WeilVolumes}, la théorie des n\oe uds \cite{Knot1,Knot2}, les partitions planes \cite{Partitions, KenyonOkunkovSheffield}, les nombres de Hurwitz \cite{Okounkov2,EMS}, les invariants de Gromov-Witten \cite{Kontsevich,Okounkov2} avec la célèbre conjecture de Bouchard, Klemm, Mariño et Pasquetti \cite{BKMP} résolue dans \cite{TopologicalVertex,BKMPSolution,Panda} pour des Calabi-Yau torique de dimension 3, puis prouvée de façon plus générale dans \cite{BKMPSolution2} et évoquée dans d'autres articles \cite{TopologicalVertex,MMTopSting1,MMTopSting2}.

En parallèle, la récurrence topologique s'est également rapprochée du domaine des systèmes intégrables \cite{Iso1,Hitchin,BE10,BleherEynard,WitteForrester,CIK,BEmom,BEH4,Iso2}. En effet, il est maintenant bien connu que les intégrales de matrices hermitiennes forment des processus déterminantaux et sont des exemples de systèmes intégrables. En particulier, les fonctions de partitions des intégrales de matrices sont des fonctions tau de ces systèmes intégrables (KP, KdV) \cite{Iso1,mKDV,VieuxDyson,Iso2} et avec les fonctions de corrélations elles doivent donc obéir à certaines séries infinies d'équations (Equations d'Hirota, Formule de Sato, Relations de Plücker) qui contraignent fortement le modèle mais permettent également de le résoudre plus facilement. Ces résultats sont également intéressants car ils fournissent un exemple concret de systèmes intégrables (à l'aide des intégrales de matrices aléatoires). Plus récemment, ce lien a permis de s'intéresser aux systèmes intégrables par le prisme de la récurrence topologique et en particulier dans l'approche des systèmes intégrables pour lesquels une paire de Lax est connue \cite{ER14,WitteForrester,2m1}. Cette thématique émergente est abordée plus en détails dans la deuxième partie de cette Habilitation à Diriger des Recherches (HDR) et constitue à l'heure actuelle un champ de recherche très actif.

Devant le champ considérable d'applications de la récurrence topologique et pour éviter une présentation trop superficielle d'un grand nombre de sujets, j'ai choisi de ne présenter en détails que deux aspects spécifiques. Le premier chapitre est consacré à la présentation générale de la récurrence topologique dans le cadre des intégrales de matrices aléatoires. Ce chapitre se veut volontairement pédagogique, afin de fixer les notations et de présenter l'intérêt initial et le formalisme inhérent à la récurrence topologique. On y trouve également quelques exemples de tests de la théorie. Le second chapitre est plus proche des domaines actuels de recherche puisqu'il concerne l'application de la récurrence topologique à l'étude des systèmes intégrables définis par une paire de Lax. Plus technique, ce chapitre présente également certaines questions encore non résolues à l'heure actuelle. En revanche, les applications à la géométrie énumérative évoquées rapidement ci-dessus ne seront pas développées dans cette HDR. Ce choix s'explique par le fait que les notions géométriques nécessaires sont lourdes à introduire et à expliquer. Par ailleurs, d'excellentes synthèses sur ces sujets existent déjà comme par exemple \cite{BookEynard,Marino}.

En proposant à la fois des applications probabilistes et des liens avec les systèmes intégrables, l'objectif de cette habilitation est ainsi de souligner que la récurrence topologique, couplée à la méthode des équations de boucles, constitue un outil efficace dans ces deux domaines.

\selectlanguage{french}
\chapter{Récurrence topologique et intégrales de matrices}
 \label{chap1}
\thispagestyle{fancy}

Historiquement, la récurrence topologique a été introduite pour la résolution des intégrales de matrices hermitiennes \cite{MehtaBook,AGZ} dans le cadre de l'étude des modèles de théorie des champs en deux dimensions \cite{DGZ,DysonGas}. Elle a depuis connu de nombreux développements par le biais des systèmes intégrables, des applications à la théorie des cordes topologiques et à la géométrie énumérative. En parallèle, la récurrence topologique a également fourni des alternatives aux techniques classiques d'analyse (problèmes de Riemann-Hilbert \cite{KMcL,NotesBleher,Kap}, asymptotiques de polynômes orthogonaux \cite{MarcoPaths,BEH4}, déterminants de Toeplitz \cite{Szego,Widom,Widom2,BS,BW,FishHart,BM,BO,DJ,Krasovsky}, méthodes des grandes déviations, etc.) ainsi que pour l'étude du spectre de matrices aléatoires. Avec la découverte du phénomène d'universalité locale en lien avec les systèmes intégrables (Cf. \cite{MehtaBook,TT,TracyWidom,TW,Johansson,These}), la communauté mathématique s'est ainsi beaucoup intéressée aux fonctions de corrélations des valeurs propres, dans le centre de la distribution ou sur ses bords. Dans le cas des intégrales de matrices aléatoires, ces phénomènes d'universalité correspondent à l'étude des doubles limites d'échelle qui permettent de se focaliser sur le voisinage d'un point précis de la distribution limite et d'en étudier les corrélations locales. Par contrecoup, l'étude directe des fonctions de partition et des probabilités associées a été un peu délaissée bien qu'elle puisse parfois résoudre certains problèmes intéressants pour les probabilistes ou les analystes comme le calcul de développements asymptotiques de certains déterminants de Toeplitz \cite{ToeplitzNew} ou l'étude de temps de retour d'un grand nombre de particules diffusant sur le cercle et présentant une interaction Coulombienne \cite{UnitaryMarchal}. L'objectif de ce chapitre est donc d'étudier plus spécifiquement les fonctions de partition de certains modèles et leurs développements asymptotiques en $\frac{1}{N}$. Le chapitre commence naturellement avec l'étude d'intégrales de matrices hermitiennes, dont l'important cas Gaussien, puis montre que la méthode des équations de boucles et de la récurrence topologique peuvent également être utilisées pour d'autres types d'intégrales comme des intégrales hermitiennes avec bords ou des intégrales de matrices unitaires. L'objectif est ainsi de présenter les grandes possibilités de la méthode générale, ses principales difficultés et avantages avec une variété d'exemples représentatifs des possibilités offertes par la récurrence topologique dans ce contexte.

\section{Intégrales de matrices hermitiennes sans bord}
\subsection{Définition des intégrales hermitiennes}

L'étude des intégrales de matrices hermitiennes correspond à l'étude de la fonction de partition suivante:
\beq \label{IntHermitienne}\td{Z}_N=\int_{\mathcal{H}_N} dM e^{-\frac{N}{T}\Tr V(M)} \eeq
où $\mathcal{H}_N$ est l'ensemble des matrices hermitiennes de taille $N$. L'intégration est donnée par le produit des mesures de Lebesgue des entrées indépendantes des matrices hermitiennes:
\beqq dM=\left(\prod_{i<j}^N \text{dRe}(M_{i,j}) \text{dIm}(M_{i,j})\right)\left(\prod_{i=1}^N \text{dRe}(M_{i,i})\right)\eeqq
Le paramètre $T$ est communément appelé ``température'' tandis que le potentiel $x\mapsto V(x)$ est généralement choisi polynomial (le premier terme $t_1$ pouvant être fixé à $0$ par translation):
\beq V(x)=\sum_{i=2}^\infty t_i x^i\eeq
L'intégrale \eqref{IntHermitienne} n'est pas toujours convergente mais elle l'est systématiquement si le potentiel $V(x)$ est polynomial, de degré pair et avec un coefficient dominant positif qui représente la grande majorité des situations étudiées dans la littérature. Néanmoins, pour l'étude de certains problèmes issus de la géométrie énumérative, il est possible de considérer l'intégrale \eqref{IntHermitienne} pour des potentiels où l'intégrale ne converge pas et dans ce cas, seuls les développements formels autour du potentiel Gaussien (à la Feyman) sont considérés. Afin d'en faciliter la lecture, les potentiels seront toujours choisis comme donnant des intégrales convergentes dans ce document.
L'intérêt des matrices hermitiennes est qu'elles sont diagonalisables avec des valeurs propres réelles. La diagonalisation induit une mesure sur les valeurs propres qui est donnée classiquement par \cite{MehtaBook}
\beq \label{IntHermitienneDiag}\td{Z}_N=c_N\int_{\mathbb{R}^N} d\lambda_1\dots d\lambda_N \Delta(\lambda_1,\dots,\lambda_N)^2 e^{-\frac{N}{T}\underset{i=1}{\overset{N}{\sum}} V(\lambda_i)}\overset{\text{def}}{=}c_NZ_N \eeq
 où $\Delta(\lambda_1,\dots,\lambda_N)=\underset{i<j}{\overset{N}{\prod}}(\lambda_i-\lambda_j)$ est le déterminant de Vandermonde. La constante de normalisation $c_N$ peut être calculée explicitement et ne dépend pas du potentiel $V$. Elle se calcule en croisant les résultats des intégrales de Selberg avec les calculs explicites du modèle Gaussien (Cf. section \ref{Gaussien}). On peut également restreindre la définition de l'intégrale \eqref{IntHermitienneDiag} à un sous-ensemble mesurable (en pratique une réunion finie d'intervalles) de $\mathbb{R}$ et dans ce cas on appelle ``bords durs'' les points finis délimitant les intervalles du domaine d'intégration et appartenant au support de la densité limite des valeurs propres. Ces cas seront traités dans le paragraphe \ref{BordDur}.    

Physiquement, l'intégrale précédente montre que les valeurs propres sont soumises d'une part à une force d'attraction dans les minima du potentiel $V$ et d'autre part à une force de répulsion interne liée à la présence du déterminant de Vandermonde. Afin d'étudier les propriétés des valeurs propres, il est usuel de définir les fonctions de corrélation suivantes:

\begin{definition}[Fonctions de corrélation]\label{FonctionCorrelation} Les fonctions de corrélation non connexes (n.c.) associées à la fonction de partition \eqref{IntHermitienneDiag} sont définies par:
\beaa W_1^{n.c.}(x)&=&\left<\sum_{i=1}^N \frac{1}{x-\lambda_i}\right>\cr
W_n^{n.c.}(x_1,\dots,x_n)&=&\left<\sum_{i_1,\dots,i_n=1}^N \frac{1}{x_1-\lambda_{i_1}}\dots \frac{1}{x_n-\lambda_{i_n} }\right> \,,\,\forall\, n\geq 2 \cr
\eeaa
où l'on désigne la valeur moyenne d'une fonction des valeurs propres par:
\beqq \left< f(\lambda_1,\dots, \lambda_n)\right>=\frac{1}{Z_N}\int_{\mathbb{R}^N} d\lambda_1\dots d\lambda_N f(\lambda_1,\dots,\lambda_N) \Delta(\lambda_1,\dots,\lambda_N)^2 e^{-\frac{N}{T}\underset{i=1}{\overset{N}{\sum}} V(\lambda_i)}\eeqq
Par ailleurs, on définit également les fonctions de corrélation connexes par:
\bea W_1(x)&=&W_1^{n.c.}(x)=\left<\sum_{i=1}^N \frac{1}{x-\lambda_i}\right>\cr
W_2(x_1,x_2)&=&W_2^{n.c.}(x_1,x_2)-W_1^{n.c.}(x_1)W_1^{n.c.}(x_2)\cr
W_3(x_1,x_2,x_3)&=&W_3^{n.c.}(x_1,x_2,x_3)-W_1(x_1)W_2(x_2,x_3)-W_1(x_2)W_2(x_1,x_3)\cr
&&-W_1(x_3)W_2(x_1,x_2)-W_1(x_1)W_1(x_2)W_1(x_3)\cr
\normalfont{\text{etc.}}&&
\eea
ou d'une façon plus générale par la relation inverse:
\beqq W_n^{n.c.}(x_1,\dots,x_n)= \sum_{\mu\, \vdash \{x_1,\dots,x_n\}}\prod_{i=1}^{\text{l}(\mu)}  W_{|\mu_i|}(\mu_i)\eeqq
où la dernière somme est prise sur l'ensemble des partitions $\mu$ de l'ensemble $\{x_1,\dots,x_n\}$.
\end{definition}

L'intérêt des fonctions de corrélation résulte dans le fait que le développement asymptotique à $x_i\to \infty$ permet d'obtenir les différents moments et corrélations entre les valeurs propres. D'un point de vue analytique, les fonctions de corrélation ne sont a priori définies que pour des $x_i$ situés en dehors de l'axe réel (lieu où vivent les valeurs propres). Pour un potentiel $V$ quelconque, obtenir les valeurs exactes de $Z_N$, de $W_1(x)$ ou de $W_n(x_1,\dots,x_n)$ est en général impossible. En revanche, il est souvent possible de déterminer un asymptotique de ces quantités lorsque $N\to \infty$; c'est précisément le rôle de la récurrence topologique que de déterminer ces asymptotiques. L'ingrédient de départ essentiel de la récurrence topologique est la connaissance de la distribution limite des valeurs propres. Cela est résumé dans le théorème suivant (Cf. \cite{NotesBleher} pour une synthèse) issu de la théorie du potentiel:

\begin{theorem}[Densité limite d'équilibre]\label{DensiteLimite} Soit $d\nu_N(x)=\frac{1}{N}\underset{i=1}{\overset{N}{\sum}} \delta(x-\lambda_i)$ la mesure empirique des valeurs propres associée à \eqref{IntHermitienneDiag} alors $d\nu_N$ converge en loi vers une mesure appelée mesure d'équilibre $d\nu_\infty$ qui est absolument continue par rapport à la mesure de Lebesgue et supportée par un nombre fini d'intervalles: $\text{{\normalfont Supp}}(d\nu_\infty)=\underset{i=1}{\overset{d}{\bigcup}} [a_i,b_i]$. Par ailleurs, sur son support la densité d'équilibre est donnée par:
\beq d\nu_\infty(x)=\frac{1}{2\pi i} h(x)\sqrt{\prod_{i=1}^d (x-b_i)(x-a_i)}dx\overset{\text{def}}{=}\frac{1}{2\pi i} h(x)\sqrt{R(x)}dx \eeq 
où $h(x)$ est un polynôme de degré $\text{deg}(h)=\text{deg}(V)-d-1$. De plus, ses coefficients sont contraints par les relations:
\bea \label{Coeff} h(x)&=&\underset{x\to \infty}{\text{{\normalfont Pol}}}\left(\frac{V'(x)}{\sqrt{R(x)}}\right)\cr
2&=&\Res_{z\to \infty} h(z)\sqrt{R(z)}\cr
0&=&\int_{b_i}^{a_{i+1}} d\nu_\infty(x) \,\,,\,\,\forall\, 1\leq i\leq d-1
\eea
On dit enfin que la densité limite est non-critique si $h$ est strictement positif sur chacun des intervalles $[a_i,b_i]$.
\end{theorem}

Notons en particulier que \eqref{Coeff} permet d'obtenir le même nombre d'équations (non-linéaires) que le nombre de coefficients du polynôme $h$ à déterminer. Ainsi génériquement (i.e. pour des valeurs génériques des coefficients du potentiel $V$), ces conditions suffisent pour déterminer la densité d'équilibre. En revanche, rien n'exclut a priori que pour des valeurs spécifiques des coefficients du potentiel, ces conditions ne déterminent le polynôme $h$ que partiellement (par exemple, lorsque la densité d'équilibre devient critique).
Par ailleurs, en pratique si l'existence d'une densité limite d'équilibre est garantie, il est souvent compliqué de la déterminer explicitement. En effet, dès que le support est composé de plus d'un seul segment, la détermination de la densité limite même avec l'aide de \eqref{Coeff} nécessite de déterminer les ``fractions de remplissage'', c'est-à-dire la proportion limite des valeurs propres qui vont s'accumuler sur chacun des segments: $\epsilon_i=\int_{[a_i,b_i]}d\nu_\infty$. Ce calcul est équivalent à la résolution de la dernière condition de \eqref{Coeff} qui, hormis lors de l'existence de symétries supplémentaires, est souvent impossible en pratique. Ainsi la très grande majorité des cas où les calculs sont possibles correspondent au cas ``à une coupure'' c'est-à-dire au cas où la densité d'équilibre est supportée par un seul segment ($d=1$). Dans ce cas, les deux premières équations de \eqref{Coeff} permettent le calcul complet de $a_1,b_1$ et des coefficients de $h$. Notons que ce cas est loin d'être isolé puisque \textbf{si le potentiel $V(x)$ est convexe} alors il est démontré dans \cite{KuijMcLau,BorotGuionnetKoz} que la \textbf{densité limite d'équilibre est nécessairement non-critique et supportée par un seul intervalle}. 

\subsection{Méthode des équations de boucles}

La méthode des équations de boucles, connue également sous le nom d'équations de Schwinger-Dyson, constitue un ensemble d'équations satisfaites par les fonctions de corrélations. Ces équations de boucles s'obtiennent simplement en étudiant l'invariance des intégrales sous certains difféomorphismes comme par exemple les translations: $x\to x+\epsilon$. De façon alternative, elles peuvent s'obtenir en procédant à des intégrations par parties judicieusement choisies. Dans le cas où le domaine d'intégration ne comporte pas de bord, ces équations peuvent être trouvées dans \cite{EO,BEO13,Hardedges,Lie,LoopEquationBeta2MM,These}. Dans le cas hermitien, la première équation de boucles peut ainsi s'écrire:
\beq \label{Loop1SansBord}\frac{T^2}{N^2}W_1(x)^2-\frac{T}{N}V'(x)W_1(x)+\frac{T^2}{N^2}W_2(x,x)=-P_1(x)\overset{\text{def}}{=}-\frac{T}{N}\left<\sum_{i=1}^N\frac{V'(x)-V'(\lambda_i)}{x-\lambda_i}\right>\eeq

Notons qu'il s'agit d'une équation \textbf{exacte}. En revanche, pour résoudre cette équation de boucle et retrouver les résultats précédents, il faut supposer ou démontrer l'existence d'un développement en $\frac{1}{N}$ des fonctions de corrélations de la forme:

\begin{assumption}[Existence d'un développement topologique]\label{AssumpDevPertur} On suppose que les fonctions de corrélation et la fonction de partition admettent un développement en $\frac{1}{N}$ de la forme suivante:
\bea \label{sauv}W_1(x)&=&\sum_{g=0}^\infty W_1^{(g)}(x)\left(\frac{N}{T}\right)^{1-2g}\cr
W_n(x_1,\dots,x_n)&=&\sum_{g=0}^\infty W_n^{(g)}(x_1,\dots,x_n)\left(\frac{N}{T}\right)^{2-n-2g} \,\,,\,\,\forall\, n\geq 2\cr
\ln Z_N&=& \sum_{g=0}^\infty Z^{(g)} \left(\frac{N}{T}\right)^{2-2g}
\eea
Ce développement est appelé ``développement perturbatif'' ou ``développement topologique''.
\end{assumption}

\begin{remark}\label{RemarqueConvergence}La notion de développement topologique présentée ci-dessus a fait l'objet de nombreuses confusions dans la littérature. En effet, il existe deux façons opposées de considérer l'hypothèse \ref{AssumpDevPertur}. La première, utilisée en combinatoire et en physique théorique, consiste à considérer les développements de façon purement formelle (on parle alors de ``développements formels'') , c'est-à-dire à supposer qu'ils existent et à étudier les relations algébriques entre les coefficients des développements formels en $\frac{1}{x}$ des fonctions de corrélations. Dans ce cadre, \eqref{sauv} constitue donc une hypothèse (qui ne nécessite ainsi pas de démonstration) et les égalités sont à comprendre comme des égalités algébriques entre séries formelles. En revanche, lors de l'étude d'intégrales convergentes dans le cadre de problèmes issus des probabilités ou de l'analyse, les développements précédents sont à comprendre de façon convergente, i.e. comme:
\beq \label{SensConvergence} W_n(x_1,\dots,x_n)=\sum_{g=0}^m W_n^{(g)}(x_1,\dots,x_n)\left(\frac{N}{T}\right)^{2-n-2g} +o\left(N^{2-n-2m}\right) \,\,,\,\, \forall\, m>0\eeq
et on parle alors de ``développements convergents''. Dans ce cas, rien ne garantit a priori que les fonctions de corrélation et la fonction de partition admettent des développements d'une telle nature et ainsi il est nécessaire de démontrer au préalable l'existence de tels développements asymptotiques. Notons également que le terminologie de ``développements convergents'' ne correspond pas à l'existence d'un développement en série entière en $\frac{1}{N}$ des fonctions de corrélations puisque \eqref{SensConvergence} n'exclut nullement la présence de termes exponentiellement petits (i.e. en $O\left(e^{-\alpha N}\right)$). Malgré cela, la notation d'égalité telle que présentée dans \eqref{sauv} s'est imposée dans la littérature même si elle doit être comprise dans le sens de \eqref{SensConvergence} et non dans le sens d'un développement en série entière (même si le signe $``=''$ le suggèrerait).
\end{remark}

Dans ce chapitre, nous nous intéressons aux cas d'intégrales convergentes issus de problèmes probabilistes et il est donc nécessaire de démontrer que les fonctions de corrélations admettent un tel développement topologique dans le sens précisé en remarque \ref{RemarqueConvergence}. Notons que la fonction de partition ne peut admettre un développement topologique que si l'intégrale correspondante est correctement normalisée (sinon il faut rajouter un préfacteur au développement qui peut être calculé à l'aide du cas Gaussien et des intégrales de Selberg (Cf. section \ref{Gaussien})). D'une façon générale, la fonction de partition pose un problème de constantes d'intégration qui n'est que très rarement évoqué dans la littérature.
Comme indiqué précedemment, dans le cadre de problèmes analytiques ou probabilistes, l'existence d'un tel développement topologique est loin d'être garantie quel que soit le potentiel $V$. En effet, si l'existence du premier ordre $W_1^{(0)}(x)$ est en général assurée (par exemple par le théorème \ref{DensiteLimite}), l'existence d'un développement convergent (dans le sens \eqref{SensConvergence}), pair en $g$ et commençant à l'ordre $N^{2-n}$ est loin d'être triviale. Il existe néanmoins des théorèmes généraux permettant de s'assurer de la validité de l'existence d'un tel développement topologique. Par exemple, le théorème principal de \cite{BG} couvre un large éventail de situations y compris en dehors du cas hermitien comme nous le verrons plus loin. L'utilisation de ce théorème nous assure que l'hypothèse \ref{AssumpDevPertur} est valide. Néanmoins, il nécessite de vérifier les hypothèses suivantes (Hypothèses $1.1$ de \cite{BG}):

\begin{proposition}[Conditions suffisantes pour l'existence d'un développement topologique]\label{ConditionsTopologiques} Le résultat principal de \cite{BG} montre que si le modèle de matrices hermitiennes satisfait les hypothèses suivantes:
\begin{itemize}
\item \underline{Continuité du potentiel}: Le potentiel $V$ est continu sur le domaine d'intégration $\mathcal{D}\subset \mathbb{R}$.
\item \underline{Confinement}: Si $\pm \infty$ appartiennent au domaine d'intégration alors on doit vérifier que:
$\underset{x\to \pm \infty}{\liminf} \frac{V(x)}{2\ln |x|}>1$.
\item \underline{Support à une coupure}: La distribution limite d'équilibre $d\mu_{\text{eq}}$ est supportée par un seul intervalle $[\alpha_-,\alpha_+]$.
\item \underline{Non-criticalité}: La distribution limite d'équilibre $d\mu_{\text{eq}}$ est strictement positive dans l'intérieur de son support et se comporte comme $d\mu_{\text{eq}}(x)=c_\alpha\sqrt{x-\alpha}dx+ o\left(\sqrt{x-\alpha}\right)$ avec $c_\alpha\neq 0$ au voisinage d'un bord mou ou comme $d\mu_{\text{eq}}(x)=\frac{c_\alpha}{\sqrt{x-\alpha}}dx+ o\left(\frac{1}{\sqrt{x-\alpha}}\right)$ avec $c_\alpha\neq 0$ au voisinage d'un bord dur.  
\item \underline{Condition de minimalité}: La fonction $x\mapsto \frac{1}{2}V(x)-\int_{\alpha_-}^{\alpha_+} \ln|x-\xi|d\mu_{\text{eq}}(\xi)$ définie sur $\mathcal{D}\setminus [\alpha_-,\alpha_+]$ atteint ses minima uniquement aux bords $\alpha_+$ et $\alpha_-$ de la distribution limite.
\item \underline{Régularité du potentiel}: Le potentiel $V$ est holomorphe dans un voisinage du support de la distribution d'équilibre.
\end{itemize} 
alors les fonctions de corrélations et la fonction de partition (sous réserve de normalisation adéquate pour cette dernière) admettent un développement topologique de la forme \eqref{sauv} (et à comprendre dans le sens de \eqref{SensConvergence}).
\end{proposition}

Vérifier l'ensemble de ces hypothèses peut être fastidieux mais il existe une condition suffisante, satisfaite dans la plupart des cas pratiques, qui facilitent grandement la discussion. En effet, \textbf{si le potentiel $V$ est continu et convexe sur le domaine d'intégration alors la distribution limite d'équilibre est supportée par un seul intervalle, ne comporte pas de points isolés et est non-critique et la condition de minimalité est satisfaite}. Il ne reste donc qu'à vérifier les conditions de régularité du potentiel (holomorphe dans un voisinage du support de la distribution d'équilibre) et de confinement à l'infini qui sont en général relativement simples. Dans le cas où le développement topologique existe, on peut alors projeter \eqref{Loop1SansBord} à l'ordre dominant ($N^0$) pour obtenir:
\beq W_1^{(0)}(x)^2-V'(x)W_1^{(0)}(x)+P_1^{(0)}(x)=0\eeq
Il est alors usuel de translater $W_1^{(0)}(x)$ de $\frac{1}{2}V'(x)$ en définissant:
\beq \label{y} y(x)\overset{\text{déf}}{=}W_1^{(0)}(x)-\frac{1}{2}V'(x)\eeq
Pour obtenir:
\beq \label{CourbeSpectrale} y(x)^2=\frac{V'(x)^2}{4}-P_1^{(0)}(x) \eeq
Cette équation définit la ``courbe spectrale'' associée au modèle de matrice hermitien \eqref{IntHermitienneDiag}. Il s'agit d'une équation hyperelliptique où le membre de droite est un polynôme. Notons que ce résultat est équivalent aux deux premières équations de \eqref{Coeff} puisque l'on a:
\beq W_1^{(0)}(x)=\int_{z\in \text{Supp}(d\nu_\infty)} \frac{d\nu_{\infty}(z)}{x-z}=\text{Stieltjes}(d\nu_\infty)(x)\eeq
Ainsi, dans le cadre d'intégrales convergentes, la densité limite $d\nu_{\infty}(x)$ permet d'obtenir la courbe spectrale par sa transformée de Stieltjes. En revanche, dans le cadre d'intégrales et de développements formels, la densité limite n'ayant pas de sens a priori, seules les relations algébriques \eqref{CourbeSpectrale} sont disponibles ce qui ne détermine pas complètement la courbe spectrale. Dans cette approche les fractions de remplissage ne sont pas déterminées et dans le cas d'intégrales formelles il peut être intéressant de choisir ces fractions de remplissage non pas de manière ``dynamique'' (comme le fait la troisième équation de \eqref{Coeff}) mais de les supposer fixées à certaines valeurs d'intérêt.

\subsection{La récurrence topologique}
Une fois la densité limite d'équilibre déterminée (par \eqref{Coeff} ou par \eqref{CourbeSpectrale}), on peut la réécrire sous la forme $y(x)^2=\text{Pol}(x)$, c'est-à-dire la donnée d'une surface de Riemann à deux feuillets de genre $\genus=d-1$. Plus généralement, on peut définir:

\begin{definition}[Courbe Spectrale]\label{DeffCourbeSpec} On appelle courbe spectrale la donnée d'une surface de Riemann $\Sigma$ de genre $\genus$ munie d'une base symplectique de cycles non-contractibles $\left(\mathcal{A}_i\right)_{1\leq i\leq \genus}$ et $\left(\mathcal{B}_i\right)_{1\leq i\leq \genus}$, de deux fonctions $(x(z),y(z))$ ($z\in \Sigma$) méromorphes sur $\Sigma$ (vérifiant donc une équation polynomiale $P(x,y)=0$) et d'une bi-forme différentielle symétrique $\omega_2^{(0)}(z_1,z_2)$ holomorphe sur $\Sigma \times \Sigma$ à l'exception d'un pôle double sur la diagonale $z_1=z_2$ de la forme $\omega_2^{(0)}(z_1,z_2)\underset{z_1\to z_2}{=}\frac{dz_1 dz_2}{(z_1-z_2)^2}+\text{holo.}$ et normalisée sur les cycles par $\oint_{\mathcal{A}_i}\omega_2^{(0)}(z_1,z_2)=0$.
\end{definition}

A partir d'une courbe spectrale on peut alors définir la récurrence topologique:

\begin{definition}[Récurrence topologique]\label{RecTop} Soit $\mathcal{C}=(\Sigma,x(z),y(z),\omega_2^{(0)}(z_1,z_2))$ une courbe spectrale.
\begin{itemize} \item On appelle points de ramifications, notés $(a_i)_{1\leq i\leq r}$ les points pour lesquels la forme différentielle $dx=x'(z)dz$ s'annule. \footnote{Les images $x_i=x(a_i)$ des points de ramifications sont appelés points de branchements bien que la littérature, en particulier anglophone, confonde souvent les deux terminologies}. On dit que la courbe spectrale est régulière si les points de ramifications sont des zéros simples de $dx$. Dans le cas contraire, on dit que la courbe spectrale est singulière.
\item Si $\mathcal{C}$ est régulière, on définit $\bar{z}$ l'involution localement définie dans un voisinage des points de ramifications vérifiant $x(z)=x(\bar{z})$
\item Les fractions de remplissage $(\epsilon_i)_{1\leq i\leq \genus}$ par $\epsilon_i=\frac{1}{2\pi i}\oint_{\mathcal{A}_i} ydx$
\item On note $\left(\alpha_i\right)_{1\leq i\leq s}$ les pôles de $ydx$. On définit alors $t_i=\underset{z\to \alpha_i}{\Res} ydx$ parfois appelé la ``température'' du pôle.
\item On définit alors les différentielles d'Eynard-Orantin $\left(\omega_n^{(g)}\right)_{n\geq 1, g\geq 0}$ par (on note $\mathbf{z_n}=(z_1,\dots,z_n)$):
\bea \label{RecTopEq} \omega_1^{(0)}&=&0\cr
\omega_{n+1}^{(g)}(z,\mathbf{z_n})&=&\sum_{i=1}^r \Res_{q\to a_i}\frac{dE_q(z)}{(y(q)-y(\bar{q}))dx(q)}\Big[ \omega_{n+2}^{(g-1)}(q,q,\mathbf{z_n})\cr
&&+\sum_{m=0}^g \sum_{I\subset \mathbf{z_n}}\omega_{|I|+1}^{(m)}(q,I)\,\omega_{|\mathbf{z_n}\setminus I|+1}^{(g-m)}(q,\mathbf{z_n}\setminus I)\Big]\cr
&&\eea
où $dE_q(z)=\frac{1}{2}\int_q^{\bar{q}} \omega_2^{(0)}(q,z)$.
\item Les énergies libres ou invariants symplectiques $\left(F^{(g)}\right)_{g\geq 2}$ sont définis comme:
\beq F^{(g)}=\frac{1}{2-2g}\sum_{i=1}^r \Res_{q\to a_i}\Phi(q)\,\omega_1^{(g)}(q)\eeq
où $\Phi(q)=\int^q ydx$ (le choix de la borne inférieure ne jouant aucun rôle). On notera également $F^{(g)}$ comme $\omega_0^{(g)}$.
\item Les cas particuliers de $F^{(0)}$ et de $F^{(1)}$ sont obtenus par des formules spécifiques (Eq. $4.11$ et $4.13$ de \cite{EO} et également une formule alternative pour $F^{(1)}$ dans \cite{F1})
\end{itemize} 
\end{definition}

On voit donc que les différentielles d'Eynard-Orantin sont définies par récurrence à partir de $\omega_2^{(0)}$ et des divers ingrédients de la courbe spectrale. On observe également que la donnée complète d'une surface de Riemann globale et de fonctions $x(z),y(z)$ ou de $\omega_2^{(0)}$ n'est pas nécessaire pour le calcul de la récurrence en elle même. En effet, puisqu'il s'agit de calculs de résidus autour des points de ramifications, seuls des développements formels autour de chacun de ces points sont nécessaires en pratique et une reformulation uniquement à partir de ces ``germes'' est possible \cite{BEO13}. Elle a été appliquée à de nombreuses courbes spectrales (Cf. \cite{ListSpecCurve} pour une liste non exhaustive). L'intérêt des différentielles d'Eynard-Orantin et des invariants symplectiques provient du fait qu'ils satisfont de très nombreuses propriétés (Cf. \cite{EO}):

\begin{proposition}[Propriétés de la récurrence topologique]\label{PropTopRec} Les différentielles d'Eynard-Orantin satisfont les propriétés suivantes:
\begin{itemize}\item Symétrie: $\omega_{n}^{(g)}(\mathbf{z_n})$ est une fonction symétrique de ses variables.
\item Normalisation: Pour $1\leq i \leq \genus$, $n\geq 1$, $g\geq 0$: $\int_{\mathcal{A}_i}\omega_{n}^{(g)}(\mathbf{z_n})=0$
\item Formule d'inversion (également appelée formule du ``dilaton''): Pour $n\geq 0$ et $g\geq 0$ on a:
\beqq \omega_{n}^{(g)}(\mathbf{z_n})=\sum_{i=1}^r\Res_{q\to a_i} \Phi(q)\omega_{n+1}^{(g)}(q,\mathbf{z_n}) +\delta_{n=1,g=0}\, y(z_1)dx(z_1)\eeqq
\item Equations de boucle: Pour $n\geq 1$ et $g\geq 0$, la fonction définie par:
\bea \label{LoopEquations}&&P_{n}^{(g)}(x(p),\mathbf{z_n})\overset{\text{def}}{=}\frac{1}{dx(p)^2}\sum_{p^i \,/\, x(p^i)=x(p)\, ,p^i\neq p}\Big[ -2y(p^i)dx(p)\omega_{n+1}^{(g)}(p^{i},\mathbf{z_n})\cr
&&+\omega_{n+2}^{(g-1)}(p^i,p^i,\mathbf{z_n})+\sum_{m=0}^g \sum_{I\subset \mathbf{z_n}}\omega_{|I|+1}^{(m)}(p^i,I)\omega_{|\mathbf{z_n}\setminus I|+1}^{(g-m)}(p^i,\mathbf{z_n}\setminus I)\Big]\cr
&&
\eea
est une fonction rationnelle en $x$ sans singularité aux points de branchements.
\item Invariance symplectique des $\left(F^{(g)}\right)_{g\geq 0}$ (Cf. \cite{EO} et précisée dans \cite{XY}). Lorsque $z\mapsto y(z)$ et $z\mapsto dx(z)$ sont des fonctions méromorphes et que l'on définit:
\beq \label{CorrectionFg} \hat{F}^{(g)}=F^{(g)}+\frac{1}{2-2g}\sum_{i=1}^s t_i\int_*^{\alpha_i} \omega_1^{(g)}\eeq
alors les $\left(\hat{F}^{(g)}\right)_{g\geq 0}$ sont invariants par toute transformation symplectique de la courbe spectrale, i.e. tout changement $(\td{x},\td{y})=(f(x,y),g(x,y))$ vérifiant $d\td{x}\wedge d\td{y}=dx\wedge dy$.
\end{itemize}
\end{proposition}  

A priori la récurrence topologique, qui est une construction purement algébrique, n'a pas grand chose à voir avec les intégrales de matrices aléatoires. Néanmoins le lien devient évident lorsque l'on observe le fait que les équations de boucle de la proposition \ref{PropTopRec} possèdent la même structure que les équations de boucle des modèles de matrices hermitiennes (le cas général étant un modèle à deux matrices diagonalisées grâce à la formule d'Harish-Chandra-Itzykson-Zuber). Plus précisément, les équations de boucle du modèles de matrices hermitiennes impliquent celles de la récurrence topologique \cite{Eynard004,Eynard005}. Notons ici que les équations de boucle admettent en général beaucoup (voire une infinité) de solutions et que parmi ces solutions, la récurrence topologique produit une solution ayant certaines propriétés particulières (i.e. coefficients des séries n'ayant des pôles qu'uniquement aux points de ramifications et avec un développement de la forme topologique \eqref{AssumpDevPertur}). En particulier, lorsqu'un modèle de matrices possède un développement topologique, on peut montrer que les deux solutions coïncident de la façon suivante (Section $10$ de \cite{EO}):

\begin{theorem}[Récurrence topologique et modèle de matrices hermitiennes]\label{RecTopMM} Si un modèle (à une ou deux matrices) de matrices hermitiennes (par exemple \eqref{IntHermitienneDiag}) admet un développement topologique dans le sens de \ref{AssumpDevPertur}, alors les développements des fonctions de corrélation $W_n^{(g)}$ s'identifient avec les différentielles d'Eynard-Orantin calculées sur la courbe spectrale donnée par la transformée de Stieltjes de la densité limite d'équilibre de la façon suivante\footnote{Dans cette habilitation, la notation $\propto$ signifie que pour tout $g\geq 0$, il existe une constante $C_g$ ne dépendant ni du potentiel ni de $T$ telle que $Z^{(g)}=-F^{(g)}+C_g$. Ces constantes sont liées au choix de normalisation des fonctions de partition. Des exemples sont fournis dans les sections \ref{Gaussien},\ref{Seccc},\ref{Secccc}} (en notant $x_i=x(z_i)$):
\beaa W_n^{(g)}(x_1,\dots,x_n)\,dx_1\dots dx_n&=&\omega_n^{(g)}(z_1,\dots,z_n) \,\,,\,\,\forall\, n\geq 1, g\geq 0\cr
Z^{(g)}&\propto& -F^{(g)} \,\,,\,\, \forall\, g\geq 0
\eeaa
On peut ainsi reconstruire les fonctions de corrélation par:
\beaa W_n(x_1,\dots,x_n)\,dx_1\dots dx_n&=&\sum_{g=0}^{\infty} \omega_n^{(g)} \left(\frac{N}{T}\right)^{2-n-2g}\cr
\eeaa
\end{theorem}

On voit donc l'intérêt de la récurrence topologique dans l'étude des matrices aléatoires hermitiennes. Dès que le potentiel est convexe, la combinaison des résultats précédents nous assure que les fonctions de corrélations peuvent se calculer par l'application de la récurrence topologique sur la transformée de Stieltjes de la densité limite, qui est alors nécessairement une surface de Riemann de genre $0$. En d'autres termes, une fois la densité limite obtenue, toutes les corrélations des valeurs propres sont entièrement déductibles par une simple récurrence. Le cas de la fonction de partition est similaire mais il est nécessaire de normaliser correctement $c_N$ pour que le développement topologique existe (sinon il peut y avoir des termes en $N\ln N$ et $\ln N$ supplémentaires). De plus, le choix de normalisation de la fonction de partition n'autorise une identification de $Z^{(g)}$ à $-F^{(g)}$ qu'à des constantes près (des exemples sont développés dans les sections \ref{Gaussien}, \ref{Seccc} et\ref{Secccc} où l'impact de la normalisation de la fonction de partition est explicité).
\begin{remark}[Cas des courbes de genre $0$] D'un point de vue pratique, le calcul de la récurrence topologique est particulièrement simple lorsque la courbe spectrale est de genre $0$. En effet, l'absence de fonctions holomorphes non triviales sur $\Sigma$ rend le choix de $\omega_2^{(0)}(z_1,z_2)$ unique:
\beq \omega_2^{(0)}(z_1,z_2)=\frac{dz_1dz_2}{(z_1-z_2)^2}\eeq
La récurrence topologique se réduit alors à de simples calculs sur des fonctions méromorphes dans $\overline{\mathbb{C}}$ qui peuvent être effectués par n'importe quel logiciel de calcul formel. 
\end{remark}

\subsection{Normalisation: le cas Gaussien}\label{Gaussien}
La cas Gaussien est incontestablement l'exemple le plus simple et le plus connu de la théorie précédente. Dans ce cas, $V(x)=\frac{x^2}{2}$ et l'on a donc:
\beq \td{Z}_N^{\text{Gaussien}}=\int_{\mathcal{H}_N} dM e^{-\frac{N}{T} \Tr \frac{M^2}{2}}\eeq
L'intérêt de ce modèle est d'être entièrement calculable. En effet, écrit en composantes indépendantes, l'intégrale précédente n'est constituée que de produits d'intégrales Gaussiennes qui peuvent donc être calculées:
\bea \label{CalculGaussienDirect}\td{Z}_N^{\text{Gaussien}}&=&\int\dots \int \left(\prod_{i<j}^N \text{dRe}(M_{i,j}) \text{dIm}(M_{i,j})\right)\left(\prod_{k=1}^N \text{dRe}(M_{k,k})\right)\cr
&& e^{-\frac{N}{T}\left(
\frac{1}{2}\underset{r=1}{\overset{N}{\sum}} (\text{Re}M_{r,r})^2 +\underset{r<s}{\overset{N}{\sum}}\left[(\text{Re}(M_{r,s}))^2+(\text{Im}(M_{r,s}))^2\right]\right)}\cr
&=&\left(\sqrt{\frac{2\pi T}{N}}\right)^{N} \left(\sqrt{\frac{\pi T}{N}}\right)^{N(N-1)}=2^{\frac{N}{2}} \left(\frac{\pi T}{N}\right)^{\frac{N^2}{2}}
\eea
On obtient ainsi 
\beq
\ln \td{Z}_N^{\text{Gaussien}}=-\frac{N^2}{2}\ln N+\frac{N}{2}\ln 2+\frac{N^2}{2}\ln (\pi T) \eeq
La courbe spectrale associée au cas Gaussien est donnée par \eqref{CourbeSpectrale} $y(x)^2=\frac{x^2}{4}-T$. On retrouve ainsi la célèbre loi du demi-cercle $d\mu_\infty(x)=\frac{1}{2T\pi}\sqrt{4T-x^2}\,\mathds{1}_{-2\sqrt{T}\leq x\leq 2\sqrt{T}}\,dx$ que les simulations numériques confirment:

\medskip
\begin{center}
\includegraphics[width=10cm]{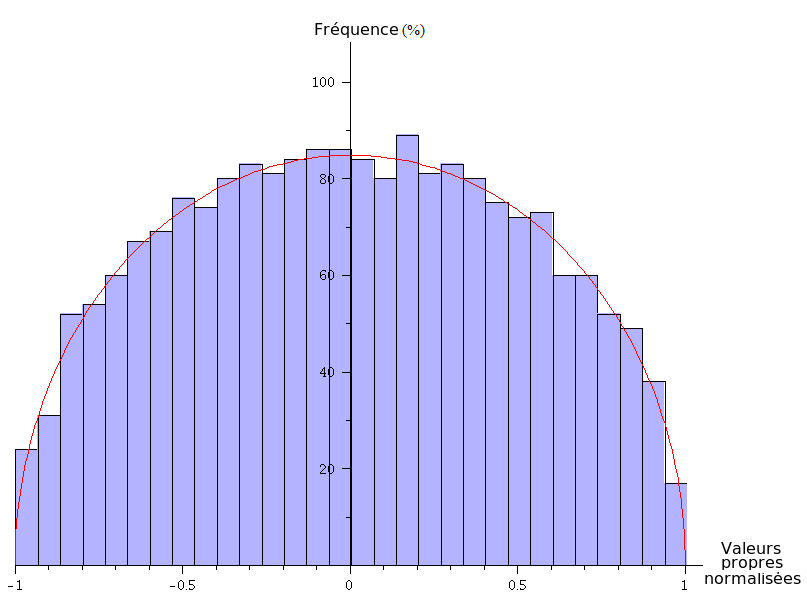}

\textit{Fig. $1$: Loi du demi-cercle de Wigner pour $T=\frac{1}{4}$. Simulations réalisées sur $500$ matrices hermitiennes aléatoires (entrées i.i.d. Gaussiennes) de taille $50\times 50$.} 
\end{center}
\medskip

Les invariants symplectiques de cette courbe spectrale sont également connus. On peut les trouver par exemple dans \cite{P2} (eq. $2.32$ avec $\theta=T$) ou \cite{Do}:
\bea \label{FgGaussien}F^{(0)}_{\text{Gaussien}}(T)&=&\frac{3T^2}{4}-\frac{T^2}{2}\ln T\cr
F^{(1)}_{\text{Gaussien}}(T)&=&-\frac{1}{12}\ln T\cr
F^{(g)}_{\text{Gaussien}}(T)&=&-\frac{B_{2g}}{2g(2g-2)T^{2g-2}} \,\,,\,\, \forall\, g\geq 2
\eea
où les nombres de Bernoulli $\left(B_{k}\right)_{k\geq 0}$ sont définis par la relation:
\beqq \frac{x}{e^x-1}=\sum_{k=0}^\infty \frac{B_k}{k!}x^k\eeqq
Ainsi, on obtient le développement de la fonction de partition issue de la récurrence topologique par:
\bea \label{ZNGauss} \ln Z_{N, \text{Top. Rec.}}^{\text{Gaussien}}&=&\sum_{g=0}^\infty F_{\text{Top. Rec.}}^{(g)}(T)\left(\frac{N}{T}\right)^{2-2g}\cr
&=& \left(\frac{3}{4}-\frac{1}{2}\ln T\right)N^2-\frac{1}{12}\ln T-\sum_{g=2}^\infty \frac{B_{2g}}{2g(2g-2)}N^{2-2g}\cr
&\overset{def}{=}&\sum_{g=0}^\infty F^{(g)}_{\text{Gaussien}}N^{2-2g}
\eea
La dépendance en $T$ est uniquement présente dans les deux premiers termes du développement de $F^{(g)}_{\text{Gaussien}}$. Cette dépendance peut facilement être retrouvée en utilisant la propriété d'invariance symplectique. Ainsi, en remplaçant $(x,y)\to \left(\sqrt{T}x,\frac{1}{\sqrt{T}}y\right)$, on obtient une courbe équivalente $y^2=T^2\frac{x^2-4}{4}$. La définition de la récurrence topologique permet alors immédiatement de vérifier que pour $g\geq 2$: $F^{(g)}(T)= T^{2-2g}F^{(g)}(T=1)$ et que $\ln Z_{N, \text{Top. Rec.}}^{\text{Gaussien}}$ ne dépendra de $T$ qu'à travers les deux premiers invariants symplectiques. Dans le cas Gaussien, ce résultat peut être comparé avec un calcul exact issu des intégrales de Selberg et de Mehta:

\begin{proposition}[Intégrales de Mehta] L'application des intégrales de Selberg \cite{MehtaBook} donne:
\beq \frac{1}{(2\pi)^{\frac{N}{2}}}\int\dots\int dt_1\dots dt_N \Delta(\mathbf{t})^2e^{-\frac{1}{2}\underset{i=1}{\overset{N}{\sum}} t_i^2}=N! \prod_{j=1}^{N-1}j!\eeq
\end{proposition}

Par changement de variables $t_i=\sqrt{\frac{N}{T}}\lambda_i$ on obtient:
\beq \label{ZNGaussien} Z_N^{\text{Gaussien}}=\frac{(N!)(2\pi)^{\frac{N}{2}} T^{\frac{N^2}{2}}}{N^{\frac{N^2}{2}}}\prod_{j=1}^{N-1}j!\eeq
Ainsi:
\beq \label{Normalisation} c_N=\frac{\td{Z}_N^{\text{Gaussien}}}{Z_N^{\text{Gaussien}}}=\frac{\pi^{\frac{N(N-1)}{2}}}{N!\underset{j=1}{\overset{N-1}{\prod}}j!}\eeq
On retrouve ainsi des résultats connus sur le volume du groupe unitaire \cite{VolumeUnitaryGroup}. Notons que cette normalisation est indépendante de $T$ et du potentiel $V$. En repartant de \eqref{ZNGaussien}, le développement asymptotique de la fonction $G$ de Barnes donne\footnote{La notation $\xi(x)$ correspond dans ce document à la fonction zêta de Riemann}:
\beq \label{Barnes} \ln \left(\prod_{i=1}^{N-1}i!\right)=\frac{N^2}{2}\ln N+\frac{N}{2}\ln(2\pi) -\frac{1}{12}\ln N +\xi'(-1)-\frac{3}{4}N^2+\sum_{g=2}^{\infty} \frac{B_{2g}}{2g(2g-2)} N^{2-2g}\eeq
et l'on trouve donc que:
\beq \label{LnZNGaussien}\ln Z_N^{\text{Gaussien}}=\ln(N!)+\frac{N^2}{2}\ln T + N\ln(2\pi)-\frac{1}{12}\ln N+\xi'(-1)-\frac{3}{4}N^2+\sum_{g=2}^{\infty} \frac{B_{2g}}{2g(2g-2)} N^{2-2g}\eeq 
Afin d'obtenir un développement en puissance de $\frac{1}{N^2}$ similaire à celui fourni par la récurrence topologique, il est nécessaire de définir:

\begin{proposition}[Normalisation de la fonction de partition]\label{PropGaussien} La fonction de partition du modèle hermitien, normalisée de la façon suivante:
\beq \hat{Z}_N^{\text{Gaussien}}=\frac{N^{\frac{1}{12}}}{T^{\frac{1}{12}}(2\pi)^N e^{\xi'(-1)}  N!}Z_N^{\text{Gaussien}}\eeq
possède un développement topologique de la forme:
\beqq \hat{Z}_N^{\text{Gaussien}}=\sum_{g=0}^\infty Z_{\text{Gaussien}}^{(g)}N^{2-2g} \text{  avec  } Z^{(g)}=-F^{(g)}_{\text{Gaussien}}\eeqq
\end{proposition}

La présence du signe $-$ dans l'identification de $Z_{\text{Gaussien}}^{(g)}$ avec $F_{\text{Gaussien}}^{(g)}$ provient d'un mauvais choix de signe (qui historiquement a toujours été un peu fluctuant) dans la définition des invariants symplectiques. Par ailleurs, le terme $N!$ provient du fait que l'intégrale sur les valeurs propres n'est pas ordonnée ce qui fait agir le groupe des permutations et produit un facteur $N!$. Enfin, le facteur $(2\pi)^N$ provient du volume du groupe $\mathcal{U}(1)^N$ qui agit (comme sous-groupe de $\mathcal{U}(N)$) par multiplication à droite sur $U$ dans la diagonalisation $M=U \text{diag}(\lambda_1,\dots,\lambda_N)U^{-1}$ tout en préservant $\text{diag}(\lambda_1,\dots,\lambda_N)$. Notons par ailleurs que comme:
\beqq \ln(N!)=\frac{1}{2}\,\ln(2\pi)+N\,\ln\, N+\frac{1}{2}\,\ln\, N -N+\sum_{k=1}^\infty \frac{B_{2k}}{2k(2k-1)N^{2k-1}}\eeqq
On obtient par \eqref{LnZNGaussien} que
\bea \ln Z_N^{\text{Gaussien}}&=&N^2\left(\frac{1}{2}\ln T-\frac{3}{4}\right)+N\,\ln\, N + N(\ln(2\pi)-1) +\frac{5}{12}\,\ln\, N+ \frac{1}{2}\ln(2\pi)+\xi'(-1)\cr
&& +\sum_{g=1}^\infty \frac{B_{2g}}{2g(2g-1)}N^{1-2g}+\sum_{g=2}^{\infty} \frac{B_{2g}}{2g(2g-2)} N^{2-2g}
\eea
En d'autres termes:
\bea \label{ZNGaussienRes} Z_N^{\text{Gaussien}}&=&N^{N+\frac{5}{12}}\,\text{exp}\left(\sum_{k=-2}^\infty Z^{(k)}_{\text{Gauss}}N^{-k}\right)\cr
Z^{(-2)}_{\text{Gauss}}&=&\frac{1}{2}\,\ln\, T-\frac{3}{4}\cr
Z^{(-1)}_{\text{Gauss}}&=&\ln(2\pi)-1\cr
Z^{(0)}_{\text{Gauss}}&=&\frac{1}{2}\,\ln(2\pi)+\xi'(-1)\cr
Z^{(2k)}_{\text{Gauss}}&=&\frac{B_{2k+2}}{2k(2k+2)} \,\,,\,\, \forall\, k\geq 1\cr
Z^{(2k+1)}_{\text{Gauss}}&=&\frac{B_{2k+2}}{(2k+1)(2k+2)} \,\,,\,\, \forall\, k\geq 0
\eea
La forme générale de ce développement est ainsi parfaitement en accord avec le cas beaucoup plus général (ensembles $\beta$, potentiel quelconque) développé dans \cite{BG}. 

\begin{remark}Notons que pour le cas de référence Gaussien, la fonction de partition peut se réexprimer à l'aide de la fonction $G$ de Barnes. D'un point de vue analytique, on peut alors préciser dans ce cas la nature du développement asymptotique \eqref{LnZNGaussien}: celui-ci est valable tant que $N$ tend vers l'infini dans un secteur du plan complexe excluant l'axe réel négatif. En particulier, il ne s'agit pas d'un développement convergent (i.e. un développement en série entière autour de $N=\infty$) dans n'importe quel voisinage de $N\to \infty$. Néanmoins, il est bien valable pour $N\in \mathbb{N}^*\subset\mathbb{R}_+$ ce qui est suffisant pour les applications en probabilités. 
\end{remark}

\subsection{Potentiel général}
Pour un potentiel général $V$, les résultats concernant la normalisation de la fonction de partition $Z_N$ sont beaucoup plus parcellaires. En effet, même dans le cas où le potentiel $V$ est convexe et la courbe spectrale régulière avec un support composé d'une seule coupure, l'existence d'un développement topologique pour les fonctions de corrélations ne garantit nullement l'existence d'un développement en $N^{2-2g}$ pour la fonction de partition $Z_N$. En effet, la définition standard \eqref{IntHermitienneDiag} ne garantit pas ce développement comme on peut le voir dans l'exemple Gaussien \eqref{ZNGaussienRes}. Les deux obstacles à l'existence de ce développement sont:
\begin{itemize} \item La présence de termes logarithmiques $\ln N$ et $N\ln N$ dans le développement de $\ln Z_N$.
\item La présence potentielle de puissances impaires dans le développement en $\frac{1}{N}$ de $\ln Z_N$.
\end{itemize}
Ces deux problèmes sont de nature et de difficulté différentes. Le premier point est résolu facilement par le résultat principal de \cite{BorotGuionnetKoz}. Pour un potentiel polynomial convexe (ou vérifiant certaines hypothèses décrites précédemment ou dans \cite{BG}) la fonction de partition possède toujours un développement de la forme:
\beq Z_N=N^{N+\frac{5}{12}}\,\text{exp}\left(\sum_{k=-2}^\infty Z^{(k)}N^{-k}\right)\eeq
On voit donc que les termes logarithmiques sont universels et identiques au cas du modèle Gaussien. En revanche, ce résultat ne permet pas d'éliminer les termes impairs du développement en $\frac{1}{N}$ ni l'identification des termes pairs avec les invariants symplectiques issus de la récurrence topologique. A ma connaissance, la normalisation adéquate de l'intégrale multiple $Z_N$ pour obtenir l'identification n'a jamais été fournie dans la littérature bien qu'elle ait un intérêt tant pour les probabilistes et les analystes, que pour la récurrence topologique elle-même (pour que la série de terme général $F^{(g)}$ puisse s'identifier avec quelque chose). Un résultat généralisant la proposition \ref{PropGaussien} pour des potentiels plus complexes serait ainsi souhaitable et ne semble pas inaccessible, même si des termes de corrections sont vraisemblablement nécessaires comme cela est décrit dans \cite{CorrectionTermsTopRec,BG2}. 

\section{Intégrales de matrices hermitiennes à bords durs \label{BordDur}}

\subsection{Le modèle exponentiel \label{Seccc}}

L'application de la récurrence topologique ne se résume pas au calcul d'intégrales sur des valeurs propres réelles. On peut ainsi restreindre le domaine des valeurs propres à un sous-espace de $\mathbb{R}$. On peut également abandonner la forme de départ matricielle pour définir des intégrales sur les valeurs propres directement et étendre ces intégrales sur un contour fermé ou ouvert du plan complexe. Dans ce cas de figure, les méthodes alternatives de polynômes orthogonaux, problème de Riemann-Hilbert, etc. deviennent fastidieuses voire inadaptées. En revanche, la méthode des équations de boucles couplée à la récurrence topologique permet toujours d'obtenir des résultats bien que la détermination de la courbe spectrale soit plus subtile puisqu'il faut tenir compte des éventuels ``bords durs'' (bornes des intervalles d'intégration). Un exemple simple d'application de la méthode générale est l'exemple suivant:

\beq\label{ModeleExp} Z_N^{\text{Exp}}=\int_{\mathbb{R}_+^*}\dots \int_{\mathbb{R}_+^*}d\lambda_1\dots d\lambda_N\, \Delta(\boldsymbol{\lambda})^2\,e^{-\frac{N}{T}\underset{i=1}{\overset{N}{\sum}} \lambda_i}\eeq
L'intégrale précédente est évidemment convergente et dans ce cas, on peut également définir l'intégrale de matrices hermitiennes associées:
\beq \label{ModeleExpMat}\td{Z}_N^{\text{Exp}}=\int_{\mathcal{M}_N^+}dM e^{-\frac{N}{T}\Tr M}=C_N\, Z_N^{\text{Exp}}\eeq
où $\mathcal{M}_N^+$ est le sous-ensemble ouvert des matrices hermitiennes ayant des valeurs propres strictement positives. Ce modèle est intéressant car les intégrales peuvent être calculées explicitement. En utilisant l'équation $3.21$ de \cite{VolumeUnitaryGroup} on obtient ainsi en notant $\mathbf{t}_N=(t_1,\dots,t_N)$:
\beq \int_{t_1>\dots>t_N>0}dt_1\dots dt_N \,\Delta(\mathbf{t}_N)^2\,e^{-\underset{i=1}{\overset{N}{\sum}}  t_i}=\left(\prod_{i=1}^{N-1}i!\right)^2\eeq
Par simple changement de variables $t_i=\frac{N}{T}\lambda_i$ et en faisant agir le groupe des permutations pour enlever l'ordre des valeurs propres, on obtient:
\bea \label{ZNExpRes}Z_N^{\text{Exp}}&=&\int_{\lambda_1,\dots,\lambda_N>0}d\lambda_1\dots d\lambda_N \,\Delta(\boldsymbol{\lambda})^2\,e^{-\frac{N}{T}\underset{i=1}{\overset{N}{\sum}} \lambda_i}\cr
&=&(N!)\left(\frac{N}{T}\right)^{-N^2}\left(\prod_{i=1}^{N-1}i!\right)^2
\eea
Par ailleurs, la constante $C_N$ issue de la diagonalisation de \eqref{ModeleExpMat} en \eqref{ModeleExp} est également connue (Cf. \cite{VolumeUnitaryGroup}):
\beq C_N=\frac{\pi^{N(N-1)}}{N!\underset{i=1}{\overset{N-1}{\prod}}i!}\eeq
ce qui permet d'obtenir: 
\beq \td{Z}_N^{\text{Exp}}=\left(\frac{N}{T}\right)^{-N^2}\pi^{N(N-1)}\left(\prod_{i=1}^{N-1}i!\right)\eeq
Pour pouvoir obtenir un lien avec la récurrence topologique, il faut au préalable déterminer la courbe spectrale associée à l'intégrale \eqref{ModeleExp}. La technique générale s'applique ici mais la présence d'un bord dur en $\lambda=0$ nécessite un certain soin. Les équations de boucles ont été écrites à divers endroits, pour des intégrales de matrices simples, des intégrales à deux matrices, les ensembles $\beta$ et avec ou sans bords durs \cite{Hardedges,HardWall,Hardedges2,Lie,TopRecBethe,CBM,CBM2,LoopEquationBeta2MM,These,UnitaryMarchal}. Dans notre cas, elles s'obtiennent facilement en regardant (on note $\boldsymbol{\lambda}=(\lambda_1,\dots,\lambda_N)$):
\beq \label{LoopBords}I_N(x)=\frac{1}{Z_N^{\text{Exp}}}\int_{\boldsymbol{\lambda}\in (\mathbb{R}_+)^N}d\boldsymbol{\lambda}\, \sum_{j=1}^N\frac{d}{d\lambda_j}\left(\frac{1}{x-\lambda_j} \Delta(\boldsymbol{\lambda})^2e^{-\frac{N}{T} \underset{i=1}{\overset{N}{\sum}} \lambda_i}\right)\eeq
L'action de la dérivée sur le produit fournit les termes de l'équation de boucle sans bord dur \eqref{Loop1SansBord}. En revanche, la présence de bords durs implique que l'intégrale précédente n'est plus nulle mais est de la forme $I=\frac{\alpha(N)}{x}$. La courbe spectrale est donc de la forme:
\beq y(x)=W_1^{(0)}(x)-\frac{1}{2} \,\,\,\Rightarrow\,\,\, y(x)^2=\frac{1}{4}+\frac{\alpha_0}{x}=\frac{x+4\alpha_0}{4x}\eeq
La détermination de $\alpha_0$ provient du fait qu'au voisinage de l'infini on a par définition $W_1(x)=\frac{1}{x}+O\left(\frac{1}{x^2}\right)$ d'où $W_1^{(0)}(x)=\frac{T}{x}+O\left(\frac{1}{x^2}\right)$. Par conséquent $y(x)^2=\frac{1}{4}-\frac{T}{x}+O\left(\frac{1}{x^2}\right)$ ce qui donne $\alpha_0=-T$. On obtient donc:
\beq \label{CourbeSpectraleExp} y(x)^2=\frac{1}{4}-\frac{T}{x}=\frac{x-4T}{4x}\eeq
La courbe spectrale précédente est de genre $0$ avec deux points de branchements en $x=4T$ et $x=0$. On peut la paramétrer par:
\beq x(z)=2T+T\left(z+\frac{1}{z}\right)=\frac{T(z+1)^2}{z} \,\,\,, \,\,\, y(z)=\frac{z-1}{2(z+1)}\eeq
Par ailleurs, on peut réaliser une transformation symplectique $(x,y)\to (Tx,T^{-1}y)$ donnant la courbe symplectiquement équivalente $y^2(x)=T^2\frac{x-4}{4x}$ qui permet d'obtenir la dépendance des fonctions de corrélations en $T$. Cette courbe donne également la densité limite d'équilibre des valeurs propres que l'on peut vérifier par des simulations numériques. Cette densité limite est donnée par $d\mu_\infty(x)=\frac{1}{T\pi}\sqrt{\frac{4T-x}{4x}}\, \mathds{1}_{0\leq x\leq 4T}\,dx$.
\medskip

\begin{center}\includegraphics[width=14cm]{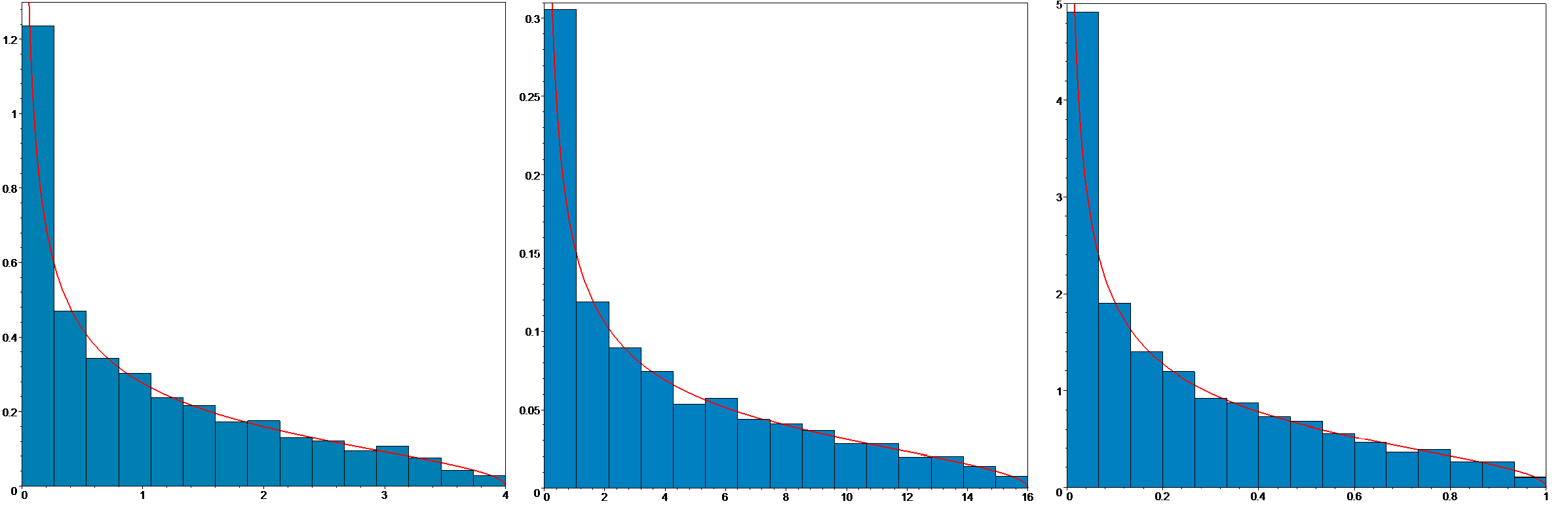}

\textit{Fig. $2$: Distribution limite d'équilibre du modèle exponentiel pour $T\in\{1,4,\frac{1}{4}\}$ (de gauche à droite). Simulations réalisées sur $50$ modèles indépendants à $N=30$ valeurs propres.} 
\end{center}

\medskip

La connaissance de l'expression exacte de \eqref{ZNExpRes} permet d'obtenir les valeurs exactes des invariants symplectiques de cette courbe:

\begin{theorem}[Invariants symplectiques du modèle exponentiel]\label{FgCourbeExpo} La normalisation adéquate du modèle exponentiel \eqref{ModeleExp} est donnée par:
\beqq \hat{Z}_N^{\text{Exp}}=\frac{N^{\frac{1}{6}}}{T^{\frac{1}{6}}(N!)(2\pi)^N\xi'(-1)^2}\, Z_N^{\text{Exp}}\eeqq
et on obtient alors:
\beqq \hat{Z}_N^{\text{Exp}}=\left(\hat{Z}_N^{\text{Gaussien}}\right)^2=\sum_{g=0}^\infty F_{\text{Exp}}^{(g)}\,N^{2-2g} \eeqq
En d'autres termes, en utilisant \eqref{ZNGaussienRes} et \eqref{FgGaussien}, les invariants symplectiques de la courbe $y^2=\frac{x-4T}{4x}$ sont:
\bea\label{FgExpo} F_{\text{Exp}}^{(0)}&=&\frac{3}{2}\,T^2 -T^2\,\ln\, T\cr
F_{\text{Exp}}^{(1)}&=&-\frac{1}{6}\,\ln\, T\cr
F_{\text{Exp}}^{(g)}&=&-\frac{2B_{2g}}{2g(2g-2)T^{2g-2}} \,\,,\,\, \forall\, g\geq 2
\eea
\end{theorem}

Les premiers ordres de ces résultats peuvent être vérifiés aisément par le calcul direct de la récurrence topologique. En revanche, pour obtenir l'identification de la série, il est nécessaire de vérifier que la fonction de partition et les fonctions de corrélation du modèle exponentiel admettent un développement topologique. Cela peut être fait en vérifiant les hypothèses de la proposition \ref{ConditionsTopologiques}. Dans le cas du modèle exponentiel, le potentiel $V(x)=x$ est continu et  convexe sur $\mathbb{R}_+^*$ et il est trivialement confinant en $+\infty$. Il est évidemment holomorphe dans un voisinage de $[0,4T]$ ce qui démontre l'ensemble des conditions nécessaires à l'application de la proposition \ref{ConditionsTopologiques}. Notons que l'expression exacte des invariants symplectiques du modèle exponentiel peut se retrouver à partir de leur invariance symplectique. En effet, le changement $(X,Y)=\left(x^2,\frac{1}{2x}y\right)$ dans la courbe Gaussienne $y^2=\frac{x^2-4T}{4}$ fournit la courbe spectrale $4Y^2=\frac{X-4T}{4X}$ qui, après la transformation $\td{Y}= 2Y$, fournit la courbe spectrale du modèle exponentiel. Cette dernière transformation n'affecte les invariants symplectiques que par un facteur $2$ ce qui permet de retrouver intégralement le théorème \ref{FgCourbeExpo} à partir du cas Gaussien. Par ailleurs, en prenant une intégrale sur $\mathbb{R}_-^*$ plutôt que sur $\mathbb{R}_+^*$ et en inversant le signe du potentiel (ou en faisant $\lambda_i\to -\lambda_i$ directement), on obtient la courbe spectrale $(x,y)\to (-x,-y)$ symplectiquement équivalente $y(x)^2=\frac{x+4}{4x}$ qui possède bien les mêmes invariants symplectiques.

\medskip

L'exemple du cas Gaussien et du cas précédent montrent que l'on peut utiliser la récurrence topologique pour obtenir le développement asymptotique d'intégrales de matrices hermitiennes ou de gaz de Coulomb bi-dimensionnels (intégrales sur les valeurs propres) dont le support des valeurs propres peut être restreint à un sous-domaine de $\mathbb{R}$, mais que l'on peut également procéder en sens inverse à savoir obtenir tous les invariants symplectiques (et parfois les fonctions de corrélations) d'une courbe donnée en utilisant la valeur exacte d'une intégrale matricielle lorsqu'elle est connue (intégrales de Selberg ou autres). En combinant ces résultats ensemble, on arrive parfois à des constatations surprenantes:

\begin{proposition} En combinant \eqref{FgGaussien}, \eqref{FgExpo}, et les résultats de \cite{P2} on a:
\begin{center}
\small{\begin{tabular}{|c|c|c|c|}
\hline
{\normalfont Courbe Spectrale}& $F^{(0)}$&$F^{(1)}$& $F^{(g)}$ {\normalfont avec} $g\geq 2$\\
\hline
$y^2=\frac{x^2-4T}{4}$ & $\frac{3T^2}{4}-\frac{T^2}{2}\ln \,T$&$-\frac{1}{12}\ln \,T$& $-\frac{B_{2g}}{2g(2g-2)T^{2g-2}}$\\
\hline
$y^2=\frac{x-4T}{4x}$& $\frac{3T^2}{2} -T^2\ln \,T$ & $-\frac{1}{6}\ln\, T$& $-\frac{2B_{2g}}{2g(2g-2)T^{2g-2}}$\\
\hline
$y^2=\frac{x+T^2}{4x^2}$& $-\frac{3T^2}{4}+\frac{T^2}{2}\ln\, T+\frac{T^2}{2}\ln\, 2+\frac{i\pi}{24}$&$\frac{1}{12}\ln\, T+\frac{1}{24}\ln\, 2+\frac{i\pi T^2}{24}$& $\frac{B_{2g}}{2g(2g-2)T^{2g-2}}$\\
\hline 
\end{tabular}}\normalsize{}
\end{center}
Comme on l'a vu précédemment, les deux premières lignes peuvent être aisément déduites l'une de l'autre. De manière surprenante, la connexion entre les deux premiers cas et le dernier est plus complexe à établir (Théorème $4.3$ de \cite{P2}).
\end{proposition}

\subsection{Intégrales Gaussiennes à valeurs propres strictement positives\label{Secccc}}

Evaluer la probabilité qu'une matrice aléatoire hermitienne avec des entrées indépendantes et identiquement distribuées suivant une loi normale admette des valeurs propres toutes strictement positives est un problème qui a été récemment étudié dans \cite{Majumdar,Majumdar2} et presque intégralement résolu dans \cite{Majumdar3}. Dans sa version généralisée qui ne sera pas abordée ici, on peut également s'intéresser à ce qu'une proportion $0\leq c\leq 1$ des valeurs propres soient positives (le cas $c=1$ étant la question précédente). Ce type de problème peut également être étudié à l'aide de la récurrence topologique comme nous allons le voir. Pour cela on s'intéresse à la quantité suivante:
\beq \label{ProbaVPPositives} \mathcal{P}(T)=\frac{1}{Z_N^{\text{Gaussien}}(T)}\int_{(\mathbb{R_+})^N} d\boldsymbol{\lambda}\, \Delta(\boldsymbol{\lambda})^2 \,e^{-\frac{N}{2T}\underset{i=1}{\overset{N}{\sum}} \lambda_i^2}\overset{def}{=}\frac{1}{Z_N^{\text{Gaussien}}(T)}Z_{N,+}^{\text{Gaussien}}(T)\eeq
Le facteur de normalisation $Z_N^{\text{Gaussien}}(T)$ a déjà été calculé \eqref{LnZNGaussien} donc seul le calcul de la seconde intégrale $Z_{N,+}^{\text{Gaussien}}(T)$ reste à faire. Cette intégrale est une intégrale hermitienne standard avec bords durs en $x=0$ et un potentiel $V(x)=\frac{x^2}{2}\mathds{1}_{x\geq 0}$. L'application des équations de boucles à bord durs (\eqref{CourbeSpectrale} mélangée avec \eqref{LoopBords}) fournit une courbe spectrale de la forme:
\beqq y^2(x)=\frac{V'(x)^2}{4}-P_1^{(0)}(x)+\frac{\alpha_0}{x}=\frac{x^2}{4}-T+\frac{\alpha_0}{x}\eeqq
Malheureusement $W_1^{(0)}(x)=\frac{T}{x}+O\left(\frac{1}{x^2}\right)$ implique $y(x)^2=\frac{x^2}{4}-T+O\left(\frac{1}{x}\right)$ et ne permet plus de déterminer $\alpha_0$ de façon directe. La courbe spectrale précédente définit une surface de Riemann qui génériquement possède $4$ points de branchements: un en $x=0$ et trois correspondant aux zéros complexes de $P_3(x)=\frac{x^3}{4}-Tx+\alpha_0$. D'un point de vue probabiliste en revanche, on attend une densité limite d'équilibre supportée suivant un intervalle de type $[0,a]$ puisque les valeurs propres ne peuvent s'aggréger qu'autour du seul minimum du potentiel en $x=0$. Cette analyse heuristique est confortée par le résultat général de \cite{BorotGuionnetKoz}: le fait que le potentiel soit strictement convexe sur $\mathbb{R}_+^*$ implique que la densité limite d'équilibre ne peut être supportée que par un seul intervalle. Ainsi on doit nécessairement avoir $y(x)^2=\frac{(x-a)^2(x-b)}{4x}$ avec $b>0$. La comparaison avec la courbe initiale donne:
\beqq \alpha_0=-4\left(\frac{T}{3}\right)^{\frac{3}{2}} \,\,,\,\, a=-2\left(\frac{T}{3}\right)^{\frac{1}{2}} \text{ et } b=4\left(\frac{T}{3}\right)^{\frac{1}{2}}
\eeqq
et au final:
\beq\label{CourbeSpectraleGaussienPlus} y(x)^2=\frac{\left(x+2\left(\frac{T}{3}\right)^{\frac{1}{2}}\right)^2\left(x-4\left(\frac{T}{3}\right)^{\frac{1}{2}}\right)}{4x}=\frac{\frac{x^3}{4}-Tx-4\left(\frac{T}{3}\right)^{\frac{3}{2}}}{x}\eeq
qui se paramétrise sur $\overline{\mathbb{C}}$ par:
\beqq x(z)=\left(\frac{T}{3}\right)^{\frac{1}{2}}\left(2+z+\frac{1}{z}\right) \text{  et  } y(z)=\left(\frac{T}{3}\right)^{\frac{1}{2}}\left(4+z+\frac{1}{z}\right)\frac{z-1}{2(z+1)}\eeqq
et donne la densité limite d'équilibre:
\beq d\mu_\infty(x)=\frac{1}{2\pi T}\left(x+2\left(\frac{T}{3}\right)^{\frac{1}{2}}\right)\sqrt{\frac{\left(4\left(\frac{T}{3}\right)^{\frac{1}{2}}-x\right)}{x}} \,\,\mathds{1}_{0\leq x\leq 4\left(\frac{T}{3}\right)^{\frac{1}{2}}}\, dx\eeq

\medskip

\begin{center}\includegraphics[width=14cm]{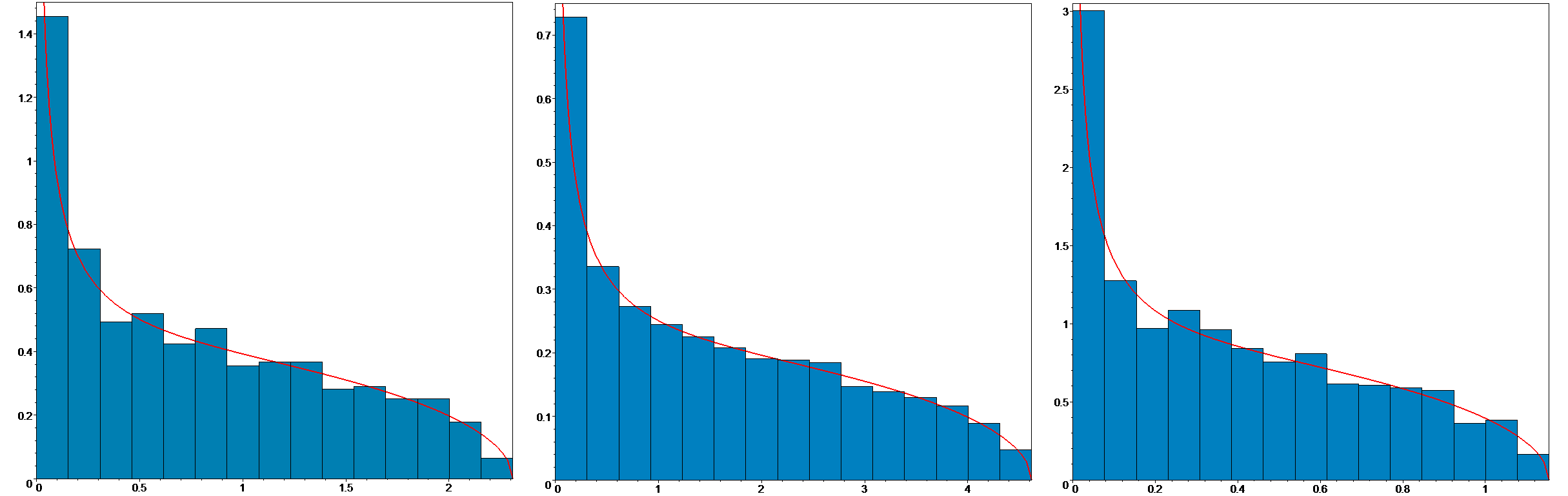}

\textit{Fig. $3$: Distribution limite d'équilibre du modèle Gaussien à valeurs propres strictement positives pour $T\in\{1,4,\frac{1}{4}\}$ (de gauche à droite). Simulations réalisées sur $50$ modèles indépendants à $N=30$ valeurs propres.} 
\end{center}

\medskip

Notons ici les limites usuelles de la méthode des équations de boucles: elles laissent très souvent des coefficients indéterminés en lien avec le nombre de coupures de la densité limite d'équilibre. Dans beaucoup de situations comme celle ci-dessus, des arguments d'analyse ou des symétries supplémentaires du problème permettent de compléter la détermination complète. Lorsque cela n'est pas possible, la seule solution reste alors la résolution d'un problème de minimisation d'énergie fonctionnelle par la résolution d'un problème de Riemann-Hilbert et la condition de Boutroux \cite{Kap,MarcoPaths}. Les invariants symplectiques de la courbe spectrale \eqref{CourbeSpectraleGaussienPlus} ne sont pas connus à l'heure actuelle. Néanmoins, on peut très facilement calculer les premiers termes qui sont ceux nécessaires pour obtenir les ordres dominants de l'asymptotique de la probabilité recherchée. On trouve:
\bea\label{FgGaussienPlus} F^{(0)}_{\text{Gaussien, +}}(T)&=&\frac{3T^2}{4}-\frac{T^2}{2}\,\ln\, T +\frac{T^2}{2}\,\ln \,3\cr
  F^{(1)}_{\text{Gaussien, +}}(T)&=&\frac{1}{3}\,\ln\, 2-\frac{1}{8}\,\ln\, 3\cr
	F^{(2)}_{\text{Gaussien, +}}(T)&=&\frac{29}{5760\,T^2}\cr
	F^{(3)}_{\text{Gaussien, +}}(T)&=&-\frac{4855}{1161216\,T^4}
\eea

La valeur de $F^{(1)}$ dans le cas de bords durs est obtenue grâce à une formule de L. Chekhov \cite{HardWall}:
\beq d\mu_\infty= -\frac{M(x)}{T\pi}\sqrt{\frac{x-b}{x-a}} \,\, \Rightarrow F^{(1)}=\frac{1}{24}\ln \left(M(a)^3M(b)\frac{(a-b)^4}{T^4}\right)\eeq

Notons que la courbe spectrale n'est pas singulière car la densité limite d'équilibre ne s'annule pas sur son support (le zéro est sur l'axe réel négatif en $x=-2\left(\frac{T}{3}\right)^{\frac{1}{2}}$) et son comportement au voisinage des extrémités est bien en racine-carrée. On peut par ailleurs vérifier les conditions de la proposition \ref{ConditionsTopologiques}. En effet, le potentiel $V(x)=\frac{x^2}{2}$ est continu, strictement convexe sur $R_+^*$, trivialement confinant en $+\infty$ et holomorphe sur $\mathbb{C}$. On obtient ainsi que la fonction de partition (modulo une normalisation adéquate) admet un développement de type topologique qui peut se calculer à l'aide des invariants symplectiques de la courbe précédente. Le calcul des invariants symplectiques permet de définir un asymptotique topologique de la forme (rappelons que le développement topologique introduit dans ce chapitre est en puissance de $\frac{T}{N}$ et avec un signe $-$ global):
\bea \ln Z_{N,+}^{\text{Top. Rec.}}(T)&=&\left(-\frac{3}{4}+\frac{1}{2}\,\ln \,T -\frac{1}{2}\,\ln \,3\right)N^2-\frac{1}{3}\,\ln\, 2+\frac{1}{8}\,\ln\, 3\cr
&&-\frac{29}{5760N^2}+\frac{4855}{1161216N^4}+O\left(\frac{1}{N^6}\right)\cr
&&\eea
Compte tenu de la forme de courbe spectrale, il est également facile de vérifier que le développement précédent ne fait intervenir la température $T$ que dans les deux premiers termes. Le calcul du développement de la probabilité recherchée $\mathcal{P}(T)$ \eqref{ProbaVPPositives} nécessite alors de faire attention aux questions de normalisation. En effet, comme il l'a été vu dans le cas simple Gaussien (proposition \ref{PropGaussien}), la récurrence topologique fournit une fonction de partition qui n'est égale à l'intégrale de départ que si cette dernière est correctement normalisée. Dans le cas présent, la normalisation n'est pas immédiate car l'intégrale avec bords ne peut se ramener à une intégrale de Selberg. Néanmoins, en utilisant des intégrales de Selberg de type Laguerre la normalisation a été étudiée dans \cite{Majumdar3} (équation $A.8$):
\bea \ln Z_{N,+}^{\text{Gaussien}}(T)&=& -N^2\,\ln \,N+\frac{3}{2}N^2+\frac{1}{6}\,\ln\, 2+\ln(N!)+ 2\,\ln\left(\prod_{i=1}^{N-1}i!\right)\cr
&&-\sum_{g=0}^\infty F^{(g)}_{\text{Gaussien, +}}\left(\frac{T}{N}\right)^{2g-2}\eea
En combinant avec le résultat de l'intégrale Gaussienne \eqref{ZNGaussien} on arrive à:
\bea \ln \,\mathcal{P}(T)&=& -\frac{N^2}{2}\ln\, N+\frac{3}{2}N^2 -\frac{N^2}{2}\ln\, T-\frac{N}{2}\ln(2\pi) +\frac{1}{6}\ln\, 2\cr
&&+ \ln\left(\prod_{i=1}^{N-1}i!\right)-\sum_{g=0}^\infty F^{(g)}_{\text{Gaussien, +}}\left(\frac{T}{N}\right)^{2g-2}\eea
d'où un asymptotique final en utilisant \eqref{Barnes}:
\bea \label{AsymptTheoFinal} \ln \,\mathcal{P}(T)&=& \frac{3}{4}N^2 -\frac{N^2}{2}\,\ln\, T -\frac{1}{12}\,\ln\, N+\xi'(-1)+\frac{1}{6}\,\ln\, 2\cr
&&-\sum_{g=0}^\infty F^{(g)}_{\text{Gaussien, +}}\left(\frac{T}{N}\right)^{2g-2}+\sum_{g=2}^{\infty} \frac{B_{2g}}{2g(2g-2)} N^{2-2g}\cr
\ln \,\mathcal{P}(T)&=&-\frac{\ln\, 3}{2}N^2-\frac{1}{12}\ln \,N-\frac{1}{6}\,\ln\, 2+\frac{1}{8}\,\ln\, 3+\xi'(-1)-\frac{53}{5760N^2}\cr
&&+\frac{6007}{1161216N^4} +\sum_{g=4}^\infty\left(\frac{B_{2g}}{2g(2g-2)}- F^{(g)}_{\text{Gaussien, +}}(T=1)\right)N^{2-2g}\cr
&&\eea

Notons qu'au final on retrouve le fait que la probabilité d'avoir des valeurs propres positives ne dépend pas de $T$. Cela est évident par la définition même de cette probabilité puisque le changement de variables $\lambda_i\to \sqrt{T}\lambda_i$ dans $Z_{N,+}^{\text{Gaussien}}(T)$ et $Z_N^{\text{Gaussien}}(T)$ produit la même puissance de $T$ et élimine donc sa dépendance dans $\mathcal{P}(T)$. Une fois le résultat théorique obtenu, il est intéressant de le comparer aux simulations numériques:

\medskip

\begin{center}\includegraphics[width=16cm]{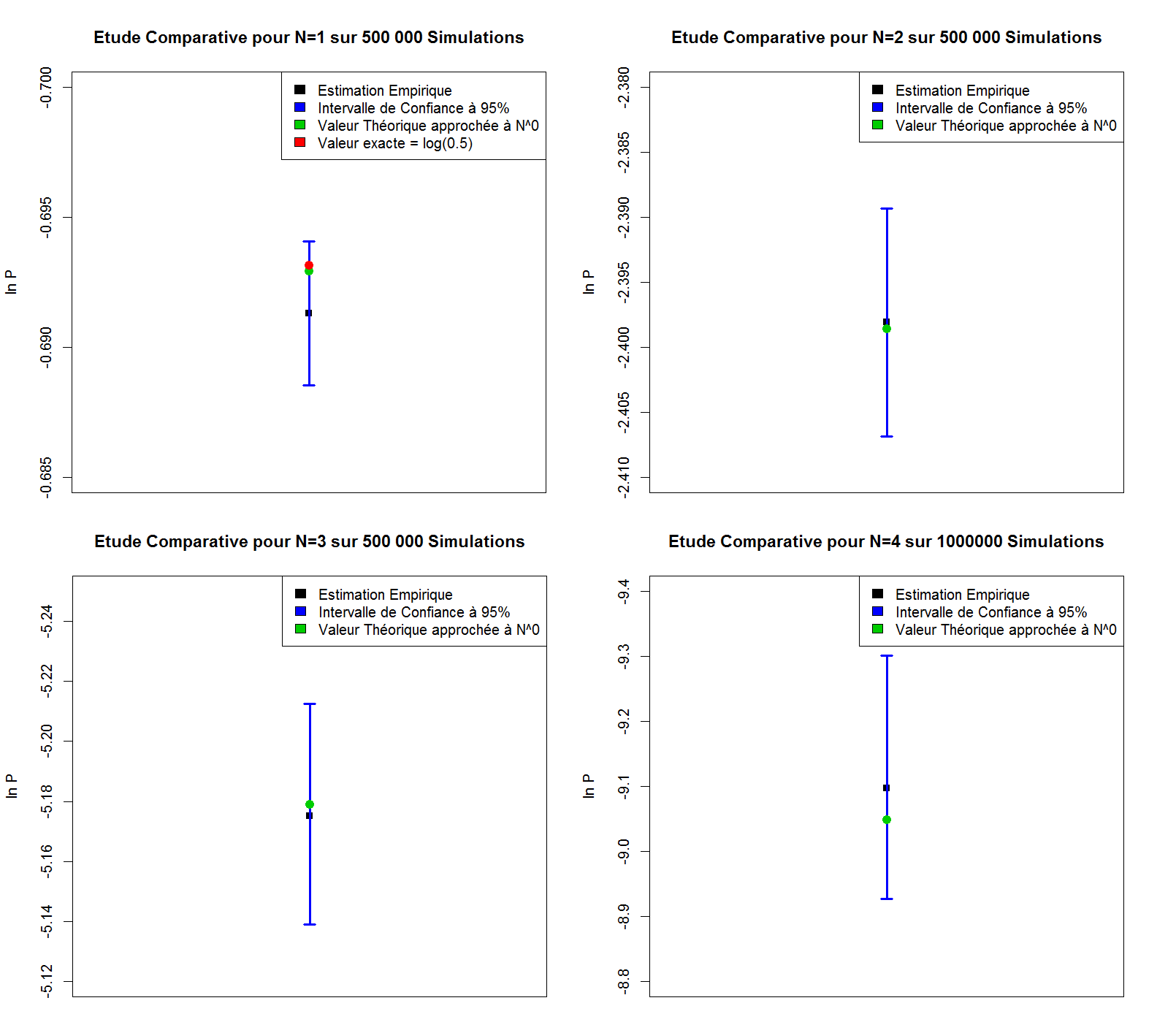}

\textit{Fig. $4$: Etude comparative sur $\mathcal{P}(T=1)$ entre la valeur asymptotique calculée à l'ordre $N^0$ et les simulations numériques pour $N\in\{1,2,3,4\}$ }
\end{center} 

\medskip

Les simulations numériques précédentes ont été réalisées en simulant un grand nombre $n$ ($n=5\times 10^5$ ou $n=10^6$ typiquement) de matrices hermitiennes Gaussiennes aléatoires et en comptant empiriquement le nombre de matrices présentant toutes ses valeurs propres strictement positives. Devant la baisse rapide des probabilités avec l'augmentation de $N$, qui nécessite donc pour obtenir une estimation correcte d'augmenter le nombre de simulations $n$, cette méthode n'est possible que pour des petites valeurs de $N$. Les intervalles de confiance utilisés correspondent aux intervalles de confiance standards utilisés pour l'estimation d'une probabilité:
\beqq  \text{IC}_{n,\alpha}(p)= \left[ f_n -n_{1-\frac{\alpha}{2}}\sqrt{\frac{f_n(1-f_n)}{n}},f_n +n_{1-\frac{\alpha}{2}}\sqrt{\frac{f_n(1-f_n)}{n}}\right]\eeqq
où $f_n$ est la fréquence empirique observée sur les $n$ simulations et $n_{1-\frac{\alpha}{2}}$ est la quantile de la loi normale centrée et réduite d'ordre $1-\frac{\alpha}{2}$. On observe ainsi que la valeur théorique approchée (calculée par \eqref{AsymptTheoFinal}) se situe dans tous les cas à l'intérieur de l'intervalle de confiance, ce qui signifie que le résultat théorique calculé ci-dessus est bien compatible avec les simulations numériques.

On voit donc que la méthode des équations de boucles et la récurrence topologique permettent, même dans des situations d'intégrales hermitiennes difficiles, d'obtenir des développements asymptotiques $N\to \infty$. Il faut néanmoins reconnaître trois faiblesses inhérentes à cette technique:

\begin{itemize}\item Cette technique n'est efficace que lorsque la densité limite d'équilibre possède un support composé d'un seul intervalle. Dans le cas contraire, il existe une forme générale de développement perturbatif plus compliquée que le simple développement topologique \cite{BG2} mais qui d'un point de vue calculatoire devient difficile en dehors de l'ordre dominant.
\item Même dans le cas favorable d'une densité limite à une coupure, la question de la normalisation de la fonction de partition pour l'identifier au développement topologique reste en général non triviale. En effet, la normalisation exacte de $Z_N$ ne joue aucun rôle dans le calcul des équations de boucles ni dans le calcul de la récurrence topologique. Cet aspect est à la fois un avantage (pour calculer les corrélations on peut se passer d'une normalisation précise de $Z_N$) mais aussi un inconvénient pour identifier les invariants symplectiques avec le développement de $\ln Z_N$. En particulier, il est nécessaire au préalable de montrer que $\ln Z_N$ admet bien un développement de type topologique (ce qui suppose de l'avoir normalisé convenablement) avant de pouvoir l'identifier. Comme on l'a vu dans les exemples précédents, cela est surtout possible lorsqu'on peut se ramener à une intégrale calculable explicitement (Selberg ou autres) pour au moins une valeur particulière du potentiel ou des bornes d'intégration. Dans le cas contraire, on peut toujours utiliser des simulations numériques pour chercher les premières corrections de façon empirique ou essayer d'appliquer des résultats d'homogénéité.
\item Il est nécessaire de vérifier par des arguments d'analyse qu'un développement topologique existe bel et bien pour la fonction de partition et les fonctions de corrélation. Cela n'est pas toujours le cas même si le potentiel n'a qu'un seul minimum et que la distribution limite possède un support composé d'un unique segment. L'exemple du potentiel $V(x)=|x|$ sans bord dur (i.e. intégré sur $\mathbb{R}$ tout entier) est ainsi typique. L'intégrale sur les valeurs propres est bien convergente et la distribution limite d'équilibre existe et est composée d'un seul intervalle symétrique autour de $0$ (qui est l'unique miminum de $V(x)=|x|$ sur l'axe réel). Néanmoins, il n'est pas attendu de développement topologique car la distribution limite est singulière et le potentiel non holomorphe. A l'heure actuelle, de telles intégrales restent mal comprises et le rôle de la récurrence topologique reste à préciser. On peut néanmoins espérer que la récurrence topologique puisse calculer une partie du développement asymptotique, qui pourrait être complétée par d'autres termes.
\end{itemize}

\section{Intégrales de matrices unitaires}

La technique de la récurrence topologique ne se limite pas à des calculs sur les matrices hermitiennes mais peut également être utilisée dans le cas des matrices unitaires. J'ai par exemple développé cet aspect dans l'étude des temps de récurrence forts de Poincaré pour une application à la physique théorique des mesures en mécanique quantique \cite{UnitaryMarchal}. L'approche de la récurrence topologique et des courbes spectrales permet ainsi d'améliorer certains résultats dus à Widom sur l'asymptotique des déterminants de Toeplitz \cite{Widom,Widom2} et également de compléter ou de retrouver certains résultats concernant les puissances de matrices aléatoires unitaires \cite{DJ}. D'un point de vue pratique, l'avantage des matrices unitaires est double:
\begin{itemize} \item Le groupe des matrices unitaire est compact et il possède donc la mesure de Haar comme mesure naturelle. La question de la convergence des intégrales est donc triviale en général.
\item Les valeurs propres sont toujours situées sur le cercle unité ce qui permet d'avoir un exemple de contour sans bord dans $\mathbb{C}$ qui ne soit pas l'axe réel.
\end{itemize} 
L'idée principale développée dans \cite{UnitaryMarchal} et dans \cite{ToeplitzNew} est que la mesure induite sur les valeurs propres de matrices unitaires aléatoires peut se réécrire sous plusieurs formes complémentaires données dans la proposition ci-dessous:

\begin{proposition}[Diverses reformulations des intégrales unitaires]\label{Mee} Si l'on définit $\mathcal{I}=\underset{j=1}{\overset{r}{\bigcup}} [\alpha_j,\beta_j] \subset[-\pi,\pi]$, $\mathcal{T}=\underset{j=1}{\overset{r}{\bigcup}} \left\{e^{it}\,,\,t\in [\alpha_j,\beta_j]\right\}$ et $\mathcal{J}=\underset{j=1}{\overset{r}{\bigcup}} [\tan \frac{\alpha_j}{2},\tan\frac{\beta_j}{2}]$, les quantités suivantes sont égales:
\begin{enumerate}\item Une intégrale de Toeplitz de symbole $f=\mathds{1}_{\mathcal{I}}$:
\bea \label{IntExp} Z_N(\mathcal{I})&=&\frac{1}{(2\pi)^NN!}\int_{[-\pi,\pi]^N} d\theta_1\dots d\theta_N \left(\prod_{k=1}^N f(e^{i\theta_k})\right)  \prod_{1\leq i<j\leq N}\left|e^{i \theta_i}-e^{i\theta_j}\right|^2\cr
&=&\frac{1}{(2\pi)^NN!}\int_{\mathcal{I}^N} d\theta_1\dots d\theta_N  \prod_{1\leq i<j\leq N}\left|e^{i \theta_i}-e^{i\theta_j}\right|^2
\eea
\item Le déterminant d'une matrice de Toeplitz:
\beq \label{ToepDet} Z_N(\mathcal{I})=I_N(\mathds{1}_{\mathcal{I}})= \det \left(T_{i,j}=t_{i-j}\right)_{1\leq i,j\leq N} \eeq
où les coefficients de Fourier de $f=\mathds{1}_{\mathcal{I}}$ sont donnés par:
\bea \label{FourierCoeff} t_0&=&\frac{1}{2\pi}\sum_{j=1}^r(\beta_j-\alpha_j)\cr
t_k&=&\frac{1}{2\pi}\sum_{j=1}^re^{ik\frac{\alpha_j+\beta_j}{2}}(\beta_j-\alpha_j)\sin_c\frac{k(\beta_j-\alpha_j)}{2} \,\,\,\,,\,\,\,\,  \forall\, k\neq 0
\eea
avec la notation $\sin_c(x)=\frac{\sin x}{x}$.
\item Une intégrale de type ``gaz de Coulomb'' avec un potentiel logarithmique $x\mapsto \ln(1+x^2)$:
\bea \label{RealInt} Z_N(\mathcal{I})&=&\frac{2^{N(N-1)}}{(2\pi)^N N!}\int_{\mathcal{I}^N} d\theta_1\dots d\theta_N  \prod_{1\leq i<j\leq N}\sin^2\left(\frac{\theta_i-\theta_j}{2}\right)\cr
&=& \frac{2^{N^2}}{(2\pi)^N N!}\int_{\mathcal{J}^N} dt_1\dots dt_N\,  \Delta(t_1,\dots,t_N)^2\,e^{-N\underset{k=1}{\overset{N}{\sum}} \ln(1+t_k^2)}
\eea
où $\Delta(t_1,\dots,t_N)$ est le déterminant de Vandermonde usuel.
\item Une intégrale de matrices Hermitiennes avec un support donné:
\beq \label{IntMatHerm}Z_N(\mathcal{I})= c_N\int_{\mathcal{N}_N(\mathcal{J})} \frac{dM_N}{\left(\det(I_N+M_N^2)\right)^N}\eeq
où $\mathcal{N}_N(\mathcal{J})$ est l'ensemble des matrices Hermitiennes dont les valeurs propres sont dans $\mathcal{J}$. La constante de normalisation $c_N$ est donnée par le volume du groupe unitaire:
\beqq c_N=\frac{1}{(2\pi)^NN!} \frac{1}{\text{Vol } \mathcal{U}_N}=\frac{1}{2^N\pi^{\frac{N(N+1)}{2}}}\underset{j=1}{\overset{N-1}{\prod}}j!\eeqq
\item Une intégrale complexe sur certains segments du cercle unité:
\beq \label{ComplexInt} Z_N(\mathcal{I})=(-1)^{\frac{N(N+1)}{2}}i^N \int_{\mathcal{T}^N} du_1\dots d u_N\, \Delta(u_1,\dots,u_N)^2 \,e^{-N\underset{k=1}{\overset{N}{\sum}} \ln \,u_k}\eeq  
\end{enumerate}
\end{proposition}

Notons que le choix des angles $(\alpha_j)_{1\leq j\leq r}$ et $(\beta_j)_{1\leq j\leq r}$ n'est défini qu'à un choix modulo $2\pi$. Par simplicité (pour pouvoir appliquer la fonction $x\mapsto \tan \frac{x}{2}$), nous supposerons désormais que ces angles sont choisis dans l'intervalle $[-\pi,\pi]$. La démonstration de la proposition précédente est immédiate et réalisée en détail dans \cite{ToeplitzNew}. Les différentes intégrales s'obtiennent par changement de variables ($u_i=e^{i\theta_i}$ et $t_i=\tan\left(\frac{\theta_i}{2}\right)$ tandis que la reformulation de l'intégrale de Toeplitz en termes de déterminants est bien connue \cite{BS}. On rappelle ainsi:

\begin{definition}[Déterminants de Toeplitz]\label{Toeplitz} Soit $f$ une fonction mesurable sur le cercle unité alors l'intégrale:
\beqq \mathcal{I}_N(f)=\frac{1}{(2\pi)^N N!}\int_{[-\pi,\pi]^N} \prod_{k=1}^N f(e^{i\theta_k}) \prod_{i<j} |e^{i\theta_i}-e^{i\theta_j}|^2 d\theta_1\dots d\theta_N\eeqq
s'appelle une intégrale de Toeplitz. Elle est égale au déterminant de la matrice de Toeplitz associée suivante:
\beqq \mathcal{I}_N(f)=\det\left((T_N(f))_{i,j}=t_{i-j}\right)_{1\leq i,j\leq N} \,\,\normalfont{\text{ où }}\,\, t_k=\frac{1}{2\pi}\int_0^{2\pi}f(e^{i\theta})e^{-ik\theta}d\theta \,\,\text{ où }\,\, -N\leq k\leq N\eeqq
Inversement, pour une matrice de Toeplitz $T_N$ donnée, on peut recontruire sur le cercle unité la fonction $f$, appelée ``symbole'', par sa série de Fourier:
\beqq f(e^{i\theta})=\sum_{k=-N}^N t_k e^{ik\theta}\eeqq
\end{definition}

Chacune des reformulations précédentes possèdent ses avantages et ses inconvénients. L'intégrale \eqref{ComplexInt} permet de relier le problème à des intégrales sur des matrices unitaires dont les valeurs propres sont les $(u_k)_{k\geq 1}$. La reformulation angulaire \eqref{IntExp} permet de faire le lien avec les déterminants de Toeplitz \eqref{ToepDet} qui peuvent se calculer numériquement de façon très rapide y compris pour des valeurs de $N$ assez élevées (Cf. simulations numériques). Enfin la reformulation en termes de gaz de Coulomb \eqref{RealInt} et de valeurs propres de matrices hermitiennes \eqref{IntMatHerm} permet d'appliquer les résultats de la théorie issue des matrices Hermitiennes et en particulier la récurrence topologique ainsi que les résultats de \cite{BG,BG2}. 

\medskip

Il est à noter également que, contrairement au cas hermitien, la question de la normalisation est ici relativement simple. En effet, il est facile de montrer que:
\beq \label{NormalisationUnitaire} Z_N([-\pi,\pi])=\frac{1}{(2\pi)^NN!}\int_{[-\pi,\pi]^N} d\theta_1\dots d\theta_N \left(\prod_{i<j}^N |e^{i\theta_i}-e^{i\theta_j}|^2\right)=1\eeq
permettant ainsi une normalisation naturelle. Ce résultat peut se montrer par le développement de $\Delta(e^{i\theta_1},\dots,e^{i\theta_N})^2$ à l'aide de sommes sur les permutations (Cf. Annexe A de \cite{UnitaryMarchal}). Il est également évident par la correspondance avec les déterminants de Toeplitz (qui donne le déterminant de la matrice identité). Les déterminants de Toeplitz et le calcul de leurs asymptotiques est un domaine des mathématiques qui a été étudié par de nombreuses personnes et en particulier par Szegö et Widom qui ont obtenu de nombreux résultats.

\subsection{Déterminants de Toeplitz et asymptotique de Widom}

Compte tenu des résultats précédents, une première question naturelle est d'évaluer la probabilité qu'une matrice unitaire tirée uniformément suivant la mesure de Haar possède toutes ses valeurs propres dans un secteur angulaire $[\alpha,\beta]\subset [-\pi,\pi]$ du cercle unité. Le problème étant manifestement invariant par rotation, il est évident que cette probabilité ne dépend pas du choix de la détermination des angles mais seulement de la largeur du domaine d'intégration $|\beta-\alpha|$. En pratique cette probabilité est donnée en utilisant \eqref{NormalisationUnitaire} par:
\beq \label{ProbaUnitaire} \mathcal{P}_{N,\alpha,\beta}=\frac{1}{(2\pi)^N N!} \int_{[\alpha,\beta]^N} d\theta_1\dots d\theta_N \left(\prod_{i<j}^N |e^{i\theta_i}-e^{i\theta_j}|^2\right)=Z_N([\alpha,\beta]) \eeq 

Cette probabilité est ainsi directement en lien avec le calcul d'un déterminant de Toeplitz. Malheureusement, si la réécriture du problème intégral en un déterminant de taille $N\times N$ laisse penser que le problème a été considérablement simplifié, il n'en est rien. En effet, le calcul des asymptotiques $N\to \infty$ des déterminants de Toeplitz est notoirement difficile suivant les propriétés du symbole $f$ (Cf. \cite{Szego,Widom,Widom2,BS,BW,FishHart,BM,BO,Krasovsky}). Les résultats les plus connus concernant ces asymptotiques sont ainsi:

\begin{proposition}[Asymptotiques connus des déterminants de Toeplitz]\label{ResultatsToep} Les cas classiques des déterminants de Toeplitz sont donnés par:
\begin{itemize} \item Si $f$ est continue et strictement positive sur le cercle unité alors (Szegö \cite{Szego}):
\beqq \frac{1}{N}\ln\, \det \,T_N(f)\overset{N\to \infty}{\to} \frac{1}{2\pi}\int_0^{2\pi}\ln(f(\lambda)) d\lambda\eeqq
\item Si $f$ est continue par morceaux et strictement positive en dehors de points isolés sur le cercle unité (on parle de singularités de type Fisher-Hartwig, i.e. de sauts isolés et dénombrables de la fonction $f$) alors $\frac{1}{N}\,\ln\, \det\, T_N(f)$ admet une limite lorsque $N\to \infty$ (Cf. \cite{FishHart,BM} pour la formule explicite de la limite en fonction du symbole).
\end{itemize}
\end{proposition}

Ainsi, le cas de la normalisation $f=1$ tombe dans la première catégorie classique étudiée par Szegö et reste cohérent avec \eqref{NormalisationUnitaire}. En revanche, on s'aperçoit que l'intégrale \eqref{ProbaUnitaire} ne tombe pas dans ces deux grandes catégories (qui ont fait l'objet d'études détaillées incluant les premières corrections aux limites présentées ci-dessus) puisque le symbole est une fonction indicatrice $\mathds{1}_{[\alpha,\beta]}$ qui est donc nulle sur des segments entiers. Le seul cas similaire étudié dans la littérature est dû à H. Widom \cite{Widom}:

\begin{proposition}[Asymptotique de Widom]\label{ResultatsToepWidom} Pour $f=\mathds{1}_{[\alpha,\beta]}$ on a l'asymptotique suivant \cite{Widom}:
\beqq \lim_{N\to \infty} \frac{1}{N^2}\,\ln\, \det \,T_N(\mathds{1}_{[\alpha,\beta]})= \ln \left(\sin \left(\frac{\left|\beta-\alpha\right|}{4}\right)\right)\eeqq
\end{proposition} 

Il existe donc clairement une différence de comportement asymptotique entre le cas où le symbole $f$ ne s'annule qu'au plus en un nombre dénombrable de points isolés (Proposition \ref{ResultatsToep}) et le cas où il s'annule sur des segments complets (Proposition \ref{ResultatsToepWidom}). En effet, l'ordre dominant est modifié: on passe de $\frac{1}{N}\ln Z_N$ convergent à $\frac{1}{N^2}\ln Z_N$ convergent. Comme nous allons le voir, la technique des équations de boucles et la récurrence topologique permettent d'aller beaucoup plus loin que l'ordre dominant dans ce dernier cas. En effet, on peut tout à fait appliquer la théorie générale de \cite{BG,BG2} à la reformulation \eqref{RealInt} comme cela a été réalisé dans \cite{ToeplitzNew}.

\subsection{Courbe spectrale dans le cas d'un seul intervalle}  

En reprenant la reformulation \eqref{RealInt} dans le cas à un seul intervalle ($r=1$), nous obtenons l'intégrale:
\beq Z_N([\alpha,\beta])= \frac{2^{N^2}}{(2\pi)^N N!}\int_{\left[\tan \frac{\alpha}{2},\tan \frac{\beta}{2}\right]^N} dt_1\dots dt_N \, \Delta(t_1,\dots,t_N)^2\,e^{-N\underset{k=1}{\overset{N}{\sum}} \ln(1+t_k^2)}
\eeq
Dans le cas d'un seul intervalle $[\alpha,\beta] \subset(-\pi,\pi)$ (le cas $[-\pi,\pi]$ étant donné par \eqref{NormalisationUnitaire}), on peut utiliser l'invariance par rotation pour se ramener au cas où $\beta=-\alpha\equiv \gamma$:
\beq \label{RealInt1Cut} Z_N([-\gamma,\gamma])= \frac{2^{N^2}}{(2\pi)^N N!}\int_{[-a,a]^N} dt_1\dots dt_N \, \Delta(t_1,\dots,t_N)^2\,e^{-N\underset{k=1}{\overset{N}{\sum}} \ln(1+t_k^2)}
\eeq
où l'on a noté $\mathbf{a}\boldsymbol{=}\text{tan}\frac{\gamma}{2}$. L'existence de la densité limite des valeurs propres de \eqref{RealInt1Cut} suit la théorie standard évoquée dans les paragraphes précédents. Son calcul explicite peut se réaliser par la méthode des équations de boucles. Pour cela, notons que \eqref{RealInt1Cut} est une intégrale de type gaz de Coulomb avec deux bords durs en $t=\pm a$. Les équations de boucles des intégrales Hermitiennes avec bords durs se retrouvent à de multiples endroits dans la littérature \cite{Hardedges,HardWall,Hardedges2}. Dans notre cas on obtient en notant $W_{n,a}$ les fonctions de corrélations connexes issues de l'intégrale \eqref{RealInt1Cut}:
\beq \label{FirstLoop} W_{1,a}^2(x)+W_{2,a}(x,x)-\frac{N}{x}W_{1,a}(x)+N\left<\sum_{i=1}^N\frac{V'(x)-V'(t_i)}{x-t_i}\right>_a=\frac{c_1}{x-a}+\frac{c_2}{x+a}\eeq
où l'on a noté $\left<g(t_1,\dots,t_N)\right>_a$ la valeur moyenne par rapport à la mesure issue de \eqref{RealInt1Cut}.
L'équation précédente est exacte (valable quel que soit $N$ donné) et les coefficients $c_1$ et $c_2$ sont donnés par:
\beq c_1=-\left<\sum_{i=1}^N \frac{1}{a+t_i}\right>_a \text{ et } c_2=\left<\sum_{i=1}^N \frac{1}{a-t_i}\right>_a\eeq
Les résultats généraux de \cite{BG} nous garantissent alors que $W_{1,a}(x)\underset{N\to \infty}{\sim} NW_{1,a}^{(0)}(x)$ et $W_{2,a}(x_1,x_2)\underset{N\to \infty}{=} O(N\ln N)$. En définissant $y(x)=W_{1,a}^{(0)}(x)-\frac{1}{2}V'(x)=W_{1,a}^{(0)}(x)-\frac{x}{1+x^2}$ on obtient l'équation:
\bea \label{LoopProject} y(x)^2&=&\frac{x^2}{(1+x^2)^2}-\frac{2}{1+x^2}\left(\underset{N\to \infty}{\lim}\left<\frac{1}{N}\sum_{i=1}^N\frac{1}{1+t_i^2}\right>_a-x\underset{N\to \infty}{\lim}\left<\frac{1}{N}\sum_{i=1}^N\frac{t_i}{1+t_i^2}\right>_a\right)\cr
&&+\frac{c(a)}{x-a}-\frac{c(a)}{x+a}\eea
où la constante $c(a)$ est donnée par $c(a)=-\underset{N\to \infty}{\lim}\left<\frac{1}{N}\underset{i=1}{\overset{N}{\sum}} \frac{1}{a+t_i}\right>_a$. En utilisant l'invariance $\mathbf{t}\mapsto -\mathbf{t}$ de l'intégrale \eqref{RealInt1Cut} on obtient $\left<\underset{i=1}{\overset{N}{\sum}}\frac{t_i}{1+t_i^2}\right>_a=0$ et donc:
\beq \label{LoopProject2} y(x)^2=\frac{x^2}{(1+x^2)^2}-\frac{2d(a)}{1+x^2}+\frac{c(a)}{x-a}-\frac{c(a)}{x+a}\eeq
où $d(a)$ et $c(a)$ sont des constantes (dans le sens indépendantes de $x$) indéterminées pour le moment. Par définition de la fonction de corrélation $W_1^{(0)}(x)$ on doit avoir l'asymptotique suivant:
\beq W_{1,a}^{(0)}(x)=\frac{1}{x}-\underset{N\to \infty}{\lim}\left<\frac{1}{N}\sum_{i=1}^Nt_i\right>_a\frac{1}{x^2}+O\left(\frac{1}{x^3}\right)=\frac{1}{x}+O\left(\frac{1}{x^3}\right)
\eeq
puisque l'intégrale \eqref{RealInt1Cut} est invariante par $\mathbf{t}\mapsto -\mathbf{t}$. Cela implique en particulier que $y^2(x)=O\left(\frac{1}{x^6}\right)$ ce qui permet de déterminer de façon unique les constantes $(c(a),d(a))$. On trouve ainsi $c(a)=\frac{1}{2a(1+a^2)}$ et $d(a)=\frac{2+a^2}{2(1+a^2)}$ ce qui donne la courbe spectrale:
\beq \label{SpectralCurve1}y^2(x)=\frac{1+a^2}{(1+x^2)^2(x^2-a^2)}=\frac{1}{\cos^2(\frac{\gamma}{2})(1+x^2)^2(x^2-\tan^2(\frac{\gamma}{2}))}\eeq
et donc par transformée de Stieltjes inverse la densité limite des valeurs propres:
\beq \label{mu}d\mu_\infty(x)= \frac{dx}{\pi\cos(\frac{\gamma}{2})(1+x^2)\sqrt{\tan^2(\frac{\gamma}{2})-x^2}}\,\mathds{1}_{\left[-\tan \frac{\gamma}{2},\tan \frac{\gamma}{2}\right]}(x)\eeq
On peut vérifier numériquement ce résultat par des simulations de type Monte-Carlo suivant la mesure de \eqref{RealInt1Cut}: 

\begin{center}
\includegraphics[width=8cm]{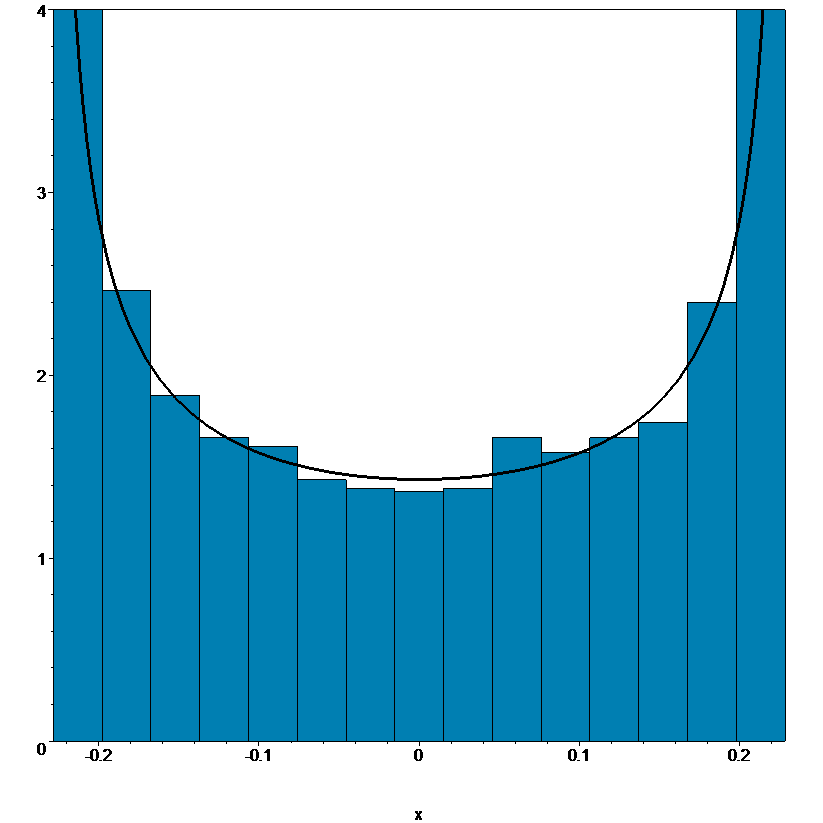}

Fig. $5$: Densité empirique des valeurs propres obtenue pour $100$ simulations Monte-Carlo indépendantes de l'intégrale \eqref{RealInt1Cut} dans le cas où $\gamma=\frac{\pi}{7}$ et $N=20$. La courbe noire est la densité limite théorique donnée par \eqref{mu}.
\end{center}

\subsection{Existence d'un développement en $\frac{1}{N}$ pour $Z_N([-\gamma,\gamma])$}

Une fois la densité limite des valeurs propres obtenue, il est nécessaire de montrer l'existence d'un développement en $\frac{1}{N}$ des fonctions de corrélations et de la fonction de partition $\ln Z_N([-\gamma,\gamma])$. Pour cela, on peut utiliser les résultats principaux de \cite{BG,BG2} grâce à la proposition suivante:

\begin{proposition} Les conditions suivantes (Hypothèses $1.1$ de \cite{BG}) sont réalisées pour $Z_N([-\gamma,\gamma])$ donnée par \eqref{RealInt1Cut}:
\begin{itemize}\item (Regularité): Le potentiel $V$ est continu sur le domaine d'intégration $[b_-,b_+]$. Dans notre cas, cela revient à dire que la fonction $x\mapsto \ln(1+x^2)$ est continue sur $[b_-,b_+]=[-a,a]$ avec $a=\tan\frac{\gamma}{2}$. Notons également que dans notre cas, le potentiel $V$ ne dépend pas de $N$ ce qui simplifie certaines hypothèses de \cite{BG}.
\item (Confinement du potentiel): Non requis ici puisque le contour d'intégration $[-a,a]$ est un compact de $\mathbb{R}$.
\item (Régime à une coupure): Le support de la densité limite est composé d'un unique intervalle $[\alpha_-,\alpha_+]$ non réduit à un point. Dans notre cas, ce résultat est trivial d'après l'expression explicite \eqref{mu} de $d\mu_{\infty}$.
\item (Contrôle des grandes déviations): La fonction $x\mapsto \frac{1}{2}V(x)+\int |x-\xi|d\mu_\infty(\xi)$ définie sur $[b_-,b_+]\setminus(\alpha_-,\alpha_+)$ atteint son minimum uniquement en $\alpha_-$ ou $\alpha_+$. Dans notre cas, ceci est trivial puisque $\alpha_-=b_-=-a$ et $\alpha_+=b_+=a$ (la densité limite possède un support qui couvre l'ensemble du domaine d'intégration initial).
\item (Non-Criticalité): La densité limite est non-critique dans le sens où elle est strictement positive sur l'intérieur de son support et se comporte aux extrémités comme $O\left(\frac{1}{\sqrt{x-b_\pm}}\right)$ si $b_\pm$ est un bord dur et comme $O\left(\sqrt{x-\alpha_\pm}\right)$ si $\alpha_\pm$ est un bord mou. Dans notre cas, nous avons seulement deux bords durs et le comportement aux bords donné par \eqref{mu} est bien celui requis. Il est également trivial d'après l'expression explicite \eqref{mu} de $d\mu_\infty$ de vérifier que la densité limite est strictement positive sur son support.
\item (Analyticité): Le potentiel $V$ peut être prolongé en une fonction analytique dans un voisinage de $[\alpha_-,\alpha_+]$. Dans notre cas, cela est trivial puisque  $x\mapsto \ln(1+x^2)$ est analytique dans un voisinage de $[-a,a]$.
\end{itemize}
\end{proposition}

Ainsi toutes les hypothèses de \cite{BG,BG2} sont vérifiées et l'on obtient donc l'existence d'un développement topologique pour les fonctions de corrélations:

\begin{theorem}[Existence d'un développement topologique]\label{LargeNExp} Les fonctions de corrélations et $Z_N([-\gamma,\gamma])$ admettent un développement (dit topologique) pour $N\to \infty$ de la forme:
\beaa \label{RRR} W_{p,a}(x_1,\dots,x_p)&=&\sum_{g=0}^{\infty} W_{p,a}^{\{2-p-2g\}}(x_1,\dots,x_p)N^{2-p-2g}\cr
 Z_N([-\gamma,\gamma])&=&\frac{N^{N+\frac{1}{4}}}{N!}\,\exp\left(\sum_{k=-2}^\infty \td{F}^{\{k\}}(a)N^{-k} \right)\cr
\ln\, Z_N([-\gamma,\gamma])&=&-\frac{1}{4}\,\ln\, N+\sum_{k=-2}^\infty F^{\{k\}}(a)N^{-k}
\eeaa
\end{theorem} 

Par ailleurs les coefficients de ces développements peuvent être obtenus par l'application de la récurrence topologique à la courbe spectrale \eqref{SpectralCurve1}:
\begin{corollary}[Reconstruction des coefficients via la récurrence topologique] Pour $p\geq 1$ et $g\geq 0$ on a:
\beqq W_{p,a}^{^{\{2-p-2g\}}}(x_1,\dots,x_p)dx_1\dots dx_p=\omega_p^{(g)}(x_1,\dots,x_p)\eeqq
où $\omega_p^{(g)}(x_1,\dots,x_p)$ est la différentielle d'Eynard-Orantin \cite{EO} obtenue par l'application de la récurrence topologique sur la courbe spectrale de genre $0$ \eqref{SpectralCurve1}. Les coefficients $F^{\{k\}}(a)$ du développement de $\ln\, Z_N([-\gamma,\gamma])$ sont reliés aux invariants symplectiques (aussi appelés ``énergies libres'') $\left(F^{(g)}\right)_{g\geq 1}$ produit par la récurrence topologique appliquée à la courbe spectrale \eqref{SpectralCurve1} par:
\beaa F^{\{2k+1\}}(a)&=&f^{\{2k+1\}} \text{ indépendant de } a\cr
F^{\{2k\}}(a)&=&-F_{\text{Top. Rec.}}^{(k+1)}(a)+ f^{\{2k\}} \text{ avec } f^{\{2k\}}\text{indépendant de } a
\eeaa 
\end{corollary}

Comme dans les cas précédents (Gaussien, exponentiel ou Gaussien avec valeurs propres positives), l'identification des coefficients du développement de $\ln\, Z_N$ avec les invariants symplectiques issus de la récurrence topologique n'est valable qu'à des constantes (i.e. indépendantes de $\gamma$) près. Afin de fixer ces constantes, il est nécessaire de se ramener à un cas où l'intégrale est explicitement connue. Dans le cas présent, il n'est pas possible d'utiliser $\gamma\to \pi$ puisque le développement topologique diverge trivialement (en effet $\ln\, Z_N([-\pi,\pi])=0$ qui n'admet pas de développement en $\frac{1}{N}$). On peut en revanche utiliser la limite $\gamma\to 0$ (i.e. $a\to 0$) qui peut être reliée directement à une intégrale de Selberg.

\subsection{Normalisation à l'aide d'une intégrale de Selberg}
On a l'intégrale de Selberg suivante:
\beq \label{Selberg} S_N(1,1,1)=\int_{[-1,1]^N} \prod_{1\leq i<j\leq N} (u_i-u_j)^2\,du_1\dots du_N=\frac{2^{N^2}(N!)\left(\underset{j=1}{\overset{N-1}{\prod}}j!\right)^4}{\underset{j=1}{\overset{2N-1}{\prod}}j!}\eeq
donc
\bea \label{CasRef} Z_N([-\gamma,\gamma])&=&\frac{2^{N^2}}{(2\pi)^N N!}\int_{[-a,a]^N} \Delta(t_1,\dots,t_N)^2\, e^{-N\underset{i=1}{\overset{N}{\sum}}\ln(1+t_i^2)}\,dt_1\dots dt_N\cr
&=&\frac{a^{N^2}2^{N^2}}{(2\pi)^N N!}\int_{[-1,1]^N} \Delta(u_1,\dots,u_N)^2\,e^{-N\underset{i=1}{\overset{N}{\sum}}\ln(1+a^2u_i^2)}\,du_1\dots du_N\cr
&\overset{\text{def}}{=}&\frac{a^{N^2}2^{N^2}}{(2\pi)^N N!}S_a
\eea
avec:
\beq S_a=\int_{[-1,1]^N} \Delta(u_1,\dots,u_N)^2\,e^{-N\underset{i=1}{\overset{N}{\sum}}\ln(1+a^2t_i^2)}\,du_1\dots du_N\eeq
$S_a$ est une intégrale Hermitienne sur $I=[-1,1]$ avec potentiel $V_a(x)=\ln(1+a^2x^2)$. Elle est donc continue en $a$ et en particulier pour $a=0$ on trouve:
\beq S_0=S_N(1,1,1)=\frac{2^{N^2}(N!)\left(\underset{j=1}{\overset{N-1}{\prod}}j!\right)^4}{\underset{j=1}{\overset{2N-1}{\prod}}j!}\eeq
La normalisation des coefficients de $\ln Z_N([-\gamma,\gamma])$ peut donc être obtenue par la condition:
\beq \label{NormalizationEq}\ln \,Z_N([-\gamma,\gamma])-N^2\ln(a)\underset{a\to 0}{\to} \ln\left(S_N(1,1,1)\right)+2N^2\,\ln\, 2-N\,\ln(2\pi)-\ln(N!)\eeq
où le membre de droite possède un développement $N\to \infty$ explicitement calculable. La comparaison des deux membres de \eqref{NormalizationEq} donne ainsi:
\bea \label{Constantes} f^{\{-2\}}&=&\ln\, 2\cr
f^{\{-1\}}&=&0\cr
f^{\{0\}}&=&F^{(1)}(a=0)+3\,\xi'(-1)+\frac{1}{12}\ln \,2=3\,\xi'(-1)+\frac{1}{12}\,\ln\, 2\cr
f^{\{2g\}}&=&F^{(g+1)}(a=0)+\frac{4(1-2^{-2g-2})B_{2g+2}}{2g(2g+2)} \text{  for  } g\geq 1\cr
f^{\{2k+1\}}&=&0 \text{  for  } k\geq 0
\eea
Notons en particulier que $\ln Z_N([-\gamma,\gamma])$ possède un développement qui n'admet que des puissances paires en $N$ puisque les constantes potentielles $\left(f^{\{2k+1\}}\right)_{k\geq -1}$ sont nulles d'après la normalisation. 

\subsection{Conclusion sur le cas à une coupure}

Ainsi, avec le calcul explicite des premiers invariants symplectiques de la courbe spectrale \eqref{SpectralCurve1} (réalisé dans \cite{ToeplitzNew}) on obtient le théorème suivant (Cf. \cite{ToeplitzNew} pour les détails):

\begin{theorem}[Asymptotique des déterminants de Toeplitz de symbole $f=\mathds{1}_{[\alpha,\beta]}$ \label{Toeplitz1Cut}] Pour $(\alpha,\beta)$ tels que $0<|\beta-\alpha|<2\pi$, le déterminant de Toeplitz $T_N(f)$ de symbole $f=\mathds{1}_{[\alpha,\beta]}$ admet un développement $N\to \infty$ de la forme:
\small{\beaa &&\ln Z_N([\alpha,\beta])=\ln \,\det T_N(\mathds{1}_{[\alpha,\beta]})=N^2\,\ln\left(\sin\left(\frac{|\beta-\alpha|}{4}\right)\right)-\frac{1}{4}\,\ln\, N -\frac{1}{4}\,\ln\left(\cos\left(\frac{|\beta-\alpha|}{4}\right)\right)\cr 
&&+3\,\xi'(-1)+\frac{1}{12}\,\ln\, 2+ \sum_{g=1}^\infty \left(F^{(g+1)}(a=0)-F^{(g+1)}(a)+\frac{4(1-2^{-2g-2})B_{2g+2}}{2g(2g+2)}\right)N^{-2g}\eeaa}\normalsize{}
où $a=\tan\left(\frac{|\beta-\alpha|}{4}\right)$ et les coefficients $\left(F^{(g)}(a)\right)_{g\geq 2}$ sont les invariants symplectiques de la récurrence topologique d'Eynard-Orantin de la courbe spectrale: 
\beqq y^2(x)=\frac{1}{\cos^2\left(\frac{|\beta-\alpha|}{4}\right)(1+x^2)^2\left(x^2-\tan^2\left(\frac{|\beta-\alpha|}{4}\right)\right)}\eeqq
En particulier, les premiers ordres sont donnés par:
\beaa &&\ln Z_N([\alpha,\beta])=\ln\, \det T_N(\mathds{1}_{[\alpha,\beta]})=N^2\,\ln\left(\sin\left(\frac{|\beta-\alpha|}{4}\right)\right)\cr
&&-\frac{1}{4}\ln N-\frac{1}{4}\ln\left(\cos\left(\frac{|\beta-\alpha|}{4}\right)\right)+3\,\xi'(-1)+\frac{1}{12}\ln\, 2+\frac{1}{64N^2}\left(2\tan^2\left(\frac{|\beta-\alpha|}{4}\right)-1\right)\cr
&&+ \frac{1}{256N^4}\left(1+2\tan^2\left(\frac{|\beta-\alpha|}{4}\right)+10\tan^4\left(\frac{|\beta-\alpha|}{4}\right)\right)+O\left(\frac{1}{N^6}\right)
\eeaa
\end{theorem}

Ce théorème peut être vérifié de façon très précise à l'aide de simulations numériques:

\begin{center}
\includegraphics[width=12cm]{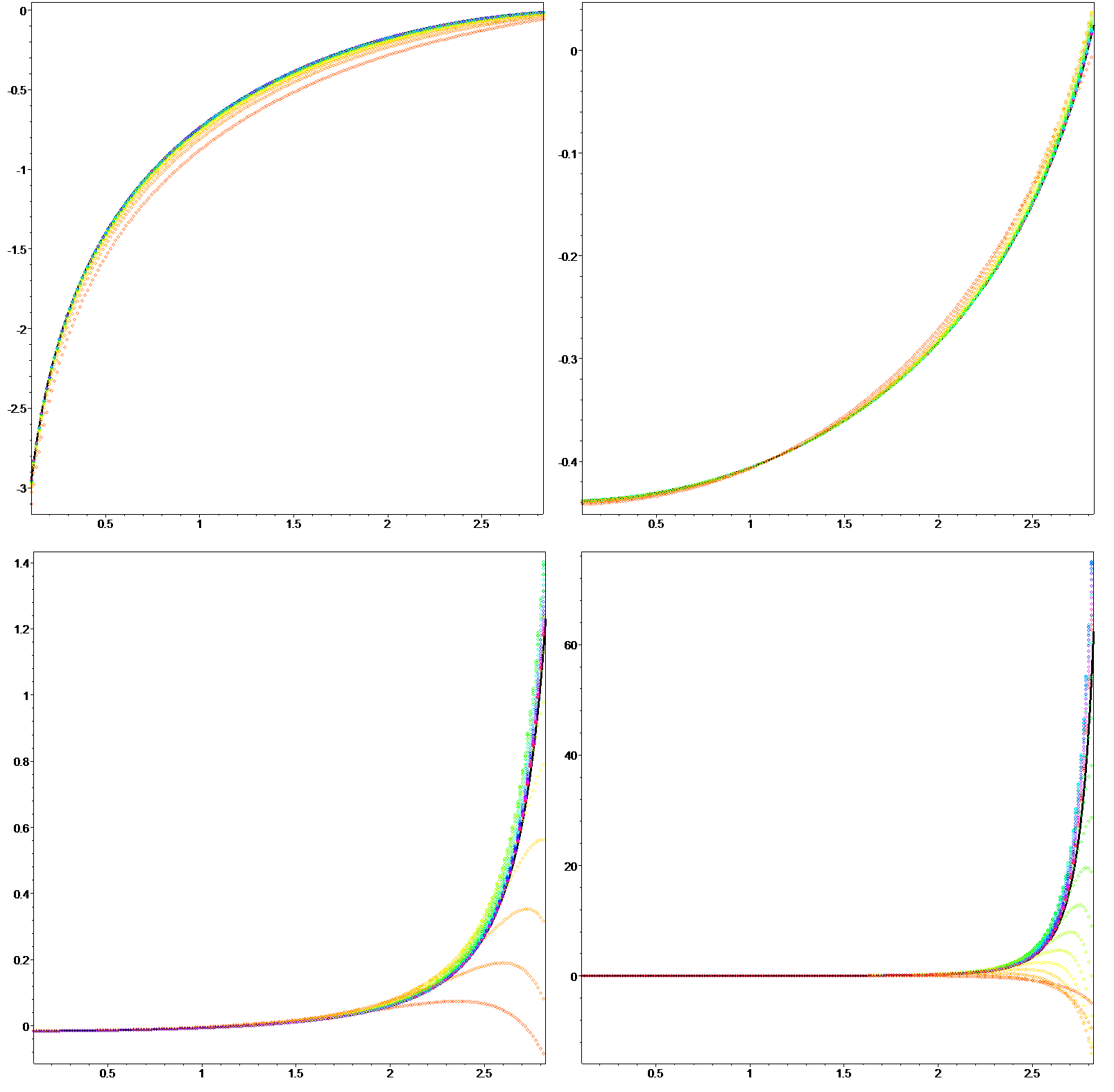}

Fig. $6$: Calculs exacts (à l'aide de la reformulation en termes de déterminants de Toeplitz \eqref{ToepDet}) de $\gamma\mapsto \ln\, Z_N([-\gamma,\gamma])$ pour $2\leq N\leq 35$ avec soustraction des premiers ordres du développement asymptotique donnés par le Théorème \ref{Toeplitz1Cut} (Points colorés: d'orange à jaune, puis vert, et violet à mesure que $N$ augmente). Les courbes noires représentent les prédictions théoriques (de droite à gauche et de haut en bas): $\gamma\mapsto \ln\left(\sin\left(\frac{\gamma}{2}\right)\right)$, $\gamma\mapsto -\frac{1}{4}\ln\left(\cos\left(\frac{\gamma}{2}\right)\right)+3\,\xi'(-1)+\frac{1}{12}\ln\, 2$, $\gamma\mapsto \frac{1}{64}\left(2\tan^2\left(\frac{\gamma}{2}\right)-1\right)$ et $\gamma \mapsto \frac{1}{256}\left(1+2\tan^2\left(\frac{\gamma}{2}\right)+10\tan^4\left(\frac{\gamma}{2}\right)\right)$.
\end{center}

Notons que la reformulation en termes de déterminants de Toeplitz est particulièrement intéressante d'un point de vue numérique car elle permet de faire des calculs exacts jusqu'à des valeurs de $N$ relativement élevées. Ainsi les calculs ci-dessus, réalisés à l'aide du logiciel Maple, n'ont pris que quelques minutes sur un ordinateur standard.

\medskip

Les simulations numériques montrent ainsi une concordance parfaite avec les prédictions théoriques établies jusqu'à l'ordre $O\left(\frac{1}{N^6}\right)$. Ces résultats permettent également d'améliorer le résultat de Widom (Théorème \ref{ResultatsToepWidom}) et de fournir un moyen efficace et complet de calculer l'asymptotique à tout ordre de ces déterminants de Toeplitz.

\medskip

Soulignons ici que dans le cas d'un support restreint à un seul intervalle $[\alpha,\beta]$, la méthode présentée ci-dessus pourrait parfaitement s'appliquer avec des symboles $f$ plus généraux de la forme $f=g(e^{i\theta_k})\mathds{1}_{[\alpha,\beta]}$ avec $g$ une fonction mesurable strictement positive sur $[\alpha,\beta]$. Dans ce cas, la courbe spectrale est différente mais reste potentiellement calculable ainsi que les déterminants de Toeplitz et les théorèmes généraux sur la forme de l'asymptotique issus de \cite{BG,BG2}. A ma connaissance, cette approche n'a jamais été effectuée et mériterait une certaine attention puisque certains éléments des développements sont amenés à changer. Un point de départ est ainsi abordé dans le paragraphe suivant.

\section{Résumé des cas précédents}
Une partie des résultats précédents peut être regroupée dans le tableau suivant:
\medskip

\leftskip -2cm\footnotesize{\begin{tabular}{|c|c|c|c|c|}
\hline
Problème& Gaussien& Exponentiel& Gaussien positif& Indicatrice Toeplitz\\ \hline 
$\begin{array}{c}\text{Potentiel}\\ V(x)\end{array}$& $\frac{x^2}{2T}$& $\left\{\begin{array}{ll}\frac{x}{T} & \mbox{si } x>0 \\+\infty & \mbox{sinon}\end{array}\right.$& $\left\{\begin{array}{ll}\frac{x^2}{2T} & \mbox{si } x>0 \\+\infty & \mbox{sinon}\end{array}\right.$&$\left\{\begin{array}{ll}\ln(1+x^2) & \mbox{si } x\in \left[\tan \frac{\alpha}{2},\tan \frac{\beta}{2}\right] \\+\infty & \mbox{sinon}\end{array}\right.$\\
\hline
$\begin{array}{c}
\text{Courbe}\\ \text{spectrale}\end{array}$& $y^2=x^2-4T$& $y^2=\frac{x-4T}{4x}$& $y^2=\frac{\left(x+2\sqrt{\frac{T}{3}}\right)^2\left(x-4\sqrt{\frac{T}{3}}\right)}{4x}$& $y^2=\frac{1}{\cos^2\left(\frac{|\beta-\alpha|}{4}\right)\,(1+x^2)^2(x^2-\tan^2\frac{|\beta-\alpha|}{4})}$	\\
\hline
$\begin{array}{c}\text{Support}\\ \text{densité}\\ \text{limite}\end{array}$& $[-2\sqrt{T},2\sqrt{T}]$&$[0,4T]$&$\left[0,4\sqrt{\frac{T}{3}}\right]$&$\left[\tan \frac{\alpha}{2},\tan \frac{\beta}{2}\right]$\\
\hline
$\begin{array}{c}\text{Densité}\\ \text{limite}\end{array}$&$\frac{\sqrt{4T-x^2}}{2\pi T}$& $\frac{1}{\pi T}\sqrt{\frac{4T-x}{4x}}$& $\frac{\left(x+2\sqrt{\frac{T}{3}}\right)}{2\pi T}\sqrt{\frac{4\sqrt{\frac{T}{3}}-x}{x}}$& $\frac{1}{\pi\cos\frac{|\beta-\alpha|}{4}(1+x^2)\sqrt{x^2-\tan^2\frac{|\beta-\alpha|}{4}}}$\\
\hline
$F^{(0)}$&$\frac{3T^2}{4}-\frac{T^2}{2}\ln \,T$&$\frac{3T^2}{2}-T\,\ln \,T$& $\frac{3T^2}{4}-\frac{T^2}{2}\ln\frac{T}{3}$& $\ln\, 2-\ln\left(\sin \frac{|\beta-\alpha|}{4}\right)$\\
\hline
$F^{(1)}$& $-\frac{1}{12}\ln\, T$& $-\frac{1}{6}\ln \,T$& $\frac{1}{3}\ln \, 2 -\frac{1}{8}\ln\, 3$& $\frac{1}{4}\,\ln \left(\cos\frac{|\beta-\alpha|}{4}\right)$\\
\hline
$\begin{array}{c}F^{(g)}\text{ avec}\\ g\geq 2\end{array}$&$-\frac{B_{2g}}{2g(2g-2)T^{2g-2}}$&$-\frac{2B_{2g}}{2g(2g-2)T^{2g-2}}$& ?& ?\\
\hline
\end{tabular}}\normalsize{}

\leftskip 0cm
\section{Autres applications}

Beaucoup de chercheurs ont récemment travaillé à la généralisation de la récurrence topologique à d'autres situations. Ce secteur est actuellement en plein développement et de nombreuses questions restent ouvertes. Sans rentrer dans les détails techniques, on peut néanmoins dresser une liste des thématiques émergentes et un état des lieux des dernières avancées.

\subsection{Généralisation à des symboles quelconques de supports égaux à des unions d'intervalles}
\noindent Soit $\mathcal{I}=\underset{j=1}{\overset{d}{\bigcup}} [\alpha_j,\beta_j] \subset(-\pi,\pi)$, $\mathcal{T}=\underset{j=1}{\overset{d}{\bigcup}} \left\{e^{it}\,,\,t\in [\alpha_j,\beta_j]\right\} \subset{\mathcal{C}}$ et $\mathcal{J}=\underset{j=1}{\overset{d}{\bigcup}} [\tan \frac{\alpha_j}{2},\tan\frac{\beta_j}{2}] \subset \mathbb{R}$. Soit $f$ une fonction mesurable, strictement positive sur $\mathcal{T}$ et s'annulant sur $\mathcal{C}\setminus \mathcal{T}$. On peut alors généraliser les résultats précédents de la façon suivante:

\begin{proposition}[Reformulations équivalentes du problème]\label{TheoReformnew} Il y a égalité entre les quantités suivantes:
\begin{enumerate}\item Une intégrale de Toeplitz de symbole $f$:
\beq \label{IntToepnew} Z_n(f)=\frac{1}{(2\pi)^nn!}\int_{[-\pi,\pi]^n} d\theta_1\dots d\theta_n \left(\prod_{k=1}^n f(e^{i\theta_k})\right)  \prod_{1\leq i<j\leq n}\left|e^{i \theta_i}-e^{i\theta_j}\right|^2
\eeq
\item Le déterminant d'une matrice de Toeplitz de taille $n\times n$:
\beq \label{ToepDetnew} Z_n(f)= \det \left(T_{i,j}=\td{t}_{i-j}\right)_{1\leq i,j\leq n} \eeq
dont les coefficients de Fourier discrets sont:
\beq \label{FourierCoeffnew} \td{t}_k=\frac{1}{2\pi}\int_{-\pi}^{\pi} f(e^{i\theta})e^{-ik\theta}d\theta\,\,\,\,,\,\,\,\,  \forall\, k\in \mathbb{Z}
\eeq
\item Une intégrale réelle $n$-dimensionnelle avec un potentiel logarithmique et des interactions de Vandermonde:
\bea \label{RealIntnew} Z_n(f)&=&\frac{2^{n(n-1)}}{(2\pi)^n n!}\int_{[-\pi,\pi]^n} d\theta_1\dots d\theta_n \left(\prod_{k=1}^n f(e^{i\theta_k})\right) \prod_{1\leq i<j\leq n}\sin^2\left(\frac{\theta_i-\theta_j}{2}\right)\cr
&=& \frac{2^{n^2}}{(2\pi)^n n!}\int_{\mathbb{R}^n} dt_1\dots dt_n \,\left(\prod_{k=1}^n f(e^{2i\arctan t_k})\right) \Delta(\mathbf{t})^2\,e^{-n\underset{k=1}{\overset{n}{\sum}} \ln(1+t_k^2)}\cr
&=& \frac{2^{n^2}}{(2\pi)^n n!}\int_{\mathcal{J}^n} dt_1\dots dt_n\, \Delta(\mathbf{t})^2\,e^{-n\underset{k=1}{\overset{n}{\sum}} V_n(t_k)}
\eea
où le potentiel $V_n$ est donné par:
\beq V_n(x)=\ln(1+x^2)-\frac{1}{n}\ln\left(f(e^{2i\arctan x})\right)\overset{\text{déf}}{=}\ln(1+x^2)-\frac{1}{n}g(x) \,\,,\,\,\forall \, x\in\mathcal{J}\eeq
\item Une intégrale complexe $n$-dimensionnelle sur des segments du cercle unité et avec des interactions de Vandermonde:
\bea \label{ComplexIntnew} Z_n(f)&=&(-1)^{\frac{n(n+1)}{2}}i^n \int_{\mathcal{C}^n} du_1\dots d u_n\, \Delta(u_1,\dots,u_n)^2\,\left(\prod_{k=1}^n f(u_k)\right) e^{-n\underset{k=1}{\overset{n}{\sum}}\ln \,u_k}\cr
&=&(-1)^{\frac{n(n+1)}{2}}i^n \int_{\mathcal{T}^n} du_1\dots d u_n\, \Delta(u_1,\dots,u_n)^2\, e^{-n\underset{k=1}{\overset{n}{\sum}}\left(\ln\, u_k-\frac{1}{n}\ln(f(u_k))\right)}\cr
&=&(-1)^{\frac{n(n+1)}{2}}i^n \int_{\mathcal{T}^n} du_1\dots d u_n\, \Delta(u_1,\dots,u_n)^2\, e^{-n\underset{k=1}{\overset{n}{\sum}}\td{V}_n(u_k)}
\eea  
dont le potentiel $\td{V}_n$ est donné par $\td{V}_n(x)=\ln(x)-\frac{1}{n}\ln(f(x))$. 
\end{enumerate}
\end{proposition} 

Dans cette généralisation naturelle, on comprend immédiatement le rôle singulier d'un symbole donné par une fonction indicatrice: le terme en $\frac{1}{n}\ln(f(x))$ disparait alors et le potentiel $V_n(x)$ (ou de façon équivalente $\td{V}_n(x)$) ne dépend plus de $n$ et vaut $\ln(1+x^2)$ (resp. $\ln x$). Dans le cas où le symbole n'est pas une fonction indicatrice, la situation se complique ainsi puisque le potentiel possède un terme en $\frac{1}{n}$. Notons que cette complexification reste modérée puisque le développement en puissances de $\frac{1}{n}$ aurait pu contenir une infinité de termes mais que dans notre cas, il ne présente qu'un seul terme supplémentaire. Ce terme supplémentaire implique néanmoins de profonds changements dans la nature de la solution. Observons tout d'abord que puisque le terme supplémentaire est d'ordre $\frac{1}{n}$, il ne modifie pas la courbe spectrale ni le calcul de l'ordre dominant. Ainsi, les hypothèses de la proposition \ref{ConditionsTopologiques} restent vérifiées pour les mêmes conditions sur les bords $\alpha_1$ et $\beta_1$ du support. Néanmoins, les conclusions de la proposition \ref{ConditionsTopologiques} sont modifiées. En effet, les résultats de \cite{BG} montrent que dans ce cas, les fonctions de corrélations admettent un développement en $\frac{1}{n}$ mais que tous les ordres peuvent être présents (et non plus ceux de même parité):
\bea\label{DevPuissanceNnew} W_1(x)&=&\sum_{g=0}^\infty W_1^{(\frac{g}{2})}(x)n^{1-g}=\sum_{g=0}^\infty W_1^{[g-1]}(x)n^{-(g-1)}\cr
W_p(x_1,\dots,x_p)&=&\sum_{g=0}^\infty W_p^{(\frac{g}{2})}(x_1,\dots,x_p)n^{2-p-g}=\sum_{g=0}^\infty W_p^{[p+g-2]}(x_1,\dots,x_p)n^{-(p+g-2)}\cr
\ln \,Z_N(f)&=& \sum_{g=0}^\infty Z^{(\frac{g}{2})} n^{2-g}=\sum_{g=0}^\infty Z^{[g-2]}n^{-(g-2)}
\eea
Comme annoncé ci-dessus, le calcul de l'ordre dominant reste identique à celui d'une fonction indicatrice. Ainsi pour le cas d'un support composé d'un seul intervalle, la courbe spectrale est donnée par:
\beq y(x)=\frac{\cos\frac{\alpha_1+\beta_1}{4}+x\sin\frac{\alpha_1+\beta_1}{4}}{\sqrt{\cos\frac{\alpha_1}{2}\cos\frac{\beta_1}{2}} (1+x^2)\sqrt{\left(x-\tan\frac{\alpha_1}{2}\right)\left(x-\tan \frac{\beta_1}{2}\right)}}\eeq
ce qui correspond à une densité d'équilibre donnée par:
\beq d\mu_\infty(x)=\frac{\left(\cos\frac{\alpha_1+\beta_1}{4}+x\sin\frac{\alpha_1+\beta_1}{4}\right)dx}{\pi\sqrt{\cos\frac{\alpha_1}{2}\cos\frac{\beta_1}{2}} (1+x^2)\sqrt{\left(\tan\frac{\alpha_1}{2}-x\right)\left(x-\tan \frac{\beta_1}{2}\right)}}\,\mathds{1}_{\left[\tan\frac{\alpha_1}{2},\tan \frac{\beta_1}{2}\right]}(x)\eeq
La différence majeure avec le cas présenté dans la proposition \ref{Mee} se situe donc dans le calcul des ordres sous-dominants. Puisque toutes les puissances de $\frac{1}{n}$ sont présentes dans le développement, il semble donc que la récurrence topologique dans sa version habituelle nécessite des modifications substantielles afin de permettre le calcul des ordres sous-dominants. D'une façon plus générale, ce problème correspond à la généralisation de la récurrence topologique (ou tout du moins d'avoir une méthode récursive de calculs) dans le cas d'interactions de type $\Delta(\boldsymbol{\lambda})^2$ mais dont le potentiel possède un développement en $\frac{1}{n}$. A ma connaissance, des formules générales collectant tous les termes issus de l'application de la récurrence topologique à une courbe spectrale possédant un développement en puissances de $\frac{1}{n}$ n'ont jamais été explicitées. Notons, que dans le cas d'un seul terme supplémentaire en $\frac{1}{n}$, les équations de boucles peuvent être facilement explicitées. En particulier, dans le cas où $g(x)=\frac{\nu}{2}x^2$ avec $\nu>0$, elles forment un système fermé d'équations qui contient donc toute l'information pour sa résolution. Il reste ainsi dans ce domaine beaucoup de pistes à explorer et en particulier la généralisation de la récurrence topologique dans le cas de potentiels admettant un développement en $\frac{1}{n}$ si elle peut être réalisée permettrait de nombreuses applications. Une piste parallèle dévelopée dans \cite{BorotGuionnetKoz} serait d'appliquer formellement la récurrence topologique sur une courbe spectrale dépendant de $N$. En effet, la construction étant purement algébrique, ce calcul récursif est possible et fournira des différentielles d'Eynard-Orantin $\left(W_n^{(g)}\right)_{n\geq 1,g\geq 0}$ (et des énergies libres $\left(F^{(g)}\right)_{g\geq 0}$) dépendantes de $N$ et dont on pourrait alors obtenir le développement à $N$ grand. Par la suite, on pourrait reconstruire les développements asymptotiques des fonctions de corrélation $\left(W_n\right)_{n\geq 1}$, en collectant les différents ordres des développements et en les réordonnant (après avoir montré que ce réordonnement est valable) convenablement. Il semble alors raisonnable de penser, tout du moins dans les ``bons'' cas où la courbe est non-critique et de genre 0, que les développements obtenus correspondent à ceux obtenus par la résolution des équations de boucle incluant les termes correctifs en $N$ à l'aide d'une récurrence topologique nouvelle qui reste à définir.

\subsection{Modèle à deux matrices unitaires}
Dans la lignée de l'étude des intégrales de matrices hermitiennes, l'étude des ``modèles à deux matrices'' de la forme:
\beq \label{2MM}Z_N(V_1,V_2,\tau)=\int_{\mathcal{H}_N} dM_1 \int_{\mathcal{H}_N} dM_2 \,\,\text{exp}\left(-N \,\Tr\left(V_1(M_1)+V_2(M_2)-\tau M_1 M_2\right)\right)\eeq
a été réalisée par la récurrence topologique (\cite{FreeEnergy}) au début des années 2000 pour des potentiels $V_1$ et $V_2$ polynomiaux. Notons que, tout comme le modèle à une matrice, les modèles hermitiens à deux matrices ont également été étudiés initialement par une réécriture en termes de polynômes bi-orthogonaux \cite{BEH4,KMcL,Kap,MarcoPaths}. La diagonalisation et la reformulation en un problème d'intégrales sur les valeurs propres se fait au moyen de l'intégrale d'Harish-Chandra-Itzykson-Zuber \cite{IZ,Johansson2}:
\beq \label{2MMdiag} Z_N(V_1,V_2,\tau)\propto \int_{\mathbb{R}^N}d\mathbf{X} \int_{\mathbb{R}^N}d\mathbf{Y}\,\, \Delta(\mathbf{X})^2\Delta(\mathbf{Y})^2 \,e^{-N\left(\underset{i=1}{\overset{N}{\sum}} V_1(X_i)+\underset{i=1}{\overset{N}{\sum}} V_2(Y_i)-\tau \underset{i=1}{\overset{N}{\sum}} X_iY_i\right)}\eeq
et l'on peut également définir une série de fonctions de corrélation (pour chaque matrice $M_1$ et $M_2$ et des corrélateurs mixtes). On peut alors obtenir de façon formelle le développement topologique de ces fonctions de corrélation par la récurrence topologique appliquée à la courbe spectrale du problème. Cette courbe spectrale se calcule de manière similaire au cas à une matrice à l'aide des équations de boucles \cite{FreeEnergy}. La récurrence topologique est alors rigoureusement identique au cas à une matrice et finalement la seule différence notable entre le cas hermitien à une matrice et le cas hermitien à deux matrices est que la courbe spectrale ne se limite plus à une courbe hyperelliptique $y(x)^2=P(x)$ mais permet à présent d'obtenir l'ensemble des courbes algébriques:
\beqq P(x,y)=0 \text{  où  } P \text{ est un polynôme.}\eeqq

Bien évidemment, on peut également introduire dans le modèle aux valeurs propres \eqref{2MMdiag} des bords durs en restreignant le domaine d'intégration permettant, comme dans le cas à une matrice, d'obtenir des comportements au voisinage de ces bords durs en $(x-a)^{-\frac{1}{q}}$ avec $q\geq 2$. Même si ces modèles ne modifient pas la forme de la récurrence topologique, les questions d'existence de développements topologiques et la démonstration de l'existence d'une mesure d'équilibre restent ouvertes et nécessiteront vraisemblablement des outils différents de ceux mis en place dans \cite{BG,BG2} pour le cas des modèles à une matrice. Par ailleurs, les problématiques de normalisation de la fonction de partition et la détermination précise de la courbe spectrale en genre non nul deviennent également beaucoup plus techniques. Notons également que pour les modèles à plusieurs matrices avec potentiels polynomiaux, les travaux de A. Guionnet et E. Maurel-Segala \cite{MaurelGuionnet} démontrent l'existence d'un développement asymptotique en $\frac{1}{N}$ jusqu'à $o(1)$ lorsque le couplage entre les matrices n'est pas trop grand. Il ne semble pas difficile de propager leur technique pour obtenir tous les ordres en $\frac{1}{N}$ dans ce cadre.

\subsection{Ensembles $\beta$}

Les ensembles $\beta$ sont apparus assez tôt dans la littérature des matrices aléatoires puisqu'ils permettent de faire la synthèse formelle entre les trois grands ensembles de matrices aléatoires: Hermitien, réel symétrique et quaternionique auto-dual (respectivement $\beta=1,\frac{1}{2}$ et $2$ dans les conventions typographiques modernes utilisées ci-dessous). En effet, il est connu depuis longtemps que la version diagonalisée de ces $3$ ensembles donne lieu à des intégrales du type:
\beq\label{Ensemblesbeta} Z_N(\beta)=\int_{A^N} d\lambda_1\dots d\lambda_N\, |\Delta(\lambda_1,\dots,\lambda_N)|^{2\beta} \,e^{-\frac{N\beta}{T}\underset{i=1}{\overset{N}{\sum}} V(\lambda_i)} \text{  où  } A\subset \mathbb{R}\eeq
En partant directement de cette intégrale, il est naturel de considérer sa généralisation à $\beta$ arbitraire (et non plus seulement à trois valeurs particulières) en prenant par exemple $\beta\in \mathbb{Q}_+$ quelconque ou $\beta\in \mathbb{R}_+$ quelconque. Cependant cette généralisation fait a priori perdre l'interprétation simple des valeurs propres de matrices aléatoires ainsi que la méthode des polynômes (bi)-orthogonaux. Il a fallu attendre \cite{DE} pour avoir une version type matrices aléatoires (tri-diagonales) fournissant une telle diagonalisation. Cette reformulation a permis à partir des années 2006 \cite{RRV,Virag2,Virag3} de faire le lien avec des problèmes de diffusion en potentiel aléatoire et d'utiliser les outils de l'analyse stochastique pour démontrer ou redémontrer des résultats d'universalité pour les ensembles $\beta$. Il est à noter que j'ai également développé durant mon doctorat une version $\beta$ du modèle à deux matrices \cite{LoopEquationBeta2MM,These}. 

Si la méthode des polynômes (bi)-orthogonaux est perdue pour $\beta$ quelconque, la méthode des équations de boucles reste utilisable. Ainsi les versions à une matrice et à deux matrices des équations de boucles pour les ensembles $\beta$ sont connues maintenant depuis une dizaine d'années \cite{TopRecBethe,CBM,CBM2}. La première différence apparait alors dans la forme même du développement topologique (si on peut prouver analytiquement qu'il existe) qui devient:
\bea \label{DevTopBeta}W_n(x_1,\dots,x_n)&=&\sum_{k=0}^{\infty} W_n^{(k)}N^{n-2-k}\cr
\ln\, Z_n&=&\sum_{k=0}^{\infty} \omega_n^{(k)}N^{2-k}\eea
Ainsi, des termes de parités différentes apparaissent dans les développements des fonctions de corrélations ce qui n'étaient pas le cas dans le cas Hermitien $\beta=1$. Un point positif est que la courbe spectrale obtenue (i.e. la projection de la première équation de boucle à l'ordre dominant) est identique à celle du cas hermitien $\beta=1$. Autrement dit, la courbe spectrale est indépendante de $\beta$. Cela implique nécessairement que la récurrence topologique doit être modifiée pour obtenir les autres ordres des fonctions de corrélations. A ce jour, une généralisation de la récurrence topologique pour les ensembles $\beta$ existe \cite{CBM,CBM2}, mais elle ne possède pas une forme aussi pratique que la version du cas hermitien. En effet, à la différence du cas hermitien dont la récurrence se formule en termes de résidus autour des points de ramifications, la généralisation actuelle au cas des ensembles $\beta$ ne s'exprime qu'à l'aide d'intégrales sur des contours de Stokes qui en pratique se révèlent difficile à effectuer de façon analytique. 

Par exemple, même le simple cas du modèle Gaussien $\beta$:
\beq\label{EnsemblesbetaGaussien} Z_N^{\text{Gaussien}}(\beta)=\int_{\mathbb{R}^N} d\lambda_1\dots d\lambda_N \,\,|\Delta(\lambda_1,\dots,\lambda_N)|^{2\beta} e^{-\frac{N\beta}{2T}\underset{i=1}{\overset{N}{\sum}} \lambda_i^2}\eeq
n'est pas résolu complètement à ce jour, bien que les équations de boucles soient alors complètement explicites et que les intégrales de Selberg correspondantes à \eqref{ZNGauss} ou \eqref{LnZNGaussien} soient également connues \cite{MehtaBook,Spreafico} permettant de résoudre les questions de normalisation de la fonction de partition. Certaines conjectures \cite{WF} ont été émises concernant les coefficients impliqués dans les développements topologiques des fonctions de corrélation et certaines de ces conjectures ont été résolues et précisées dans \cite{M,GaussienBeta} mais d'autres restent toujours d'actualité.

Par ailleurs, certains résultats intéressants ont été obtenus récemment pour le cas général. On citera par exemple \cite{BG,BG2} qui permettent d'avoir la forme générale du développement de $\ln Z_N(\beta)$ dans le cas à une ou à plusieurs coupures ainsi que les moyens d'en calculer les premiers termes. Certains résultats d'universalité sont également connus \cite{Bour,Bour2} et les applications à la physique théorique et à la théorie conforme \cite{MS15} semblent prometteuses. Une réécriture en termes d'équations différentielles stochastiques a aussi été dévelopée dans \cite{RRV} en partant de la formulation en termes de valeurs propres de matrices tri-diagonales. Il ne fait nul doute que ce domaine va connaître une croissance importante dans les années à venir et que les formalismes fastidieux développés aujourd'hui vont se simplifier progressivement.

\subsection{Interactions générales}

L'étude des corrélations locales des valeurs propres de matrices aléatoires donnent lieu à des phénomènes d'universalité décrits par des processus déterminantaux de Fredholm dont les noyaux sont explicitement connus pour les trois valeurs usuelles de $\beta$ correspondant aux cas Hermitien, réel symétrique et quaternionique auto-dual (Cf. \cite{MehtaBook} ainsi que ma thèse \cite{These} où un résumé des résultats est présenté). Cette universalité étant locale et ne dépendant donc que de la forme des interactions locales (en $|\Delta(\boldsymbol{\lambda})|^{2\beta}$) mais pas directement du potentiel, on peut légitimement s'intéresser à savoir si ce résultat se maintient pour des interactions plus générales. Comme évoqué dans la discussion des modèles $\beta$ ci-dessus, certains résultats en ce sens sont connus pour les interactions avec $\beta$ quelconque. Du point de vue de la récurrence topologique, on peut également légitimement se poser la question. En effet par définition même de la récurrence topologique, celle-ci reconstruit (dans les situations favorables) toutes les fonctions de corrélations des valeurs propres à l'aide uniquement de la courbe spectrale (équivalente à la densité limite d'équilibre) et de l'ordre dominant de la fonction à deux points $\omega_2^{(0)}(x_1,x_2)$. En particulier, si l'on choisit la courbe spectrale et $\omega_2^{(0)}(x_1,x_2)$ correspondant à des doubles limites d'échelle au voisinage d'un point de la distribution limite donné, comme cela a été fait dans \cite{BleherEynard,P5,2m1} on obtient les fonctions $\tau$ ainsi que les fonctions de corrélations des modèles universels correspondants qui se trouvent être des modèles intégrables classiques. Il semble donc que la procédure de récurrence topologique soit intimement liée au phénomène d'universalité dans le cas Hermitien. Il est donc intéressant de se demander si la récurrence topologique pourrait ou non être utile dans le cas d'interaction plus générales que des interactions en $|\Delta(\boldsymbol{\lambda})|^{2\beta}$. Les résultats de \cite{BorotGuionnetKoz} montrent l'existence d'un développement asymptotique en puissances de $\frac{1}{N}$ dans le cas d'intégrales de matrices convergentes avec interactions généralisées $f(\lambda_i,\lambda_j)$ mais restant localement de la forme $f(\lambda_i,\lambda_j)\underset{\lambda_i\to \lambda_j}{\sim} |\lambda_i-\lambda_j|^{2\beta}$ sous des hypothèses similaires aux hypothèses de la proposition \ref{ConditionsTopologiques}. Par ailleurs, le fait que ces développements satisfassent la récurrence topologique est montré dans \cite{BEO13}. Compte-tenu des enjeux relatifs à l'utilisation d'interactions générales en physique théorique, ce domaine devrait également rapidement évoluer dans les prochaines années.

\section{Conclusion du chapitre}

Dans ce chapitre nous avons vu que la récurrence topologique, couplée à la méthode des équations de boucles, s'avère très utile pour le calcul explicite de certaines fonctions de partition issues de modèles de matrices Hermitiennes ou unitaires aléatoires. Même si la généralisation aux ensembles $\beta$ n'est à ce jour pas encore complète, la souplesse de la méthode, par rapport à d'autres méthodes comme les polynômes (bi)-orthogonaux, permet d'espérer une généralisation à des cas d'interactions plus complexes. Par ailleurs, comme exposé dans le prochain chapitre, la récurrence topologique et les intégrales de matrices hermitiennes ont aussi des liens extrêmement forts avec les modèles intégrables où elles permettent des calculs effectifs simples. 
\selectlanguage{french}
\chapter{Récurrence topologique et systèmes intégrables}
 \label{chap2}
\thispagestyle{fancy}

L'étude des systèmes intégrables est un domaine ancien et qui a pris de multiples formes. En effet, contrairement à d'autres domaines des mathématiques, la notion de systèmes intégrables possède de nombreux sens différents parfois très éloignés. Historiquement, un système mécanique était dit ``intégrable'' lorsqu'il possèdait un nombre suffisant (c'est-à-dire égal à son nombre de degrés de liberté) de quantités conservées au cours du temps. Ainsi par exemple la conservation de l'énergie du système, de la quantité de mouvement, etc. dictées par les lois de la physique permettent souvent d'obtenir des quantités conservées au cours de l'évolution (en général temporelle en mécanique) du système étudié. Les symétries du système par le théorème de Noether (``À toute transformation infinitésimale qui laisse invariante l'intégrale d'action correspond une grandeur qui se conserve'') permettent également d'obtenir en pratique de nombreuses quantités conservées. Une fois un nombre suffisant de quantités conservées obtenues, le système peut être ``intégré'' au sens historique, c'est-à-dire que l'on peut trouver des formules explicites permettant de décrire les variables du système (position, vitesse, énergie, entropie, etc.) au cours de son évolution à partir de ses conditions initiales. Typiquement, chaque quantité conservée donne lieu à une équation différentielle sur les variables d'intérêt et en combinant l'ensemble des équations différentielles obtenues, on parvient à obtenir des équations différentielles ordinaires sur chacune des variables d'intérêt que l'on peut résoudre analytiquement ou au moins numériquement. Cette approche est ainsi très claire dans le cas de systèmes possédant un nombre fini de degrés de liberté, c'est-à-dire de dimension finie, mais elle commence à devenir plus floue lorsque le système étudié possède une infinité de degrés de liberté, c'est-à-dire que la dimension du système devient infinie. Dès lors le nombre de quantités conservées doit également être infini, mais d'un infini ``suffisamment grand'' pour pouvoir rendre le système intégrable. Comme nous allons le voir dans ce chapitre, l'approche des systèmes intégrables que l'on va utiliser dans ce document est assez similaire à cette approche mécanique historique des systèmes intégrables. Ainsi, après avoir défini des ``temps'' $(t_k)_{k\geq 0}$ (toujours en nombre fini dans ce document) permettant ``l'évolution'' du système, un système sera dit intégrable s'il obéit à un certain nombre (une infinité en général) d'équations différentielles partielles par rapport à ces temps. En revanche, on ne demandera pas nécessairement que le système soit ``résoluble'' c'est-à-dire qu'il existe des formules explicites décrivant ses variables d'intérêt valables quelque soit la valeur des temps $(t_k)_{k>0}$.

\section{Hamiltoniens et paires de Lax}

Cette partie a pour but de présenter la version ``d'intégrabilité de Liouville'' d'abord dans le cadre d'un exemple simple qui permettra d'arriver à la définition de la notion d'une ``paire de Lax''. Ensuite l'exemple plus général de la hiérarchie KP sera introduit par sa paire de Lax et la relation avec la récurrence topologique sera précisée. L'approche des systèmes intégrables développée ici est axée sur les systèmes intégrables présentant une paire de Lax puisque c'est à partir de ces dernières que l'approche des formules déterminantales fait le lien avec la récurrence topologique. Néanmoins, il est important de mentionner qu'il existe des versions plus abstraites et plus générales de l'approche des systèmes intégrables (qui ne peuvent pas tous être ramenés à des paires de Lax) comme par exemple l'approche par les Grassmaniennes, des approches liées aux groupes et algèbres de Lie ou aux algèbres de Frobenius. Ces points étant éloignés des considérations sur les formules déterminantales, ils ne seront pas abordés en détail, mais d'excellentes revues existent dans la littérature à ce propos \cite{IntegrableSys,ReviewIntSys}. 

\subsection{Système de Calogero-Moser}
Considérons $\mathbb{C}^N$ comme une variété symplectique de dimension $N$ avec la forme symplectique:
\beqq \omega=\sum_{i=1}^N dp_i\wedge dq_i\eeqq
où $(q_i,p_i)_{1\leq i\leq N}$ forment un système de coordonnées symplectiques. Un Hamiltonien de cette variété est la donnée d'une fonction $H(p_1,\dots,p_N,q_1,\dots,q_N,t)$ qui définit le système dynamique par le flot:
\beq \label{flot}\dot{q}_i=\frac{\partial H}{\partial p_i} = \text{   et   } \dot{p}_i=-\frac{\partial H}{\partial q_i}   \,\,\,\,,\,\,\,\, \forall \,1\leq i\leq N\eeq
Dans tout ce qui suit la notation $\dot{\,}$ désigne la dérivée par rapport au temps $t$: $\dot{f}=\frac{d f}{dt}$. On peut ainsi penser le système précédent d'un point de vue mécanique où la variable $q_i$ représente la position (uni-dimensionnelle) de la $i^{\text{ième}}$ particule tandis que la variable $p_i$ représente sa quantité de mouvement. L'exemple du Hamiltonien de Calogero-Moser est ainsi donné par:
\beq \label{HamiltonienCaloMoser} H_{CM}=\sum_{i=1}^N p_i^2-\sum_{i, j\neq i}^N\frac{1}{(q_i-q_j)^2}\eeq
donnant 
\beqq \dot{q}_i=p_i \text{   et   } \dot{p}_i=-\sum_{j\neq i}^N\frac{2}{(q_i-q_j)^3}  \,\,\,\,,\,\,\,\, \forall \,1\leq i\leq N\eeqq

D'une façon générale, les systèmes Hamiltoniens peuvent être définis sur n'importe quelle variété symplectique. On arrive alors à la définition d'un système intégrable de Liouville:

\begin{definition}[Intégrabilité de Liouville] Un Hamiltonien $H$ sur une variété symplectique $\mathcal{M}$ de dimension $N$ est dit intégrable au sens de Liouville s'il existe $N$ Hamiltoniens $(H_i)_{1\leq i\leq N}$ tel que $\dot{H_i}=0$, $\{H_i,H_j\}=0$ pour $i\neq j$ et les formes $(dH_i)_{1\leq i\leq N}$ sont linéairement indépendantes en tout point de $\mathcal{M}$.
\end{definition} 

Rappelons que dans le cas de $\mathbb{C}^N$, le crochet de Poisson est défini par:
\beqq \{A,B\}=\sum_{i=1}^N \left(\frac{\partial A}{\partial q_i}\frac{\partial B}{\partial p_i} -\frac{\partial A}{\partial p_i}\frac{\partial B}{\partial q_i}\right)\eeqq
Dans le cas où le Hamiltonien est indépendant de $t$ (on parle d'un système ``autonome''), le Hamiltonien lui-même est une quantité conservée. Dans le cas du système de Calogero-Moser, il existe une méthode générale pour trouver les Hamiltoniens $(H_i)_{1\leq i\leq N}$. Définissons les matrices $P$ et $Q$ de taille $N\times N$ par:
\bea Q&=&\text{diag}\,(q_1,\dots,q_N) \cr
 P_{i,i}&=&p_i \text{  et  } P_{i,j}=\frac{1}{q_i-q_j} \text{ pour }i\neq j 
\eea
En définissant $H_k=\Tr(P^k)$ pour $1\leq k\leq N$ on obtient des Hamiltoniens indépendants et qui commutent caractérisant ainsi l'intégrabilité du système. On obtient ainsi par exemple:
\beqq H_1=\sum_{i=1}^N p_i \text{  et  } H_2=H_{CM}\eeqq
On peut vérifier aisément que $H_1$ est bien conservé ainsi que $H_2$ (ce dernier point étant trivial car $H_{CM}$ ne dépend pas de $t$):
\beqq \dot{H}_1=\sum_{i=1}^{N}\dot{p}_i=\sum_{i,j\neq i}^N\frac{2}{(q_i-q_j)^3}=0 \eeqq
L'intérêt des quantités conservées réside dans le fait qu'elles permettent de définir les ``temps'' pour lesquels l'évolution du système est triviale. Soit donc $(t_k)_{1\leq k\leq N}$ le flot associé à $H_k$ de telle sorte que 
\beqq \omega=\sum_{i=1}^N dH_k\wedge dt_k\eeqq
Dans le cas de Calogero-Moser, on observe alors que l'évolution de la matrice $Q(\mathbf{t})$ est triviale:
\beqq Q(\mathbf{t})=Q(\mathbf{0})+\sum_{k=1}^N kt_k P^{k-1}\eeqq
En faisant agir un changement de base $Q(\mathbf{t})\to \Psi(\mathbf{t})Q(\mathbf{t})\Psi(\mathbf{t})^{-1}$, on obtient ainsi une représentation matricielle de la forme suivante:

\begin{proposition}[Equations de compatibilité pour Calogero-Moser] En définissant les matrices $N\times N$ par \cite{IntegrableSys,Etingof}:
\bea\label{LM} L_{i,i}&=&-\sum_{k\neq i} L_{i,k} \text{  et  } L_{i,j}=\frac{1}{q_i-q_j} \text{ pour } i\neq j\cr
M^{(k)}_{i,j}&=&\frac{k}{q_i-q_j} \left[L^{k-1}\right]_{i,j} \text{  pour } i\neq j \text{ et } M^{(k)}_{i,i}=-\sum_{j\neq i}M_{i,j}^{(k)}
\eea
on obtient le système différentiel ($\forall \, 1\leq k\leq N$):
\bea \label{CompaCalo} \partial_{t_k}\Psi(\mathbf{t})&=&M^{(k)} \Psi(\mathbf{t})\cr 
 \partial_{t_k}L&=&\left[M^{(k)}(\mathbf{t}),L\right]
\eea
\end{proposition}
On obtient ainsi une représentation en termes d'équations différentielles de système Hamiltonien. En particulier si l'on note $z=t_1$, $\mathbf{t}=(t_2,\dots,t_N)$ et $M^{(1)}=D(z,\mathbf{t})$ on obtient alors:
\bea \label{LaxCalo} \partial_z \Psi(z,\mathbf{t})&=&D(z,\mathbf{t})\Psi(z,\mathbf{t})\cr
\partial_{t_k} \Psi(z,\mathbf{t})&=&M^{(k)}(z,\mathbf{t})\Psi(x,\mathbf{t}) \,\,,\,\, \forall\, k\geq 2
\eea
qui est appelée ``une paire de Lax'' associée au système.


\medskip

Notons que si l'on se donne des matrices $D(z,\mathbf{t})$ et $M^{(k)}(z,\mathbf{t})$ quelconques, des systèmes différentiels comme \eqref{LaxCalo} ne sont en général pas compatibles et n'admettent donc pas de solutions $\Psi(z,\mathbf{t})$ inversibles. Il est donc indispensable que des équations de compatiblité de type \eqref{CompaCalo} soient vérifiées ce qui contraint énormément le système et en définit l'intégrabilité. Dans la suite, on s'intéressera à des sytèmes de Lax dont la définition est la suivante:

\begin{definition}[Paire de Lax] On appelle paire de Lax, la donnée de matrices $\mathcal{D}(x,\mathbf{t})$ et $\mathcal{R}^{(k)}(x,\mathbf{t})$ où $\mathbf{t}=(t_k)_{k\geq 1}$ est une suite (finie ou infinie) de ``temps'' telle que le système différentiel:
\beq \label{LaxPair}\partial_x \Psi(x,\mathbf{t})=\DD(x,\mathbf{t})\Psi(x,\mathbf{t}) \normalfont{\text{  et  }}   \partial_{t_k} \Psi(x,\mathbf{t})=\RR^{(k)}(x,\mathbf{t})\Psi(x,\mathbf{t})\,\,,\,\, \forall \,k\geq 1\eeq
soit compatible (i.e admette au moins une solution $\Psi(x,\mathbf{t})$ inversible). En particulier il est nécessaire que les conditions de compatibilité suivantes soient réalisées:
\bea \label{CompaLax} 0&=&\partial_{t_k}\DD(x,\mathbf{t})-\partial_x \RR^{(k)}(x,\mathbf{t})+[\DD(x,\mathbf{t}),\RR^{(k)}(x,\mathbf{t})]\cr
0&=&\partial_{t_i}\RR^{(j)}(x,\mathbf{t})-\partial_{t_j} \RR^{(i)}(x,\mathbf{t})+[\RR^{(j)}(x,\mathbf{t}),\RR^{(i)}(x,\mathbf{t})] \,\,,\,\, \forall \, i\neq j\cr
&&\eea
\end{definition}

Bien que la compatibilité des systèmes différentiels soit assez contraignante, il existe de nombreuses paires de Lax associées à des ``hiérarchies intégrables'' différentes. On citera ainsi les hiérarchies KP et KdV (qui est une réduction de KP) ainsi que leurs généralisations: BKP, MKP. Ces dernières sont les plus utilisées dans le domaine de la récurrence topologique et des intégrales de matrices aléatoires. 

\subsection{Paires de Lax pour les équations de Painlevé}

Les six équations de Painlevé constituent des équations différentielles privilégiées dans l'étude des systèmes intégrables. Elles furent initialement introduites par Painlevé et Gambier dans l'étude des solutions des équations différentielles du second ordre:
\beqq y''=R(y',y,x) \text{ avec } R \text{ rationnelle}\eeqq
qui possèdent la propriété de Painlevé, c'est-à-dire que les seules singularités dépendant des conditions initiales (``movable singularities'' en anglais) sont des pôles. Cette propriété est rare pour les équations différentielles non linéaires car des singularités essentielles ou des branchements apparaissent en général et dépendent des conditions initiales (par exemple $y'(x)=\frac{1}{2}y(x)^3$ donne $y(x)=\pm(x-C)^{\frac{1}{2}}$ qui possède un branchement en $x=C$ dépendant de la condition initiale). Ainsi les équations de Painlevé apparaissent comme les réductions de ces équations différentielles qui ne peuvent être résolues à l'aide des fonctions élementaires usuelles. Cette approche a été complétée de manière indépendante par R. Fuchss en 1905 qui fut le premier à obtenir l'équation de Painlevé $6$ en étudiant les équations scalaires du second ordre admettant $5$ singularités dont une apparente; puis par L. Schlesinger en 1912 en considérant les déformations isomonodromiques des équations différentielles linéaires avec un nombre quelconque de singularités régulières (pôle simple, on parle alors de modèles Fuchsiens). Elle a ensuite été étendue par M. Jimbo, T. Miwa et K. Ueno aux cas de pôles d'ordres supérieurs (on parle alors de singularités irrégulières avec une extension des données de monodromies). On peut trouver une excellente revue sur les équations de Painlevé dans \cite{PainleveHandbook}. Rappelons l'idée générale des déformations isomodromiques: dans le cas des systèmes Fuchsiens (la notation $\,'$ désignera dans ce document la dérivée par rapport à $x$: $\frac{d}{dx}$):
\beq\label{SystFuchsien} Y'=\sum_{i=1}^r \frac{A_i}{x-\lambda_i} Y\eeq
les solutions $x\mapsto Y(x)$ exhibent un phénomène de monodromie lors du prolongement analytique le long d'un contour fermé entourant l'un des pôles. Ainsi le prolongement analytique d'une solution $Y(x)$ le long d'un contour fermé autour d'un pôle ne redonne pas la valeur initiale $Y(x_0)$ mais on obtient $\hat{Y}(x_0)=Y(x_0) M_i$ où la matrice $M_i$ est appelée ``matrice de monodromie''. L'action du groupe fondamental (les matrices de monodromies ne dépendant que de la classe du lacet dans le groupe fondamental) permet alors de caractériser l'ensemble du groupe des matrices de monodromies. On peut rapidement se convaincre que pour un ensemble de matrices de monodromies donné, il existe de nombreux systèmes de type \eqref{SystFuchsien} qui possèdent de telles monodromies. En fait, pour un ensemble donné de matrices de monodromies on peut déformer le système \eqref{SystFuchsien} (i.e. modifier les valeurs des $A_i$ en les faisant dépendre de paramètres $(\lambda_i)_{i\geq 1}$) de façon à préserver ces monodromies. On parle alors de ``déformations isomonodromiques''. Dans le cas des systèmes Fuchsiens, les déformations isomonodromiques satisfont les équations de Schlesinger:
\bea \frac{\partial A_i}{\partial \lambda_j}&=& \frac{[A_i,A_j]}{\lambda_i-\lambda_j}  \,\,\,,\,\,\, \forall \,j\neq i\cr
\frac{\partial A_i}{\partial \lambda_i}&=&-\sum_{j\neq i}^r\frac{[A_i,A_j]}{\lambda_i-\lambda_j}\,\,\,,\,\,\, \forall \,i\geq 1
\eea
Dans le cas où les matrices $A_i$ sont choisies dans l'algèbre de Lie $\mathfrak{sl}_2(\mathbb{C})$, on parle de systèmes de Garnier et en se restreignant au cas de $4$ pôles maximum ($r\leq 4$) on obtient l'équation de Painlevé $6$ (puis par confluence des paramètres les autres équations de Painlevé) qui peut donc être vue comme un cas particulier de systèmes non-linéaires de Schlesinger dans l'étude des déformations isomonodromiques. Notons que dans le cas de systèmes Fuchsiens linéaires ou sans points singuliers branchants, Malgrange et Miwa \cite{Miwa,Malgrange1,Malgrange2} ont montré que les solutions satisfont toujours la propriété de Painlevé. En particulier, contrairement à ce que la terminologie laisse penser, les équations de Painlevé échappent à cette situation et ce n'est que très récemment \cite{MattiaSuperNew} en 2017 que la propriété de Painlevé a été démontrée pour les $6$ équations de Painlevé. Dans le cas des systèmes Fuchsiens, il existe également une structure symplectique sous-jacente qui permet la réécriture du problème en termes Hamiltoniens développée par Okamoto \cite{Okamoto}. 
Comme nous le verrons par la suite, le lien entre la récurrence topologique et les systèmes intégrables n'a été précisé pour l'instant que pour les équations de Painlevé pour lesquelles les paires de Lax et les Hamiltoniens \cite{JM81,JMII81,JMIII} sont les suivants:

\begin{center}
\begin{itemize}
\item  Paire de Lax de $\PI$: 
\beq \label{P1}\DD_{\rm I}(x,t) = \begin{pmatrix} -p&x^2+qx+q^2+\frac{t}{2}\\4(x-q)&p\end{pmatrix} \,\,,\,\, \RR_{\rm I}(x,t) = \begin{pmatrix} 0&\frac{x}{2}+q\\2 &0\end{pmatrix}\eeq
\item Paire de Lax de $\PII$: 
\small{\beq\label{P2}\DD_{\rm II}(x,t) = \begin{pmatrix} x^2+p+\frac{t}{2}&x-q\\-2\left(xp+qp+\theta\right)& -\left(x^2+p+\frac{t}{2}\right)\end{pmatrix} \,\,,\,\, 
\RR_{\rm II}(x,t) = \begin{pmatrix} \frac{x+q}{2}& \frac{1}{2}\\-p& -\frac{x+q}{2}\end{pmatrix}
\eeq}\normalsize{}
\item Paire de Lax de $\PIII$ (type $D_6$)\footnote{Les versions dégénérées $D_7$ et $D_8$ de l'équation de Painlevé III, dont les propriétés sont résumées dans \cite{OKSO,SvdP}, n'ont pas été étudiées dans \cite{IwakiMarchal} en lien avec la récurrence topologique. Même s'il est vraisemblable que la propriété de type topologique soit vérifiée pour ces deux cas, une démonstration rigoureuse, utilisant par exemple les résultats généraux de \cite{TTProperty} reste à faire.}:
\small{\bea \label{P3} \DD_{\rm III}(x,t)&=&\begin{pmatrix} \frac{t}{2}-\frac{\theta_\infty}{2x}+\frac{p-\frac{t}{2}}{x^2}& -\frac{pq}{x}-\frac{p}{x^2}\\ \frac{-(p-t)q-\theta_\infty+\frac{t(\theta_0+\theta_\infty)}{2p}}{x}+\frac{p-t}{x^2}& -\left(\frac{t}{2}-\frac{\theta_\infty}{2x}+\frac{p-\frac{t}{2}}{x^2}\right)\end{pmatrix}\cr
\RR_{\rm III}(x,t)&=&\begin{pmatrix} \frac{x}{2}-\frac{p-\frac{t}{2}}{tx}+\frac{\theta_0+\theta_\infty}{2p}+q-\frac{\theta_\infty}{2t}& -\frac{pq}{t}+\frac{p}{tx}\\ \frac{-(p-t)q-\theta_\infty+\frac{t(\theta_0+\theta_\infty)}{2p}}{t}-\frac{p-t}{tx}&-\left(\frac{x}{2}-\frac{p-\frac{t}{2}}{tx}+\frac{\theta_0+\theta_\infty}{2p}+q-\frac{\theta_\infty}{2t}\right)\end{pmatrix}\cr
&&\eea}\normalsize{}
\item Paire de Lax de $\PIV$: 
\bea \label{P4} \DD_{\rm IV}(x,t)&=&\begin{pmatrix} x+t+\frac{pq+\theta_0}{x}& 1-\frac{q}{x}\\ -2(pq+\theta_0+\theta_\infty)+\frac{p(pq+2\theta_0)}{x}& -\left(x+t+\frac{pq+\theta_0}{x}\right)\end{pmatrix}\cr
\RR_{\rm IV}(x,t)&=&\begin{pmatrix} x+q+t& 1\\ -2(pq+\theta_0+\theta_\infty)&-(x+q+t)\end{pmatrix}
\eea
\item Paire de Lax de $\PV$:
\small{\bea\label{P5} \DD_{\rm V}(x,t)&=&\begin{pmatrix}d(x,t)& -\frac{pq+\theta_0}{x}+\frac{p+\frac{\theta_0-\theta_1+\theta_\infty}{2q}}{x-1}\\
\frac{pq}{x}-\frac{pq^2+q\frac{\theta_0+\theta_1+\theta_\infty}{2}}{x-1}&-d(x,t)\end{pmatrix}\cr
\RR_{\rm V}(x,t)&=&\begin{pmatrix}r(x,t) & -\frac{1}{t}\left(p(q-1)+\theta_0-\frac{\theta_0-\theta_1+\theta_\infty}{2q}\right)\\
-\frac{q}{t}\left(p(q-1)+\frac{\theta_0+\theta_1+\theta_\infty}{2}\right)& -r(x,t)\end{pmatrix}\cr
&&\eea
}\normalsize{où} 
\small{\bea d(x,t)&=&\frac{t}{2}+\frac{1}{x}\left(pq+\frac{\theta_0}{2}\right)-\frac{1}{x-1}\left(pq+\frac{\theta_0+\theta_\infty}{2}\right)\cr
r(x,t)&=&\frac{x}{2}-\frac{1}{2t}\left(p(q-1)^2-\theta_0+\frac{\theta_0-\theta_1+\theta_\infty}{2q}+q\frac{\theta_0+\theta_1+\theta_\infty}{2}\right)\eea}\normalsize{}
\item Paire de Lax de $\PVI$: 
\beq \label{P6}\DD_{\rm VI}(x,t) = \frac{A_0(t)}{x}+\frac{A_1(t)}{x-1}+\frac{A_t(t)}{x-t}\,\,,\,\, \RR_{\rm VI}(x,t) =-\frac{A_t(t)}{x-t}-\frac{(q-t)(\theta_\infty-1)}{2t(t-1)}\sigma_3 \eeq
où $\sigma_3=\text{diag}\,(1,-1)$ et 
\bea  A_0&=&\begin{pmatrix} z_0+\frac{\theta_0}{2}&-\frac{q}{t}\\ \frac{tz_0(z_0+\theta_0)}{q}&-\left(z_0+\frac{\theta_0}{2}\right)\end{pmatrix} \,\,,\,\, A_1=\begin{pmatrix} z_1+\frac{\theta_1}{2}&\frac{q-1}{t-1}\\ -\frac{(t-1)z_1(z_1+\theta_1)}{q-1}&-\left(z_1+\frac{\theta_1}{2}\right)\end{pmatrix}\cr
A_t&=&\begin{pmatrix} z_t+\frac{\theta_t}{2}&-\frac{q-t}{t(t-1)}\\ \frac{t(t-1)z_t(z_t+\theta_t)}{q-t}&-\left(z_t+\frac{\theta_t}{2}\right)\end{pmatrix}\,\,,\,\,
A_\infty=\begin{pmatrix} \frac{\theta_\infty}{2}&0\\0&-\frac{\theta_\infty}{2}\end{pmatrix}=-(A_0+A_1+A_t)\cr
&&
\eea
Ici, $z_0(t), z_1(t)$ et $z_t(t)$ sont des fonctions auxiliaires de $t$ (le seul ``temps'' présent dans le système) qui peuvent être exprimées en termes de $q(t)$ et d'une fonction $p(t)$ définie par:
\beq p=\frac{z_0+\theta_0}{q}+\frac{z_1+\theta_1}{q-1}+\frac{z_t+\theta_t}{q-t}\eeq
Les expressions explicites de $(z_0,z_1,z_t)$ en termes de $(p,q)$ sont données par:
\small{\bea \label{z0z1zt} z_0&=&\frac{1}{\theta_\infty t}\Big[ q^2(q-1)(q-t)p^2+\frac{1}{4}(\theta_0+\theta_1+\theta_t-\theta_\infty)^2q^2\cr
&&-pq((\theta_0+\theta_1+\theta_t-\theta_\infty)q^2-((\theta_0+\theta_1-\theta_\infty)t+\theta_0+\theta_t-\theta_\infty)q+(\theta_0-\theta_\infty)t)\cr
&&-\frac{1}{4}(\theta_0+\theta_1+\theta_t-\theta_\infty)( (\theta_0+\theta_1-\theta_t-\theta_\infty)t+\theta_0-\theta_1-\theta_\infty+\theta_t)q-t\theta_0\theta_\infty\Big]\cr
z_1&=&\frac{1}{(t-1)\theta_\infty}\Big[ -q(q-1)^2(q-t)p^2+p(q-1)( (\theta_0+\theta_1+\theta_t-\theta_\infty)^2q^2\cr
&&-q((\theta_0+\theta_1-\theta_\infty)t+\theta_0+\theta_t)+\theta_0 t)-\frac{1}{4}(\theta_0+\theta_1+\theta_t-\theta_\infty)^2(q-1)^2\cr
&&+\frac{1}{4}(\theta_0+\theta_1+\theta_t-\theta_\infty)(t(\theta_0+\theta_1-\theta_t-\theta_\infty)+2\theta_\infty-2\theta_1)(q-1)-\theta_1\theta_\infty(t-1)\Big]\cr
z_t&=&-z_0-z_1-\frac{1}{2}(\theta_0+\theta_1+\theta_t+\theta_\infty)
\eea}\normalsize{}
\end{itemize}
\end{center}

Dans les paires de Lax précédentes, les paramètres de monodromies $\theta, \theta_0, \theta_1, \theta_\infty, \theta_t$ sont supposés être génériques ce qui permet d'obtenir des solutions transcendantes et non des solutions particulières élémentaires (qui peuvent être complètement classifiées (Cf. \cite{LT} qui contient une bibliographie relativement exhaustive sur le sujet) et ont fait l'objet d'études spécifiques en rapport avec d'autres domaines des mathématiques). Dans tous les cas, la fonction $q(t)$ satisfait l'équation de Painlevé correspondante tandis que la fonction $p(t)$ est reliée au formalisme Hamiltonien par \eqref{flot}. Notons que ces paires de Lax diffèrent légèrement de celles de Jimbo-Miwa \cite{JM81} mais les détails permettant de relier les unes aux autres peuvent être trouvés dans l'article \cite{IwakiMarchal}. En particulier, la paire de Lax dépend d'un choix de jauge $\Psi(x,t)\to U(x,t)\Psi(x,t)$ qui aura son importance dans le lien avec la récurrence topologique puisque si  $\td{\Psi}(x,t)=U(x,t)\Psi(x,t)$ alors $\td{\Psi}(x,t)$ satisfait un système de Lax de la forme $\partial_x \td{\Psi}=\td{\mathcal{D}} \td{\Psi}, \partial_t \td{\Psi}=\td{\mathcal{R}}\td{\Psi}$ avec:
\bea \label{Gauge}\td{\mathcal{D}}(x,t)&=&U(x,t)\mathcal{D}(x,t)U^{-1}(x,t)+\frac{\partial U}{\partial x}(x,t)U^{-1}(x,t)\cr
\td{\mathcal{R}}(x,t)&=&U(x,t)\mathcal{R}(x,t)U^{-1}(x,t)+\frac{\partial U}{\partial t}(x,t)U^{-1}(x,t)
\eea
qui est différent du système initial. Dans le cas des équations de Painlevé, les systèmes Hamiltoniens équivalents aux paires de Lax ci-dessus et les équations de Painlevé associées sont donnés par:
\begin{itemize} \item Painlevé I:
\bea \label{P1eq} \ddot{q}&=&6q^2+t\cr
 \text{H}_{\text{I}}(p,q,t)&=&\frac{1}{2}p^2-2q^3-tq\eea
\item Painlevé II:
\bea \label{P2eq} \ddot{q}&=&2q^3+tq+\frac{1}{2}-\theta\cr
 \text{H}_{\text{II}}(p,q,t)&=&\frac{1}{2}p^2+(q^2+\frac{t}{2})p+\theta q\eea
\item Painlevé III:
\bea \label{P3eq}\ddot{q}&=&\frac{1}{q}\dot{q}^2-\frac{1}{t}\dot{q}+\frac{4}{t}\left(\theta_0 q^2-\theta_\infty+1\right)+4 q^3-\frac{4}{q}\cr
\text{H}_{\text{III}}(p,q,t)&=&\frac{1}{t}\Big[2q^2p^2+2(-tq^2+\theta_\infty q+t)p-(\theta_0+\theta_\infty)tq-t^2\cr
&&-\frac{1}{4}(\theta_0^2-\theta_\infty^2)- pq\Big]
\eea
\item Painlevé IV:
\bea \label{P4eq}\ddot{q}&=&\frac{1}{2q}\dot{q}^2+2\left(3q^3+4tq^2+\left(t^2-2\theta_\infty+1\right)q-\frac{\theta_0^2}{q}\right)\cr
 \text{H}_{\text{IV}}(p,q,t)&=&qp^2+2(q^2+tq+\theta_0)p+2(\theta_0+\theta_\infty)q
\eea
\item Painlevé V:
\bea \label{P5eq}\ddot{q}&=&\left(\frac{1}{2q}+\frac{1}{q-1}\right)(\dot{q})^2-\frac{\dot{q}}{t}+\frac{(q-1)^2}{t^2}\left(\alpha q+\frac{\beta}{q}\right)+\frac{\gamma q}{t}+\frac{\delta q(q+1)}{q-1}\cr
&\text{où}& \, \alpha=\frac{(\theta_0-\theta_1-\theta_\infty)^2}{8}\,,\,\beta=\frac{-(\theta_0-\theta_1+\theta_\infty)^2}{8}\,,\cr
&&\,\gamma=\theta_0+\theta_1-1 \text{ et } \delta=-\frac{1}{2}\cr
\text{H}_{\text{V}}(p,q,t)&=&\frac{1}{t}\Big[q(q-1)^2p^2+\frac{1}{2}\theta_0(\theta_0+\theta_1+\theta_\infty)q\cr
&&+\left(\frac{\theta_0-\theta_1+\theta_\infty}{2}(q-1)^2+(\theta_0+\theta_1)q(q-1)-tq\right)p\Big]
\eea
\item Painlevé VI:
\bea\label{P6eq}\ddot{q}&=&\frac{1}{2}\left(\frac{1}{q}+\frac{1}{q-1}+\frac{1}{q-t}\right)\dot{q}^2-\left(\frac{1}{t}+\frac{1}{t-1}+\frac{1}{q-t}\right)\dot{q}\cr
&&+\frac{q(q-1)(q-t)}{t^2(t-1)^2}\left[\alpha+\beta \frac{t}{q^2}+\gamma \frac{t-1}{(q-1)^2}+\delta \frac{t(t-1)}{(q-t)^2}\right]
\eea
où les paramètres sont:
\beqq \alpha=\frac{1}{2}(\theta_\infty-1)^2\,,\, \beta=-\frac{\theta_0^2}{2}\,,\, \gamma=\frac{\theta_1^2}{2} \text{ and } \delta=\frac{1-\theta_t^2}{2}\eeqq
Le Hamiltonien est:
\bea \label{H6} \text{H}_{\text{VI}}(p,q,t)&=&\frac{1}{t(t-1)}\Big[ q(q-1)(q-t)p^2\cr
&&-p\left(\theta_0(q-1)(q-t)+\theta_1q(q-t)+(\theta_t-1)q(q-1)\right)\cr
&&+\frac{1}{4}(\theta_0+\theta_1+\theta_t-\theta_\infty)(\theta_0+\theta_1+\theta_t+\theta_\infty-1)(q-t)\cr
&&+\frac{1}{2}((t-1)\theta_0+t\theta_1)(\theta_t-1)\Big]
\eea
\end{itemize}

A partir d'une paire de Lax il est souvent impossible de déterminer une solution explicite $\Psi(x,t)$ du problème. Néanmoins, on peut s'intéresser aux solutions dites WKB (Wentzel-Kramers-Brillouin) en introduisant un paramètre formel d'expansion à partir duquel on va chercher des développements en série. Dans le cas des paires de Lax précédentes, on peut introduire un paramètre $\hbar$ par changement d'échelle (``rescaling'') des différentes quantités. Par exemple, pour Painlevé II, on peut utiliser
\beqq(\td{t},\td{x},\td{q},\td{p},\td{\theta})= \left(\hbar^{\frac{2}{3}}t, \hbar^{\frac{1}{3}}x,\hbar^{\frac{1}{3}}q, \hbar^{\frac{2}{3}}p, \hbar\theta\right) \text{ et } \td{\Psi}= \begin{pmatrix} \hbar^{\frac{1}{6}}&0\\ 0 &\hbar^{-\frac{1}{6}}\end{pmatrix}\Psi \eeqq
pour obtenir une paire de Lax de la forme:
\beq \label{hLax} \hbar \partial_x \Psi(x,t,\hbar)=\DD_{\text{II}}(x,t)\Psi(x,t,\hbar) \text{  et  } \hbar \partial_t \Psi(x,t,\hbar)=\RR_{\text{II}}(x,t)\Psi(x,t,\hbar)\eeq
où les matrices $\DD_{\text{II}}(x,t)$ et $\DD_{\text{II}}(x,t)$ sont encore données par \eqref{P2}. On obtient également l'équation de Painlevé ``déformée'': 
\beq \label{P2eqhar} \hbar^2\ddot{q}=2q^3+tq+\frac{\hbar}{2}-\theta\eeq
et le système Hamiltonien correspondant:
\beq\label{H2eqhar} \hbar \dot{q} =\frac{\partial \text{H}_{\text{II}}}{\partial p} (p,q,\hbar) \text{   et   }\hbar \dot{p} =-\frac{\partial \text{H}_{\text{II}}}{\partial q} (p,q,\hbar)\eeq
où $H_{II}(p,q,t)=\frac{1}{2}p^2+(q^2+\frac{t}{2})p+\theta q$ reste inchangé par rapport au cas initial. Notons que la solution de l'équation de Painlevé après rescaling dépend de $\hbar$. Cette procédure peut être appliquée aux six équations de Painlevé et les résultats sont fournis dans mon article \cite{IwakiMarchal}. Notons que le Hamiltonien peut être amené à dépendre explicitement de $\hbar$. L'intérêt d'introduire un paramètre $\hbar$ est qu'il permet de rechercher des solutions sous la forme d'un développement en puissances de $\hbar$ de la forme suivante:

\begin{definition} Soit une paire de Lax de la forme:
\beqq \hbar \partial_x \Psi(x,t,\hbar)=\DD(x,t,\hbar)\Psi(x,t,\hbar) \text{  et  } \hbar \partial_t \Psi(x,t,\hbar)=\RR(x,t,\hbar)\Psi(x,t,\hbar)\eeqq
où les matrices $\DD(x,t,\hbar)$ et $\RR(x,t,\hbar)$ admettent un développement en $\hbar$ de la forme:
\bea \label{DevDDRR} \DD(x,t,\hbar)&=&\sum_{k=0}^\infty \DD^{(k)}(x,t)\hbar^k\cr
\RR(x,t,\hbar)&=&\sum_{k=0}^\infty \RR^{(k)}(x,t)\hbar^k
\eea
On appelle solution WKB une solution \textbf{formelle} du système différentiel précédent de la forme:
\bea \label{WKBFormelle}\Psi(x,t,\hbar)&=&\Psi_0(x,t)\left(\normalfont{\text{Id}}+\sum_{k=1}^\infty \Psi^{(k)}(x,t) \hbar^k\right)e^{\frac{1}{\hbar}\Psi^{(-1)}(x,t)}\normalfont{\text{ avec }} \Psi^{(-1)}(x,t) \normalfont{\text{ diagonale}}\cr
&=&\normalfont{\text{exp}}\left(\frac{1}{\hbar}\Psi^{(-1)}(x,t)+ \sum_{k=0}^\infty \td{\Psi}^{(k)}(x,t) \hbar^k\right)
\eea
\end{definition}

Notons que les notions précédentes peuvent tout à fait être définies pour des systèmes différentiels linéaires sans paramètre $t$ auxiliaire (dans ce cas seule la matrice $\mathcal{D}(x)$ est nécessaire). Dans le cas des équations de Painlevé, l'existence de développements en $\hbar$ de type \eqref{DevDDRR} est équivalente à rechercher une solution formelle de l'équation de Painlevé de la forme 
\beq \label{hbarq}  q(t,\hbar)=\sum_{k=0}^\infty q^{(k)}(t)\hbar^k\eeq

Les solutions WKB ou de l'équation \eqref{hbarq} sont \textbf{formelles} dans le sens où l'on ne s'intéresse pas au rayon de convergence de la série en $\hbar$ (qui en général est nul). En particulier, si la série possèdait un rayon de convergence supérieur ou égal à $1$, on pourrait obtenir en prenant $\hbar=1$ une solution de l'équation de Painlevé de départ. Les coefficients $\left(q^{(k)}(t)\right)_{k\geq 0}$ de \eqref{hbarq} peuvent s'obtenir par récurrence à partir des équations de Painlevé déformées comme \eqref{P2eqhar}. On remarque alors que $q^{(0)}(t)$ obéit à une équation algébrique tandis que les ordres suivants peuvent être facilement calculés par récurrence en termes de $q^{(0)}(t)$ et de ses dérivées. Dans l'exemple de l'équation de Painlevé $2$ on trouve ainsi une équation algébrique de degré $3$:
\beq \label{Q0P2}2(q^{(0)})^3+tq^{(0)}-\theta=0 \eeq

A ce stade, le lien entre les paires de Lax et la récurrence topologique n'est pas apparent. Il deviendra explicite dans le prochain paragraphe avec la définition d'une courbe spectrale associée à un système différentiel linéaire, ainsi que la définition de différentielles d'Eynard-Orantin possédant la même structure que les fonctions de corrélation $\omega_n^{(g)}$ définies par la récurrence topologique.

\section{Formules déterminantales associées à un système différentiel}

Cette partie introduit l'équivalent des fonctions de corrélations dans le langage des paires de Lax. L'idée développée initialement dans \cite{BE09} puis dans \cite{BBE14} consiste à définir des quantités ayant des propriétés similaires à celles des fonctions de corrélations de la récurrence topologique à partir d'un système différentiel de la forme $\hbar \partial_x \Psi(x)=\DD(x) \Psi(x)$ où $\DD(x)\in \mathcal{M}_r(\mathbb{C})$.

\subsection{Courbe spectrale associée à un système différentiel}

Puisque la notion de courbe spectrale joue un rôle fondamental dans la récurrence topologique, il apparait essentiel de trouver un équivalent dans un système différentiel linéaire. Cela a été fait dans \cite{BE09} et l'on définit donc:
\sloppy
\begin{definition}[Courbe spectrale associée à un système différentiel linéaire] \label{DefCurb}
Soit $\hbar \partial_x \Psi(x,\hbar)=\DD(x,\hbar)\Psi(x,\hbar)$ un système différentiel dépendant d'un paramètre formel $\hbar$ où $\DD(x,\hbar)$ admet un développement en $\hbar$ de la forme \eqref{DevDDRR}. On définit une courbe spectrale associée à ce système différentiel comme:
\beq \label{CourbeSpecSystDiff} P(x,y)=\det\left(y\,\text{{\normalfont Id}}- \DD^{(0)}(x)\right)=0\eeq
Dans le cas où $\DD^{(0)}(x)$ est une fonction rationnelle en $x$ on obtient ainsi une équation polynomiale, i.e. une surface de Riemann $\Sigma$. Dans le cas où la courbe spectrale est de genre non nul, cette donnée doit être complétée par le choix d'une base symplectique de cycles (correspondant à des contours reliant différents secteurs de Stokes) et d'une bi-forme différentielle $\omega_2^{(0)}(z_1,z_2)$ telle que précisée dans la définition \ref{DeffCourbeSpec}.  
\end{definition}
\fussy

Notons que cette définition est invariante par les changements de base de la forme $\Psi(x)\to U(x,\hbar)\Psi(x)$ pour $U(x,\hbar)$ admettant un développement de Taylor en $\hbar$. Par ailleurs, cette définition ne fait pas intervenir directement $\Psi(x)$ ce qui la rend indépendante de la normalisation $\Psi(x,\hbar)\to \Psi(x,\hbar)U(\hbar)$.

\subsection{Formules déterminantales}

Une fois la notion de courbe spectrale définie, il reste à trouver un équivalent des fonctions de corrélations. Cela a été fait dans \cite{BE09} puis dans \cite{BBE14} et a été généralisé pour des groupes de Lie quelconques dans \cite{Lie}. Dans le cas présent $\mathcal{G}l_n(\mathbb{C})$, cela correspond aux définitions suivantes:
\sloppy
\begin{definition}[Matrice $M$] Soit $\hbar \partial_x \Psi(x,\hbar)=\DD(x,\hbar)\Psi(x,\hbar)$ un système différentiel de la forme \eqref{DevDDRR}. On définit les matrices $E_i=\text{diag}\,(0,\dots,0,\overset{i}{1},0,\dots,0)$ pour $1\leq i\leq r$ et les matrices $M(x.E_i)$ par:
\beq \label{MatriceM} M(x.E_i)=\Psi(x)E_i\Psi(x)^{-1} \eeq
Notons que la notation $x.E_i$ permet de préciser quelle matrice $E_i$ est utilisée. Mais précisons également qu'il ne s'agit que d'une notation et que l'on aurait pu par exemple utiliser tout aussi bien la notation $M_i(x)$.
\end{definition}
\fussy

Comme on va le voir par la suite, la matrice $M$ joue un rôle fondamental dans la théorie. En particulier, elle possède de meilleures propriétés que la matrice $\Psi(x)$ qui dépend trop de la base choisie. Ainsi la matrice $M$ obéit au système différentiel suivant:
\bea\label{SystM} \hbar\partial_x M(x.E_i,t)&=&\left[\DD(x,t),M(x.E_i,t)\right]\cr
\hbar\partial_t M(x.E_i,t)&=&\left[\RR(x,t),M(x.E_i,t)\right]
\eea

La deuxième équation et la dépendance en $t$ ne sont évidemment présentes que lorsque l'on a une paire de Lax au départ. Dans le cas contraire, seule la première équation est disponible. On peut alors définir les fonctions de corrélations:
\sloppy
\begin{definition}[Fonctions de corrélation associées à un système différentiel]\label{DeterFormulas}
Soit $\hbar \partial_x \Psi(x,\hbar)=\DD(x,\hbar)\Psi(x,\hbar)$ un système différentiel. On définit les fonctions de corrélation associées par des ``formules déterminantales'':
\bea
W_n(x_1.E_{j_1},\dots,x_n.E_{j_n})
= 
\left\{
\begin{array}{lr}
\hbar^{-1} \operatorname{Tr}\left( \DD(x_1)M(x_1.E_{j_1})\right) dx_1 & n=1 \cr
\frac{1}{n}\underset{\sigma \,\text{n-cycles}}{\sum} \frac{\operatorname{Tr} \underset{i=1}{\overset{n}{\prod}} M(x_{\sigma(i)}.E_{j_{\sigma(i)}}) }{\underset{i=1}{\overset{n}{\prod}} (x_{\sigma(i)}-x_{\sigma(i+1)})} \underset{i=1}{\overset{n}{\prod}} dx_i& n\geq 2
\end{array}\right. \cr
\eea
\end{definition}
\fussy

Notons que $W_n(x_1.E_{j_1},\dots,x_n.E_{j_n})$ est une $n$-forme différentielle. La terminologie ``formules déterminantales'' provient de l'existence d'une définition alternative en termes d'un noyau $K(x_1,x_2)$ faisant intervenir un déterminant. Les détails de cette définition alternative ainsi que ceux de la version ``non connectée'' des fonctions précédentes peuvent être trouvés dans \cite{BE09}, \cite{BBE14} et \cite{Lie}. 

\medskip

L'intérêt principal de la définition précédente réside dans le fait que les fonctions $W_n(x_1.E_1,\dots,x_n.E_n)$ satisfont des équations de boucles identiques à celles obtenues dans les modèles de matrices Hermitiennes \eqref{LoopEquations}. Ces équations de boucles prennent différentes formes selon le cadre considéré. Ainsi elles sont simples \cite{BE09} lorsque les matrices sont de taille $2\times 2$, plus compliquées lorsque la taille est arbitraire \cite{BBE14} et encore plus compliquées lorsque l'on considère leur analogue pour un groupe de Lie semi-simple quelconque \cite{Lie}. Par simplicité et puisque notre étude se borne aux cas des équations de Painlevé, nous ne rappelerons ici que les équations de boucles du cas $2\times 2$ comme décrit dans \cite{BE09}. Dans ce cas, les équations de boucles sont définies par:

\begin{proposition}[Exemple\footnote{Une version beaucoup plus générale des équations de boucle faisant intervenir tous les $(E_i)_{i\leq d}$ est donnée par le théorème $4.3$ de \cite{Lie}} d'équations de boucles] Dans le cadre des paires de Lax de Painlevé \eqref{P1}-\eqref{P6}, les fonctions de corrélations satisfont les équations de boucles suivantes. Soient $(P_n)_{n\geq 1}$ les fonctions définies par:
\beq \label{eq:first-loop-equation} P_1(x)=W_2(x.E_1,x.E_1)+W_1(x.E_1)^2, \eeq
et pour $n\geq 1$, en posant $L_n=\{x_1.E_1,\dots,x_n.E_1\}$:
\bea\label{ind1}
P_{n+1}(x;L_n)&=&-W_{n+2}(x.E_1,x.E_1,L_n)-2W_1(x.E_1)W_{n+1}(x.E_1,L_n)\cr
&&- \sum_{J \subset L_n, J\notin\{ \emptyset,L_n\}} W_{1+|J|}(x.E_1,J)W_{1+n-|J|}(x.E_1,L_n\setminus J) \cr
&&- \sum_{j=1}^n \frac{d}{dx_j} \frac{ W_n(x.E_1,L_n\setminus x_j.E_1)-W_n(L_n)}{x-x_j}
\eea
alors pour tout $n\geq 0$, $x\mapsto P_{n+1}(x;L_n)$ est une fraction rationnelle en $x$ dont les seules singularités ne peuvent être que des pôles situés parmi ceux de $x\mapsto \DD(x,t)$  
\end{proposition}   

Notons que le choix de $E_1$ est purement conventionnel puisque la proposition précédente reste valable en remplaçant partout $E_1$ par $E_2$. Le résultat précédent n'est pas trivial car par définition, les fonctions $P_n$ pourraient également avoir des pôles aux points de branchement de la courbe spectrale \eqref{CourbeSpecSystDiff}. C'est justement la combinaison particulière des fonctions de corrélation qui permet d'éliminer ces singularités potentielles. Cette dernière propriété met en lumière de façon claire les similarités avec la récurrence topologique. Il reste néanmoins une dernière étape non triviale puisque pour terminer l'équivalence il est nécessaire d'avoir un développement en $\hbar$ des fonctions $W_n(x_1.E_1,\dots,x_n.E_n)$. Or comme nous allons le voir, l'existence d'un tel développement est loin d'être garantie.

\subsection{Propriété de type topologique et isomonodromies}

A partir d'un système différentiel ou de Lax de la forme $\hbar \partial_x \Psi(x,\hbar)=\DD(x,\hbar)\Psi(x,\hbar)$ nous avons pour l'instant réussi à définir une courbe spectrale et des ``fonctions de corrélation'' \eqref{DeterFormulas}. Néanmoins pour pouvoir les identifier avec les différentielles produites par la récurrence topologique, il est nécessaire d'avoir un développement en $\hbar$ adéquat de ces ``fonctions de corrélation''.
\sloppy
\begin{definition}[Propriété de type topologique]\label{TTprop}
On dit qu'un système différentiel $\hbar \partial_x \Psi(x,\hbar)=\DD(x,\hbar)\Psi(x,\hbar)$ est de type topologique si:
\begin{itemize} \item Ses ``fonctions de corrélations'' $W_n(x_1.E_{j_1},\dots,x_n.E_{j_n})$ admettent un développement en $\hbar$ de la forme:
\beq W_n(x_1.E_{j_1},\dots,x_n.E_{j_n})=\sum_{k=0}^\infty W_n^{(k)}(x_1.E_{j_1},\dots,x_n.E_{j_n})\hbar^{n+2k-2}\eeq
\item Les différentielles $W_n^{(k)}(x_1.E_{j_1},\dots,x_n.E_{j_n})$ ne peuvent avoir des singularités qu'aux points de branchement de type pôle.
\item Les différentielles $W_n^{(k)}(x_1.E_{j_1},\dots,x_n.E_{j_n})$ sont d'intégrales nulles autour des $\mathcal{A}$-cycles associés à la courbe spectrale.
\item La différentielle $W_2^{(0)}(x_1,x_2)$ s'identifie avec le noyau de Bergman (forme bi-différentielle fondamentale de seconde espèce) $\omega_2^{(0)}$ de la courbe spectrale.
\end{itemize}
\end{definition}
\fussy

Cette propriété est non-triviale pour diverses raisons qui sont de natures et de difficultés variables:

\begin{enumerate}\item Il est nécessaire de prouver l'existence d'un développement en $\hbar$ de la forme:
\beq\label{TT} \forall\, n\geq 2 \,\,:\,\, W_n(x_1.E_{j_1},\dots,x_n.E_{j_n})=\sum_{k=0}^\infty W_n^{(k)}(x_1.E_{j_1},\dots,x_n.E_{j_n})\hbar^{k}\eeq
L'existence d'un tel développement est à l'heure actuelle très difficile à montrer de façon analytique. En revanche, on peut toujours se placer dans le cadre formel et ne s'intéresser qu'à des développements de type WKB pour $\Psi(x)$ auquel cas l'existence d'un tel développement est alors immédiate. Mais on n'a alors que des résultats formels, c'est-à-dire uniquement algébriques (identification ordre par ordre en $\hbar$) mais pas analytiques.
\item Même dans le cadre formel où le développement en série existe, il reste à montrer que seuls des termes de même parité apparaissent dans chacun des développements (facteur $2k$ et non $k$ dans \eqref{TT}). Ceci n'est pas trivial en général et nécessite une argumentation spécifique.
\item Enfin, il reste à montrer que pour $n>2$, l'ordre dominant de $W_n$ commence à $\hbar^{n-2}$ et non simplement à $\hbar^0$. Cette propriété est triviale pour $n\leq 2$ par définition mais devient non triviale pour $n>2$. Elle devient même de plus en plus difficile à démontrer à mesure que $n$ augmente puisque le nombre d'ordres à annuler augmente.
\item Par définition les différentielles $W_n$ peuvent présenter des singularités aux points doubles (n\oe uds) de la courbe spectrale qu'il faut donc réussir à exclure.
\end{enumerate}

\medskip

Une fois la propriété de type topologique démontrée, on aboutit alors à l'identification complète avec la récurrence topologique:

\begin{theorem}[Reconstruction par la récurrence topologique]\label{Reconstruction} Si le système est de type topologique alors d'après les résultats de \cite{BE09} et \cite{BBE14} on a:
\beqq \forall \, n\geq 1, \forall\,  g\geq 0\,\,:\,\, W_n^{(g)}(x_1.E_{j_1},\dots,x_n.E_{j_n})=\omega_n^{(g)}(z^{j_1}(x_1),\dots,z^{j_n}(x_n) )\eeqq
où les différentielles $\omega_n^{(g)}(z^{j_1}(x_1),\dots,z^{j_n}(x) )$ sont celles calculées par la récurrence topologique appliquée à la courbe spectrale du système différentiel.\footnote{La notation $z^j(x)$ correspond à prendre la préimage du point $x$ dans le $j^{\text{ème}}$ feuillet} En d'autres termes, on peut reconstruire les fonctions de corrélation du système différentiel par l'application de la récurrence topologique à la courbe spectrale associée au système différentiel.
Par ailleurs, si le système différentiel initial est une paire de Lax dépendant également d'un paramètre $t$ alors les invariants symplectiques $\left(F^{(g)}\right)_{g\geq 0}$ calculés par la récurrence topologique \cite{EO} reconstruisent (à des constantes près) le développement de la fonction $\tau$ de Jimbo-Miwa \cite{JM81} du système intégrable:
\beqq \frac{1}{\hbar^2}\frac{d}{dt}\ln\, \tau(t,\hbar) = -\frac{d}{dt}\sum_{g=0}^{\infty} \hbar^{2g-2} F^{(g)}(t)\eeqq
\end{theorem}

Ce théorème montre le lien profond entre les systèmes intégrables et la récurrence topologique: à partir de la donnée réduite d'une courbe spectrale (qui est un calcul d'ordre dominant) on peut reconstruire l'intégralité du développement formel des solutions par l'application de la récurrence topologique. Inversement, dès la donnée de l'ordre dominant en $\hbar$ d'un système différentiel, il existe une façon canonique fournie par la récurrence topologique de construire les ordres supérieurs et d'obtenir alors un système intégrable.

\begin{remark} Il est important de noter que les solutions WKB \eqref{WKBFormelle} des paires de Lax ou les solutions de type \eqref{hbarq} ne correspondent pas à des solutions générales des équations de Painlevé. En effet, une solution générale possède deux paramètres arbitraires (reliés aux conditions initiales) tandis que les solutions \eqref{hbarq} peuvent être exprimées entièrement à l'aide de $q^{(0)}$ définie dans \eqref{Q0P2} et sont donc très particulières. Les résultats de \cite{BLMST} suggèrent que les fonctions $\tau$ associées aux solutions générales sont données, dans différents régimes, par des transformées de Fourier de blocs conformes généralisés qui eux admettent des développements en $\hbar$ de type topologique. Néanmoins, la compréhension de la reconstruction des solutions générales par la récurrence topologique reste à ce jour un problème ouvert.    
\end{remark}

La démonstration du théorème \ref{Reconstruction} (réalisée dans \cite{BBE14}) est assez typique des démonstrations impliquant la récurrence topologique. Elle consiste à montrer que les équations de boucles combinées avec une structure particulière du développement assurent l'unicité des solutions qui sont données en particulier par la récurrence topologique. Notons que les équations de boucles seules ne permettent pas d'obtenir l'unicité et que la propriété de type topologique (Définition \ref{TTprop}) est indispensable pour pouvoir démontrer l'identification.

\medskip

Pour un système différentiel donné $\DD(x)$, on voit donc maintenant que l'enjeu principal est de pouvoir démontrer la propriété de type topologique. En règle générale, cette propriété de type topologique a de fortes chances de ne pas être vérifiée. Mais dans le cas d'une paire de Lax, on peut utiliser l'équation différentielle auxiliaire en $t$ pour la prouver. C'est la stratégie utilisée dans le cas de Painlevé II dans \cite{P2} puis dans le cas des autres équations de Painlevé dans \cite{IwakiMarchal}, et enfin qui a été très récemment généralisée dans le cas général \cite{TTProperty} de courbes de genre $0$. C'est alors l'isomonodromie qui permet de prouver la propriété de type topologique du système et qui permet l'identification avec les quantités calculées par la récurrence topologique. Devant l'intérêt de cette preuve, nous allons rappeler très brièvement les étapes clés de la démonstration dans le cas de courbes spectrales de genre $0$ en les illustrant par l'équation de Painlevé II.

\subsection{Exemple de preuve de la propriété de type topologique}

Notons que dans le cadre des équations de Painlevé, nous nous sommes placés dans le cadre de solutions formelles de type WKB et que par conséquent, l'existence de développement en $\hbar$ des diverses quantités est assuré. La stratégie à mettre en place est alors la suivante:
\begin{enumerate}\item Calculer la courbe spectrale. Pour Painlevé II \eqref{P2} on trouve ainsi (en notant $q^{(0)}\equiv q_0$ pour simplifier les expressions):
\beq y^2=(x-q_0)^2\left(x^2+2q_0x+q_0^2+\frac{\theta}{q_0}\right)\eeq
qui est de genre $0$ (éliminant ainsi les questions de normalisation sur les cycles).
\item Calculer le premier ordre $M^{(0)}(x,t)$ à l'aide de l'équation 
\beqq\left[\mathcal{R}^{(0)}(x,t),M^{(0)}(x,t)\right]=0\eeqq 
qui se résout par:
\beq \label{M0}M^{(0)}(x,t)=\begin{pmatrix}\frac{1}{2}+\frac{\mathcal{R}^{(0)}_{1,1}(x,t)}{2\sqrt{-\det \mathcal{R}^{(0)}(x,t)}}& \frac{\mathcal{R}^{(0)}_{1,2}(x,t)}{2\sqrt{-\det \mathcal{R}^{(0)}(x,t)}}\\
\frac{\mathcal{R}^{(0)}_{2,1}(x,t)}{2\sqrt{-\det \mathcal{R}^{(0)}(x,t)}}&\frac{1}{2}-\frac{\mathcal{R}^{(0)}_{1,1}(x,t)}{2\sqrt{-\det \mathcal{R}^{(0)}(x,t)}}\end{pmatrix}
\eeq
Le simple calcul de $\det\, \RR(x,t)$ permet d'obtenir une formule exacte de $M^{(0)}(x,t)$. Dans le cas de Painlevé II on trouve $\det \mathcal{R}_{\rm II}^{(0)}=-\frac{1}{4}\left(x^2+2q_0x+q_0^2+\frac{\theta}{q_0}\right)$. Un calcul direct permet alors de calculer $W_2^{(0)}(x(z_1),x(z_2))$ par la définition \ref{DeterFormulas} et de vérifier qu'il est égal à $\omega_2^{(0)}(z_1,z_2)=\frac{dz_1dz_2}{(z_1-z_2)^2}$.
\item Projeter l'équation $\hbar \partial_t M(x,t)=[\RR(x,t),M(x,t)]$ à l'ordre $\hbar^k$ et obtenir une récurrence pour calculer $M^{(k+1)}(x,t)$ à l'aide des ordres inférieurs $M^{(i)}(x,t)_{0\leq i\leq k}$. En particulier, cette récurrence permet de montrer que $\left(M^{(k)}(x,t)\right)_{k\geq 0}$ (et donc de façon immédiate par la suite les $W_n^{(k)}$) ne sont singuliers qu'aux points de branchements de la courbe spectrale. Dans le cas de Painlevé II on a:
\small{\beaa &&\begin{pmatrix} 0&-\mathcal{R}^{(0)}_{2,1}&\mathcal{R}^{(0)}_{1,2}\\
-2\mathcal{R}^{(0)}_{1,2}&2\mathcal{R}^{(0)}_{1,1}&0\\
\mathcal{R}^{(0)}_{1,1}&\frac{1}{2}\mathcal{R}^{(0)}_{2,1}&\frac{1}{2}\mathcal{R}^{(0)}_{1,2}\end{pmatrix}
\begin{pmatrix} M^{(k)}(x,t)_{1,1}\\ M^{(k)}(x,t)_{1,2}\\ M^{(k)}(x,t)_{2,1}\end{pmatrix}\cr
&&=\begin{pmatrix} \partial_t M^{(k-1)}(x,t)_{1,1}-\underset{i=0}{\overset{k-1}{\sum}}\left[\mathcal{R}^{(k-i)}(x,t),M^{(i)}(x,t)\right]_{1,1}\\ 
\partial_t M^{(k-1)}(x,t)_{1,2}-\underset{i=0}{\overset{k-1}{\sum}} \left[\mathcal{R}^{(k-i)}(x,t),M^{(i)}(x,t)\right]_{1,2}\\
\sqrt{-\det \mathcal{R}^{(0)}}\,\underset{i=1}{\overset{k-1}{\sum}}\left( M^{(i)}(x,t)_{1,1}M^{(k-i)}(x,t)_{1,1}+M^{(i)}(x,t)_{1,2}M^{(k-i)}(x,t)_{2,1}\right)\end{pmatrix}\eeaa}\normalsize{}
\sloppy{Le déterminant de la matrice à gauche est toujours donné par $-2\mathcal{R}_{1,2}^{(0)}(x,t) \det \mathcal{R}^{(0)}(x,t)$ et ne s'annule donc dans le cas de Painlevé II qu'aux points de branchements permettant ainsi la récurrence.}
\item La question de la parité du développement en $\hbar$ des fonctions de corrélation peut être résolue par une condition suffisante donnée dans \cite{BBE14}. Si l'on peut trouver une matrice $\Gamma(t)$ indépendante de $x$ telle que:
\beqq \Gamma^{-1}(t) \,\mathcal{D}^{\,t}(x,t)\,\Gamma(t)=\mathcal{D}^\dagger(x,t)\eeqq
alors la parité du développement est assurée. La notation $\,^\dagger$ représente ici le changement $\hbar \to -\hbar$. Pour trouver cette matrice $\Gamma(t)$, on peut utiliser la structure Hamiltonienne du système comme cela est décrit dans \cite{IwakiMarchal}. En effet:
\beq \label{ParityDetermination} H(p^\dagger,q^\dagger,t,-\hbar)=H(p,q,t,\hbar) \text{ et } \frac{\partial H}{\partial t}(p^\dagger,q^\dagger,t,-\hbar)=\frac{\partial H}{\partial t}(p,q,t,\hbar)\eeq
permet d'exprimer $(p^\dagger,q^\dagger$) en fonction de $(p,q)$ puis ensuite d'en déduire $\DD^\dagger$ et $\RR^\dagger$ ainsi que l'action de $\,\dagger$ sur toutes les autres quantités. On trouve dans le cas de Painlevé II:
\beqq \Gamma_{\text{II}}(t)=\begin{pmatrix}-2p(t)&0\\0&1 \end{pmatrix} \eeqq
\item La dernière étape consiste à montrer que $W_n$ est d'ordre au moins $\hbar^{n-2}$. Dans \cite{P2} puis \cite{IwakiMarchal} une récurrence double et un raisonnement par l'absurde à l'aide de la structure des singularités des $W_n$ (qui doit donc être démontrée au préalable) permet de déduire cette propriété. Cette récurrence double est relativement générale et, bien que nécessitant une étude au cas par cas (pour les singularités de $P_n$ dans les équations de boucles), peut être adaptée à toutes les équations de Painlevé et à des sytèmes généraux (comme ceux décrits dans \cite{TTProperty}). Elle met en lumière à nouveau le rôle important des équations de boucles et de la structure des pôles dans les solutions de ces équations de boucles.  
\end{enumerate} 

\subsection{Perspectives de développement}

La méthode présentée ci-dessus s'applique parfaitement dans le cadre des équations de Painlevé mais on peut naturellement se poser la question de sa généralisation à d'autres paires de Lax. Ainsi il semblerait naturel d'essayer d'appliquer la démarche précédente pour toute paire de Lax de taille $2\times 2$ dont la structure Hamiltonienne est connue. Néanmoins certaines interrogations demeurent et mériteraient une étude plus approfondie:
\begin{itemize}\item L'introduction du paramètre formel $\hbar$ est pour l'instant assez mal comprise. Dans le cas des paires de Lax de Painlevé, elle se fait naturellement par rescaling et présente aussi une version équivalente directement dans le formalisme Hamiltonien. Néanmoins pour d'autres paires de Lax, il n'est pas évident a priori que cette méthode fonctionne ou soit pertinente et trouver ou insérer un paramètre formel pourrait ne pas toujours être aussi simple.
\item La généralisation aux paires de Lax de dimension arbitraire n'est pas à ce jour complètement achevée même si elle a été considérablement avancée dans \cite{TTProperty}. En effet, la structure plus compliquée des équations de boucles en dimensions supérieures rend la méthode précédente plus difficile à appliquer. Enfin la généralisation à des connexions générales sur des groupes et algèbres de Lie reste encore à préciser même si les travaux de \cite{Lie} permettent déjà une bonne compréhension des cas de dimension $2$. 
\item La construction précédente est efficace lorsque la courbe spectrale du système différentiel est de genre $0$. Dans le cas d'un genre plus grand, il est légitime de se poser la question de savoir si un développement de type WKB est encore pertinent et si l'identification avec la récurrence topologique est possible. Par ailleurs dans ce cas, les calculs explicites de la récurrence topologique deviennent beaucoup plus compliqués et même de simples vérifications sur les premiers ordres sont difficiles.
\item La question de la convergence des séries utilisées est également une question naturelle. D'un point de vue différentiel, il est rare que les développements WKB convergent et il est souvent nécessaire d'utiliser la Borel-sommabilité ou des techniques de résurgence de Voros pour donner un sens analytique aux séries obtenues. A ce jour, l'utilisation de ces notions dans ce domaine reste très mal comprise. 
\end{itemize}

\section{Quantification de la courbe spectrale}

La notion de courbe spectrale joue un rôle central dans le calcul de la récurrence topologique et des fonctions de corrélation dans le cas d'un modèle de matrice hermitien sous-jacent. Néanmoins, dans le cas d'un modèle de matrice Hermitien comme étudié dans le chapitre précédent, les fonctions de corrélation ne sont pas les seules quantités d'intérêt d'un point de vue des modèles de matrices. En effet, rappelons que (définition \ref{FonctionCorrelation}):
\bea W_n(x_1,\dots,x_n)&=&\left<\Tr\, \frac{1}{x_1-M}\Tr\,\frac{1}{x_2-M}\dots \Tr\,\frac{1}{x_n-M}\right>_c \text{  avec}\cr
\left< f(x,M)\right>&=&\frac{1}{Z_N}\int_{\mathcal{H}_N} dM f(x,M)e^{-\frac{N}{T}\Tr\, V(M)}
\eea
Ces fonctions de corrélations sont utiles pour calculer les valeurs moyennes des traces des puissances de la matrice $M$ ainsi que les corrélations:
\beq W_1(x)=\left<\Tr\, \frac{1}{x-M}\right>=\sum_{k=0}^\infty \Tr(M^k)x^{-k-1}\eeq
Ainsi les coefficients de la série de Taylor au voisinage de $x=\infty$ permettent d'obtenir les différentes valeurs moyennes des traces de $M^{k}$. De même les coefficients en $x_1^{-i_1}\dots x_n^{-i_n}$ permettent d'obtenir $c_{i_1,\dots,i_n}=\left<\Tr(M^{i_1})\dots \Tr(M^{i_n})\right>$. Ces quantités sont suffisantes pour déterminer la loi jointe complète des valeurs propres $\lambda_1,\dots,\lambda_N$ de $M$ et pas seulement la loi marginale d'une valeur propre qui ne nécessiterait que la connaissance de $W_1(x)$. Dans le cas où la récurrence topologique peut être appliquée (Cf. chapitre précédent), on peut reconstruire l'intégralité de ces moments à partir uniquement de la courbe spectrale, c'est-à-dire de la loi marginale limite. Bien qu'intéressantes, les valeurs moyennes des traces des puissances de $M$ ne constituent pas les seules variables d'intérêt d'un modèle de matrices aléatoires. En effet, toute quantité invariante par changement de base présente un intérêt comme par exemple les déterminants ou les polynômes caractéristiques. Ainsi des quantités comme:
\bea \label{Fermions} \psi(x)&=&\left<\det(x I_N-M)\right>\cr
\phi(x)&=&\left<\det((x I_N-M)^{-1})\right>\cr
\mathcal{K}(x,y)&=&\frac{1}{x-y}\left<\det\,\frac{x I_N-M}{y I_N-M}\right>
\eea
\sloppy{ont un intérêt. Ces quantités ne sont bien sûr pas indépendantes des $c_{i_1,\dots,i_n}=\left<\Tr M^{i_1}\dots \Tr M^{i_n}\right>$. Par exemple, la relation avec $\psi(x)$ est donnée par le calcul simple suivant:}
\bea \psi(x)&=&\left<\det(x I_N-M)\right>=x^N\left<\det\left(I_N-\frac{M}{x}\right)\right>\cr 
&=&x^N\left<e^{\Tr \log(I_N-\frac{M}{x})}\right>\cr
&=&x^N\left<e^{\Tr \int_\infty^x \left(\frac{dx'}{x' I_N-M}-\frac{dx'}{x'}I_N\right)}\right>\cr
&=&x^N\left<:e^{\Tr \int_\infty^x\frac{dx'}{x' I_N-M}}:\right>\cr
&=&x^N\left<:\underset{n=1}{\overset{\infty}{\sum}} \int_\infty^x \dots \int_\infty^x \Tr\,\frac{dx'}{x_1' I_N-M}\dots \Tr\,\frac{dx'}{x_n' I_N-M}:\right>\cr
&=&x^N:\underset{n=1}{\overset{\infty}{\sum}} \int_\infty^x \dots \int_\infty^x W_n^{n.c.}(x_1',\dots,x_n')dx_1'\dots dx_n':\cr
&=&x^N:\exp\left(\underset{n=1}{\overset{\infty}{\sum}}\frac{1}{n!} \int_\infty^x \dots \int_\infty^x W_n(x_1',\dots,x_n')dx_1'\dots dx_n'\right):\cr
&=&x^Ne^{\int_\infty^x \left(W_1(x')-\frac{N}{x'}\right)dx'}e^{\underset{n=2}{\overset{\infty}{\sum}}\frac{1}{n!}\int_\infty^x \dots \int_\infty^x W_n(x_1',\dots,x_n')dx_1'\dots dx_n'}
\eea
Dans ce calcul, la notation $:\,g \,:$ désigne une régularisation des quantités dont les détails peuvent être trouvés dans \cite{Lie}. Notons également qu'a priori cette série doit être comprise comme une série formelle à $N\to \infty$. Comme on le voit, le polynôme caractéristique de $M$ peut être exprimé à l'aide des fonctions de corrélation à travers la dernière ligne du calcul précédent. Ainsi, les zéros de $\psi(x)$, c'est-à-dire les valeurs moyennes des valeurs propres de $M$, peuvent être complètement reconstruits à l'aide des fonctions de corrélation. Dans le cas où les fonctions de corrélations admettent un développement topologique, on obtient alors:
\beq \label{psi}\psi(x)=x^Ne^{\frac{N}{T}\int_\infty^x \left(W_1^{(0)}(x')-\frac{T}{x'}\right)dx'}e^{\underset{(n,g)\neq(1,0)}{\overset{\infty}{\sum}}\left(\frac{T}{N}\right)^{n-2+2g}\frac{1}{n!} \int_\infty^x \dots \int_\infty^x W_n^{(g)}(x_1',\dots,x_n')dx_1'\dots dx_n'}\eeq

On note immédiatement la ressemblance de la formule précédente avec le développement WKB d'un système différentiel. D'une façon plus générale, pour une courbe spectrale $\Sigma$ donnée et un paramètre formel $\hbar$ (qui vaut $\frac{T}{N}$ dans les modèles de matrices présentés dans ce document), il semble intéressant de définir le noyau:
\beq \td{\mathcal{K}}(z,z')=\text{exp}\left({\underset{n=1,g=0}{\overset{\infty}{\sum}}\hbar^{n-2+2g}\frac{1}{n!} \int_{z'}^z \dots \int_{z'}^z \omega_n^{(g)}} \right)\eeq
Ainsi la spécialisation $z'\to \infty_j$ avec régularisation pour $\omega_1^{(0)}$ redonne une formule de type \eqref{psi}. Dans le cas de modèle de matrices, la quantité $\td{\mathcal{K}}(z,z')$ redonne le polynôme caractéristique de la matrice $M$ et il est donc légitime de se demander ce qu'il advient pour une courbe spectrale générale. Cette question est au centre de la notion de ``courbe spectrale quantique''. En effet, il a été observé dans des cas simples (Airy \cite{MS12,N15} et Painlevé I \cite{KTPainleve,Iwaki}) que $\td{\mathcal{K}}(z,\infty)$ obéissait à une version ``quantifiée'' de la courbe quantique. Par exemple, dans le cas du système d'Airy \cite{BergereEynard}, le système est donné par:
\beq \hbar\, \frac{d}{dx} \Psi(x)=\begin{pmatrix} 0&1\\ x&0\end{pmatrix}\Psi(x)\eeq
La courbe spectrale est $y^2=x$ (courbe d'Airy) dont tous les $\omega_n^{(g)}$ sont connus explicitement à l'aide des nombres de Bernoulli. On peut alors montrer que $\psi(x)=\td{\mathcal{K}}(x,\infty)$ satisfait l'équation différentielle:
\beqq \left(\hbar^2 \frac{d^2}{dx^2}-x\right) \psi(x)=0\eeqq
En d'autres termes, il s'agit de la courbe spectrale initiale $y^2-x=0$ dans laquelle $(x,y)$ est devenu ``quantique'' car obéissant désormais à $[y,x]=\hbar$, i.e. $y=\hbar \partial_x$. Dans le cas de Painlevé I et Painlevé II, des résultats similaires sont disponibles, mais la courbe quantique contient parfois des termes imprévus proportionnels à $\hbar$. D'une façon générale, on définit une courbe quantique de la manière suivante:

\begin{definition}[Courbe quantique] Soit $P(x,y)$ une équation polynomiale définissant une courbe spectrale classique. On appelle courbe quantique toute équation différentielle de la forme:
\beqq \td{P}(x,\hbar \partial_x,\hbar)=P(x,\hbar\partial_x)+\hbar f(x,\partial_x,\hbar)=0 \text{ avec }\lim_{\hbar\to 0}f(x,\partial_x,\hbar) \text{ fini }\eeqq
En d'autres termes, il s'agit d'une équation différentielle obtenue à partir de la courbe spectrale initiale en remplaçant $y\to \hbar \partial_x$ avec éventuellement des corrections qui disparaissent pour $\hbar\to 0$.
\end{definition}

Notons que si l'on peut toujours passer d'une courbe quantique à la courbe classique en prenant $\hbar\to 0$, l'opération inverse est mal définie car la définition d'une courbe quantique n'est pas unique. De cette définition il ressort la conjecture suivante:

\begin{conjecture} Soit $P(x,y)=0$ une équation polynomiale définissant une courbe spectrale $\Sigma$. Soit $\psi_j(z)=\td{\mathcal{K}}(z,\infty_j)$ une régulatisation de $\td{\mathcal{K}}(z,z')$ définie à partir des différentielles $\omega_n^{(g)}$ calculées par la récurrence topologique sur $\Sigma$. Alors il existe une courbe spectrale quantique $\td{P}_j(x,\hbar \partial_x, \hbar)$ associée à $\Sigma$ annihilant $\psi_j(x)$:
\beqq\td{P}_j(x,\hbar \partial_x, \hbar)\psi_j(x)=0\eeqq 
telle que les coefficients de $P_j$ vus comme fonction de $x$ n'ont pas de singularités aux points de branchements de la courbe spectrale $\Sigma$.
\end{conjecture}

Notons que le choix de $\infty_j$ peut influer sur la courbe quantique mais uniquement sur les termes de correction qui disparaissent dans la limite $\hbar\to 0$. A l'heure actuelle, cette conjecture a été démontrée au cas par cas pour des courbes spectrales initiales simples pour lesquelles des formules plus ou moins explicites sont connues \cite{Do,Iwaki,EynardBouchard}. En définissant un vecteur $\boldsymbol{\psi}=(\psi_1,\dots,\psi_r)$ tenant compte des différents choix possibles d'infini, on peut également se poser la question de savoir si $\boldsymbol{\psi}$ satisfait ou non le système différentiel de la forme $\hbar \partial_x \boldsymbol{\psi}=\DD \boldsymbol{\psi}$ dont les équations séculières (i.e. éliminer les autres composantes) redonneraient chacune des courbes quantiques $\left(\td{P}_j\right)_{1\leq j\leq r}$. Dans ce cas de figure, on retrouverait ainsi un système différentiel (ou une paire de Lax si un paramètre $t$ supplémentaire est présent) permettant d'achever définitivement le lien entre la récurrence topologique et les systèmes différentiels. A l'heure actuelle, cette conjecture se précise dans le cas où les courbes spectrales sont de genre $0$ mais beaucoup de mystères restent encore à éclaircir. Par ailleurs, le cas des courbes de genre strictement positif pose également la question relative aux normalisations à choisir lors d'intégrales effectuées autour des cycles d'homotopie. Quoiqu'il en soit l'activité autour de cette conjecture est très importante et de nombreux cas particuliers ont été démontrés complètement ou partiellement:
\begin{itemize}
\item Courbe d'Airy $y^2=x$ démontrée dans \cite{BergereEynard}.
\item Modèles Gaussien $y^2-xy+1=0$ démontrés dans \cite{Penkava}.
\item Nombres de Hurwitz, Invariants de Gromov-Witten sur $\mathbb{B}_1$, ``hypermaps'' démontrés dans \cite{MS15,MS12,QuantumDo,EynardBouchard}.
\item En genre strictement positif: Systèmes de Hitchin holomorphes et méromorphes de rang $2$ partiellement démontrés dans \cite{DM14,Quantum}.
\end{itemize} 
Cette conjecture est importante car elle présente un grand intérêt dans la théorie des cordes topologiques et pour les applications en géométrie énumérative. Elle permet également une approche nouvelle dans les systèmes intégrables entre les variables ``bosoniques'' (les fonctions de corrélations $\equiv$ Traces de $M$) et ``fermioniques'' (la fonction $\psi$ $\equiv$ Déterminants de $M$).

\section{Généralisation aux équations aux différences}

La question de l'utilisation de la récurrence topologique dans le cadre des systèmes différentiels intégrables donnés par des paires de Lax peut être étendue au cas des systèmes d'équations aux différences, c'est-à-dire aux systèmes de la forme:
\beq \label{equationdifference}\delta_\hbar \Psi(x;\hbar)\overset{\text{def}}{=}\left(\text{exp}\left(\hbar\frac{d}{dx}\right)\right)\Psi(x;\hbar)=\sum_{k=0}^\infty \frac{\hbar^k}{k!}\frac{d^k}{dx^k}\Psi(x;\hbar)=L(x;\hbar)\Psi(x;\hbar)\eeq
où $L(x;\hbar)$ est une matrice admettant un développement formel en $\hbar$:
\beq L(x;\hbar)=\sum_{k=0}^\infty L_k(x)\hbar^k\eeq
Dans ce cas, l'opérateur $\delta_\hbar$ remplace l'opérateur différentiel $\hbar\frac{d}{dx}$ dont il constitue une version exponentiée puisque formellement on a:
\beq \label{Lien}\delta_\hbar=\text{exp}\left(\hbar\frac{d}{dx}\right)\eeq
En conséquence, le problème \eqref{equationdifference} peut être vu comme un problème directement défini sur un groupe de Lie $G$ et non plus sur une algèbre de Lie associée $\mathfrak{g}$. En ce sens, il constitue une généralisation naturelle des problèmes développés dans les parties précédentes. Notons, que comme précédemment, l'équation \eqref{equationdifference} doit être comprise comme une équation dont on cherche les solutions admettant un développement WKB formel en $\hbar$. Ainsi les questions de convergence des séries en $\hbar$ n'ont pour l'instant jamais fait l'objet d'études approfondies. Notons également que la terminologie ``équations aux différences'' provient du fait que pour des quantités admettant un développement formel de Taylor (qui est un cas particulier de développement WKB dans lequel le terme $\text{exp}(O\left(\frac{1}{\hbar}\right))$ est nul) on a formellement:
\beq \delta_\hbar f(x)=\sum_{k=0}^\infty \frac{f^{(k)}(x)}{k!}\hbar^k=f(x+\hbar)\eeq
Notons enfin que du point de vue algébrique, l'opérateur $\delta_\hbar$ ne se comporte pas comme un opérateur différentiel puisqu'il obéit aux règles algébriques suivantes:
\bea \delta_{\hbar}(\alpha)&=&\alpha \,\,,\,\, \forall \, \alpha\in \mathbb{C}\cr
\delta_\hbar(\alpha f(x)+g(x))&=&\alpha \delta_\hbar f(x)+\delta_\hbar g(x)\cr
\delta_\hbar((fg)(x))&=&  (\delta_\hbar f(x))(\delta_\hbar g(x))
\eea

Bien qu'obéissant à des règles algébriques différentes de celui des opérateurs différentiels classiques, le lien \eqref{Lien} entre $\delta_\hbar$ et $\hbar\partial_x$ laisse entrevoir la possibilité que les formalismes de la récurrence topologique et des équations déterminantales puissent être adaptés à cette situation. Cette perspective a donné lieu à une conjecture de B. Dubrovin et D. Yang \cite{Dubrovin} concernant un des systèmes d'équations aux différences le plus simple en relation avec les invariants de Gromov-Witten de $\mathbb{P}^1$. 

\subsection{Exemple sur les invariants de Gromov-Witten de $\mathbb{P}^1$}
Définissons $\overline{\mathcal{M}}_{g,n}(\mathbb{P}^1,d)$ comme l'espace des modules des applications stables de degré $d$ d'une surface de genre $g$ à $n$ points dans $\mathbb{P}^1$ et les invariants de Gromov-Witten de $\mathbb{P}^1$ par
\beq \left< \underset{i=1}{\overset{n}{\prod}}\tau_{b_i}(\alpha_i)\right>^d_{g,n}=\int_{\left[\overline{\mathcal{M}}_{g,n}(\mathbb{P}^1,d)\right]^{\text{vir}} }\underset{i=1}{\overset{n}{\prod}} \psi_i^{\,b_i}ev_i^*(\alpha)_i\eeq
où $\left[\overline{\mathcal{M}}_{g,n}(\mathbb{P}^1,d)\right]^{\text{vir}}$ est la classe fondamentale virtuelle de l'espace de module (Cf. \cite{QuantumP1}). En utilisant le paramètre $\hbar$ comme paramètre formel, on peut regrouper ces invariants sous la forme de séries génératrices:
\beq \left<\tau_{k_1}(\omega)\dots \tau_{k_n}(\omega)\right>=\sum_{2g-2+2d=k_1+\dots+k_n} \hbar^{2g-2+n}\left<\tau_{k_1}(\omega)\dots \tau_{k_n}(\omega)\right>_{g,n}^d\eeq
Le résultat principal de \cite{QuantumP1} est de montrer qu'en définissant les séries formelles en $\frac{1}{x}$ suivantes:
\bea S_0(x)&=&x-x\,\ln\, x+\sum_{i=1}^\infty \left<-\frac{(2d-2)!\, \tau_{2d-2}(\omega)}{x^{2d-1}}\right>^d_{0,1}\cr
S_1(x)&=&-\frac{1}{2}\,\ln\, x +\frac{1}{2}\sum_{d=0}^\infty \left<\left(-\frac{\tau_0(1)}{2}-\sum_{b=0}^\infty \frac{b!\tau_b(\omega)}{x^{b+1}}\right)^2\right>^d_{0,2}\cr
F_{g,n}(x_1,\dots,x_n)&=&\left<\underset{i=1}{\overset{n}{\prod}}\left(-\frac{\tau_0(1)}{2}-\sum_{b=0}^\infty \frac{b!\tau_b(\omega)}{x_i^{b+1}}\right)\right>^d_{g,n} \,,\text{ pour } 2g-2+n>0\cr
&&\eea
alors le développement WKB formel de la fonction d'onde définie par:
\beq \label{Wave}\psi(x;\hbar)=\exp\left(\frac{1}{\hbar}S_0(x)+S_1(x)+\sum_{2g-2+n>0}\frac{\hbar^{2g-2+n}}{n!}F_{g,n}(x,\dots,x)\right)\eeq
vérifie une équation aux différences de la forme:
\beq \label{QuantumCurve} \left[\delta_\hbar+\delta_{-\hbar}-x\right]\psi(x;\hbar)=\left[\exp\left(\hbar \frac{d}{dx}\right)+\exp\left(-\hbar\frac{d}{dx}\right)-x\right]\psi(x;\hbar)=0\eeq
De plus, les fonctions $\left(F_{g,n}(x_1,\dots,x_n)\right)_{g\geq 0, n\geq 1, 2g+2-n>0}$ s'identifient avec les différentielles d'Eynard-Orantin $\omega_n^{(g)}$ calculées par la récurrence topologique appliquée à la courbe spectrale classique de genre $0$: 
\beq \label{ClassicalCurve} \forall\, z\in \overline{\mathbb{C}}\,:\, x(z)=z+\frac{1}{z} \,\,,\,\, y(z)=\ln(z) \,\,\left(\text{i.e. } y(x)=\text{cosh}^{-1}\left(\frac{x}{2}\right)\right)\eeq
via les relations:
\bea F_{g,n}(x(z_1),\dots,x(z_n))&=&\int_0^{z_1}\dots \int_0^{z_n} \omega_n^{(g)}(z_1',\dots,z_n')\cr
\Rightarrow\,\,F_{g,n}(x(z_1),\dots,x(z_n))dx_1\dots dx_n&=&\omega_n^{(g)}(z_1',\dots,z_n')\eea
En d'autres termes, les différentielles d'Eynard-Orantin sont reliées aux invariants de Gromov-Witten de $\mathbb{P}^1$ par:
\beq \label{GWwng}\omega_n^{(g)}(x_1,\dots,x_n)=\sum_{k_1,\dots,k_n\geq 0} \frac{(k_1+1)!\dots (k_n+1)!}{x_1^{k_1+2}\dots x_n^{k_n+2}}\left<\tau_{k_1}(\omega)\dots \tau_{k_n}(\omega)\right>^d_{g,n}dx_1\dots dx_n\eeq
La conjecture de B. Dubrovin et D. Yang présentée dans \cite{Dubrovin} est ainsi la suivante:

\begin{conjecture}[Conjecture de B. Dubrovin et D. Yang]\label{ConjDubro}
Les fonctions:
\bea
C_n(x_1,\dots,x_n;\hbar)&=&\sum_{k_1,\dots,k_n\geq 0} \frac{(k_1+1)!\dots (k_n+1)!}{x_1^{k_1+2}\dots x_n^{k_n+2}}\left<\tau_{k_1}(\omega)\dots \tau_{k_n}(\omega)\right>^d\cr
&=&\sum_{g=0}^\infty\hbar^{n-2-2g}\sum_{k_1,\dots,k_n\geq 0} \frac{(k_1+1)!\dots (k_n+1)!}{x_1^{k_1+2}\dots x_n^{k_n+2}}\left<\tau_{k_1}(\omega)\dots \tau_{k_n}(\omega)\right>_{g,n}^d\cr
&=& \sum_{g=0}^\infty\hbar^{n-2-2g}\frac{\omega_n^{(g)}(x_1,\dots,x_n)}{dx_1\dots dx_n}\cr
&&\text{(où $\omega_n^{(g)}$ sont les différentielles d'Eynard-Orantin associées }\cr
&&\text{ à la courbe spectrale classique \eqref{ClassicalCurve})}
\eea
sont reconstruites formellement pour $n\geq 2$ par:
\bea \label{C} C_2(x_1,x_2;\hbar)&=&\frac{\Tr\left(\td{M}(x_1;\hbar)\td{M}(x_2;\hbar)-1\right)}{(x_1-x_2)^2}\cr
C_n(x_1,\dots,x_n;\hbar)&=&\frac{(-1)^{n+1}}{n}\sum_{\sigma \in S_n}\frac{\Tr\left(\td{M}(x_{\sigma(1)};\hbar)\dots \td{M}(x_{\sigma(n)};\hbar)\right)}{(x_{\sigma(1)}-x_{\sigma(2)})\dots (x_{\sigma(n-1)}-x_{\sigma(n)})(x_{\sigma(n)}-x_{\sigma(1)})}\cr
&&\eea
où $\td{M}(x;\hbar)$ est la matrice $2\times2$ suivante:
\beq \label{ConjectureM}\td{M}(x;\hbar)=\begin{pmatrix}1&0\\ 0&0\end{pmatrix}
+ \begin{pmatrix} \alpha(x;\hbar)&\beta(x;\hbar)\\ \gamma(x;\hbar)&-\alpha(x;\hbar)\end{pmatrix}
\eeq
avec les coefficients suivants:
\small{\bea \label{ConjectureM2}\alpha(x;\hbar)&=&\sum_{j=0}^\infty \frac{1}{4^jx^{2j+2}}\sum_{i=0}^j \hbar^{2(j-i)}\frac{1}{i!(i+1)!}\sum_{l=0}^i(-1)^l(2i+1-2l)^{2j+1}\binom{2i+1}{l}\cr
\gamma(x;\hbar)&=&Q(x;\hbar)+P(x;\hbar)\cr
\beta(x;\hbar)&=&Q(x;\hbar)-P(x;\hbar)\cr
P(x;\hbar)&=&\sum_{j=0}^\infty \frac{1}{4^jx^{2j+1}}\sum_{i=0}^j \hbar^{2(j-i)}\frac{1}{(i!)^2}\sum_{l=0}^i (-1)^l (2i+1-2l)^{2j}\left[\binom{2i}{l}-\binom{2i}{l-1}\right]\cr
Q(x;\hbar)&=&\frac{1}{2}\sum_{j=0}^\infty \frac{1}{4^jx^{2j+2}}\sum_{i=0}^j\hbar^{2(j-i)+1}\frac{2i+1}{(i!)^2}\sum_{l=0}^i(-1)^l(2i+1-2l)^{2j}\left[\binom{2i}{l}-\binom{2i}{l-1}\right]\cr
&&\eea}\normalsize{}
\end{conjecture}

On voit donc que la conjecture correspond exactement à montrer la reconstruction des fonctions de corrélation par les formules déterminantales dans le cas d'un système aux différences plutôt qu'un système différentiel classique. Le résultat principal de mon article \cite{DifferenceTopRec} est ainsi de \textbf{démontrer cette conjecture}. Notons que les outils et correspondances utilisés pour cette démonstration peuvent très certainement être étendus à d'autres systèmes du même type laissant entrevoir beaucoup de projets de recherche dans cette direction. Plus généralement, ce travail, réalisé sur un exemple simple, s'inscrit dans la compréhension globale du schéma suivant:

\bigskip

\begin{center}
\begin{picture}(435,230)\small{
\thicklines 
\put(15,220){Différentielles} 
\put(5,205){d'Eynard-Orantin}
\put(8,190){$\omega_n^{(g)}(x_1,\dots,x_n)$}
\put(0,185){\line(1,0){92}}
\put(0,230){\line(1,0){92}}
\put(0,185){\line(0,1){45}}
\put(92,185){\line(0,1){45}}

\put(100,222){Récurrence} 
\put(97,210){Topologique} 
\thicklines
\put(160,205){\vector(-1,0){68}}

\put(40,130){$=$}
\put(33,122){\line(1,0){22}}
\put(33,145){\line(1,0){22}}
\put(33,122){\line(0,1){23}}
\put(55,122){\line(0,1){23}}

\put(84,138){Propriété de type}
\put(99,122){topologique}
\put(80,117){\line(1,0){91}}
\put(80,150){\line(1,0){91}}
\put(80,117){\line(0,1){33}}
\put(171,117){\line(0,1){33}}
\put(80,132){\vector(-1,0){25}} 

\put(332,215){Courbe quantique}
\put(322,198){$\tilde{\text{E}}\left(x,\text{exp}\left(\hbar \frac{d}{dx}\right)\right)\Psi=0$}
\put(318,185){\line(1,0){117}}
\put(318,230){\line(1,0){117}}
\put(318,185){\line(0,1){45}}
\put(435,185){\line(0,1){45}}

\put(248,215){Quantification}
\put(268,195){$\hbar=0$}
\put(246,210){\vector(1,0){72}}
\put(318,205){\vector(-1,0){72}}

\put(240,130){$\text{det}(yI_d-D_0(x))$}
\put(235,85){\vector(0,1){100}} 
\multiput(225,185)(0,-10){10}{\line(0,-1){5}}
\put(225,90){\vector(0,-1){5}}

\put(163,220){Courbe spectrale} 
\put(183,205){classique}
\put(176,190){$\text{E}(x,e^{\,y})=0$}
\put(160,185){\line(1,0){86}}
\put(160,230){\line(1,0){86}}
\put(160,185){\line(0,1){45}}
\put(246,185){\line(0,1){45}}

\put(208,70){Système différentiel}  
\put(237,50){linéaire} 
\put(207,30){$\hbar\Psi'(x)=D(x)\Psi(x)$}
\put(203,20){\line(1,0){102}}
\put(203,85){\line(1,0){102}}
\put(203,20){\line(0,1){65}}
\put(305,20){\line(0,1){65}}

\put(190,103){?}
\put(200,88){\line(-1,1){10}}
\put(185,103){\line(-1,1){10}}

\put(3,70){Fonctions de corrélations} 
\put(25,55){$W_n(x_1,\dots,x_n)$}
\put(3,35){$=\underset{k=0}{\overset{+\infty}{\sum}} W_n^{(k)}(x_1,\dots,x_n)\hbar^k$}
\put(0,20){\line(1,0){125}}
\put(0,85){\line(1,0){125}}
\put(0,20){\line(0,1){65}}
\put(125,20){\line(0,1){65}}

\put(145,68){\footnotesize{Formules}}
\put(130,58){\footnotesize{déterminantales}}
\put(203,53){\vector(-1,0){77}}

\put(317,65){exp}
\put(305,60){\vector(1,0){38}}
\multiput(343, 40)(-10,0){4}{\line(-1,0){5}}
\put(310,40){\vector(-1,0){5}}
\put(317,30){log}

\put(385,135){Equation}
\put(387,125){séculière}
\put(380,85){\vector(0,1){100}}
\multiput(370,185)(0,-10){10}{\line(0,-1){5}}
\put(370,90){\vector(0,-1){5}}

\put(350,70){Système linéaire}
\put(348,55){aux $\hbar$-différences} 
\put(348,40){$\left(\text{exp}\left(\hbar\frac{d}{dx}\right)\right)\Psi(x)$}
\put(360,25){$=L(x)\Psi(x)$}
\put(343,20){\line(1,0){92}}
\put(343,85){\line(1,0){92}}
\put(343,20){\line(0,1){65}}
\put(435,20){\line(0,1){65}}

\multiput(45, 86)(0,10){4}{\line(0,1){5}}
\multiput(45, 147)(0,10){4}{\line(0,1){5}}

\put(145,15){\footnotesize{Formules}}
\put(130,5){\footnotesize{déterminantales}}
\put(62,0){\line(1,0){327}}
\put(389,0){\line(0,1){20}}
\put(62,0){\vector(0,1){20}}

}\normalsize{}
\end{picture}

\smallskip

Fig. 7: Schéma général dans le cas des systèmes linéaires aux différences ou différentiels.
\end{center}

\subsection{Correspondance avec les systèmes différentiels classiques}
L'idée principale développée dans \cite{DifferenceTopRec} est d'utiliser une correspondance entre les systèmes aux différences et un système différentiel associé. Pour cela, il convient tout d'abord d'introduire une version matricielle adaptée du problème scalaire aux différences défini par \eqref{QuantumCurve}. Ainsi en partant de l'équation:
\beq \label{QuantumC}\left(\delta_\hbar+\delta_{-\hbar}-x\right)f(x)=0 \,\,\Leftrightarrow \,\, f(x+\hbar)+f(x-\hbar)-xf(x)=0 \eeq
le théorème principal de \cite{QuantumP1} nous assure que $\psi(x;\hbar)$ (défini par \eqref{Wave}) est une solution formelle de \eqref{QuantumC}. De plus par invariance $\hbar\to -\hbar$, on obtient que 
\beq \label{phi} \phi(x;\hbar)=\exp\left(-\frac{1}{\hbar}S_0(x)+S_1(x)+\sum_{2g-2+n>0}(-1)^n\frac{\hbar^{2g-2+n}}{n!}F_{g,n}(x,\dots,x)\right)\eeq
est également une autre solution linéairement indépendante de \eqref{QuantumC}. On définit alors la matrice $2\times 2$:
\beq \label{Psi}\Psi(x;\hbar)=\begin{pmatrix} \psi(x+\frac{\hbar}{2};\hbar)&\phi(x+\frac{\hbar}{2};\hbar)\\ \psi(x-\frac{\hbar}{2};\hbar)&\phi(x-\frac{\hbar}{2};\hbar)\end{pmatrix}\eeq
où $\psi(x\pm \frac{\hbar}{2})$ et $\phi(x\pm \frac{\hbar}{2})$ sont définis comme des séries WKB formelles en $\hbar$:
\bea \psi\left(x+\frac{\hbar}{2};\hbar\right)&=&\sum_{k=0}^\infty \frac{\hbar^k}{2^k k!}\frac{d^k}{dx^k}\psi(x;\hbar)=\exp\left(\frac{1}{\hbar}S_0(x)\right)\left(\sum_{k=1}^\infty \hat{S}_{k,+}(x)\hbar^k\right)\cr
\psi\left(x-\frac{\hbar}{2};\hbar\right)&=&\sum_{k=0}^\infty \frac{(-1)^k\hbar^k}{2^k k!}\frac{d^k}{dx^k}\psi(x;\hbar)=\exp\left(\frac{1}{\hbar}S_0(x)\right)\left(\sum_{k=1}^\infty \hat{S}_{k,-}(x)\hbar^k\right)\cr
\phi\left(x-\frac{\hbar}{2};\hbar\right)&=&\sum_{k=0}^\infty \frac{(-1)^k\hbar^k}{2^k k!}\frac{d^k}{dx^k}\phi(x;\hbar)=\exp\left(-\frac{1}{\hbar}S_0(x)\right)\left(\sum_{k=1}^\infty \td{S}_{k,-}(x)\hbar^k\right) \cr
\phi\left(x+\frac{\hbar}{2};\hbar\right)&=&\sum_{k=0}^\infty \frac{\hbar^k}{2^k k!}\frac{d^k}{dx^k}\phi(x;\hbar)=\exp\left(-\frac{1}{\hbar}S_0(x)\right)\left(\sum_{k=1}^\infty \td{S}_{k,+}(x)\hbar^k\right)
\eea
Matriciellement, $\Psi(x;\hbar)$ est une solution formelle du système aux différences:
\beq \label{DiffSystem} \delta_\hbar \Psi(x;\hbar)=\Psi(x+\hbar)=L(x;\hbar) \Psi(x;\hbar) \text{  avec  } L(x;\hbar)=\begin{pmatrix}x+\frac{\hbar}{2}&-1\\ 1&0\end{pmatrix}\eeq
Notons que le choix d'un système matriciel à partir d'un problème scalaire n'est pas unique et que choisir le système pertinent n'est pas toujours évident. Dans notre cas, le choix \eqref{Psi} est nécessaire pour vérifier la propriété de parité en $\hbar$ des fonctions de corrélation. Tout autre choix est bien sûr possible mais ne conduira pas à la vérification de la conjecture. Notons également que puisque $\det L(x;\hbar)$ (traduisant le fait que $L(x;\hbar)\in G=SL_2(\mathbb{C})$ ici), on peut normaliser les fonctions $\psi(x;\hbar)$ et $\phi(x;\hbar)$ de telle sorte que $\det \Psi(x;\hbar)=1$. L'étape suivante est alors d'associer un système différentiel à \eqref{DiffSystem}:

\begin{proposition}Au système aux différences \eqref{DiffSystem} on peut associer un système différentiel:
\beq \label{Compat} \hbar \frac{d}{dx} \Psi(x;\hbar)=D(x;\hbar)\Psi(x;\hbar) \eeq
compatible avec l'équation $\delta_\hbar \Psi(x;\hbar)=\Psi(x+\hbar)=L(x;\hbar)\Psi(x;\hbar)$ en posant
\beq D(x;\hbar)=\left(\hbar\frac{d}{dx} \Psi(x;\hbar)\right)\Psi(x;\hbar)^{-1} \label{DDD}\eeq
En particulier, $D(x;\hbar)$ admet automatiquement un développement formel en $\hbar$ de la forme: $D(x;\hbar)=\underset{k=0}{\overset{\infty}{\sum}}D_k(x)\hbar^k$.
\end{proposition} 

Notons que cette définition implique automatiquement que $\Tr D(x;\hbar)=0$ (conséquence de $\det L(x;\hbar)=1$) c'est-à-dire que $D(x;\hbar)\in \mathfrak{g}=\mathfrak{sl}_2(\mathbb{C})$. Notons qu'a priori il peut exister plusieurs systèmes différentiels compatibles avec un système aux différences mais que la définition \eqref{DDD} permet de faire un lien direct correspondant à un choix de matrice $\Psi(x;\hbar)$ donnée. Le but de \cite{DifferenceTopRec} est ainsi d'utiliser les différentes relations de compatibilité entre les deux systèmes pour démontrer les conditions de type topologique sur le système différentiel \eqref{Compat} et ainsi d'aboutir à une démonstration de la conjecture \ref{ConjDubro}. Les grandes étapes sont ainsi les suivantes:
\begin{enumerate}\item Etudier l'ordre dominant et la connexion entre $D_0$ et $L_0$. En particulier, on observe que la courbe spectrale classique (le polynôme caractéristique de $D_0$) est bien donné par \eqref{ClassicalCurve}.
\item Etablir des formules de récurrence reliant les entrées des matrices $(L_k)_{k\geq 0}$ avec celles de $(D_i)_{i\geq 0}$. Ces relations sont établies en toute généralité pour des systèmes $2\times 2$ dans \cite{DifferenceTopRec}. Ces relations permettent en particulier de suivre les singularités des matrices $(D_k)_{k\geq 0}$ à partir de celles des $(L_k)_{k\geq 0}$ et réciproquement (dans le cas de \eqref{DiffSystem}, $L(x;\hbar)=L_0(x)+L_1\hbar$ permet de suivre facilement la dépendance en $x$ des matrices $D_k$).
\item Introduire la matrice $M(x;\hbar)$ utilisée pour les formules déterminantales:
\beq \label{MatriceMM}M(x;\hbar)=\Psi(x;\hbar)\begin{pmatrix} 1&0\\0&0\end{pmatrix} \Psi^{-1}(x,\hbar)\eeq
En particulier, on note qu'elle vérifie les équations:
\bea \label{EqM}\hbar \frac{d}{dx} M(x;\hbar)&=&\left[D(x;\hbar),M(x;\hbar)\right]\cr
\delta_\hbar M(x;\hbar)=M(x+\hbar;\hbar)&=&L(x;\hbar)M(x;\hbar)L(x;\hbar)^{-1}
\eea
La seconde équation, spécifique au cas des systèmes aux différences, permet alors d'obtenir des équations de récurrence permettant d'exprimer les entrées des matrices $M_k(x)$ en fonction des matrices d'ordres inférieurs $(M_j){j<k}$ et des entrées des matrices$(L_i)_{i\leq k}$. Ces relations de récurrence permettent alors d'étudier les différentes singularités potentielles des matrices $(M_k(x))_{k\geq 0}$ en fonction de celles présentes dans la matrice $L(x;\hbar)$.
\item En combinant les résultats précédents, on peut ainsi contrôler les singularités potentielles des matrices $(M_k(x))_{k\geq 0}$ et démontrer la condition concernant les positions des singularités dans la propriété de type topologique. On peut également effectuer le calcul direct de $W_2^{(0)}(x_1,x_2)dx_1dx_2$ à partir de la définition \ref{DeterFormulas} et vérifier son accord avec la différentielle d'Eynard-Orantin $\omega_2^{(0)}$.
\item Vérifier la propriété de parité en étudiant l'influence du changement $\hbar\to -\hbar$. Cette propriété dépend de façon essentielle du choix de la matrice $\Psi(x;\hbar)$. En particulier, il est important de choisir $\psi(x\pm\frac{\hbar}{2};\hbar)$ et $\phi(x\pm\frac{\hbar}{2};\hbar)$ dans la définition \eqref{Psi} de $\Psi(x;\hbar)$. Par exemple, le choix de $\psi(x\pm \hbar)$ et $\phi(x\pm \hbar)$ ne permet pas de vérifier la propriété de parité même si la matrice $L(x;\hbar)$ correspondante ne dépend plus de $\hbar$. 
\item En confrontant les équations de boucles vérifiées par les formules déterminantales et la structure des singularités obtenues par l'étude des $(M_k(x))_{k\geq 0}$ on montre, par le même raisonnement que celui développé dans \cite{P2}, que l'ordre dominant de la fonction de corrélation $W_n$ est au moins $O\left(\hbar^{n-2}\right)$. Notons, que dans notre cas, le raisonnement est possible malgré l'absence d'une seconde équation sur un paramètre de temps $t$ comme dans les paires de Lax, car la dépendance en $x$ de la matrice $L(x;\hbar)$ est particulièrement simple (un polynôme de degré $1$).
\item La combinaison des résultats précédents montre que la condition de type topologique est vérifiée et qu'ainsi les fonctions de corrélations définies par les formules déterminantales sont reconstruites par les différentielles d'Eynard-Orantin calculées par la récurrence topologique. Il suffit alors de vérifier que la formule conjecturale de la matrice $\td{M}(x;\hbar)$ proposée par Dubrovin et Yang correspond bien avec la matrice $M(x;\hbar)$ définie par \eqref{MatriceMM}. Notons que cette dernière étape se réduit à montrer des égalités de sommes multiples impliquant des coefficients binomiaux. Bien que ces vérifications puissent être effectuées par récurrence, il serait intéressant de voir si ces identités possèdent des interprétations combinatoires intéressantes ou non.
\end{enumerate}

Bien que \cite{DifferenceTopRec} constitue un exemple simple, il semble raisonnable que la stratégie présentée ci-dessus puisse s'appliquer à de nombreux systèmes aux différences. Cela permettrait en particulier de mettre en lumière la correspondance existant entre les problèmes (aux différences) définis sur les groupes de Lie et ceux (différentiels) définis sur les algèbres de Lie associées. En particulier, dans le cas de systèmes aux différences où la dépendance en $x$ est plus compliquée (ici $L(x;\hbar)$ était un polynôme en $x$ de degré $1$), il semble légitime de penser que la condition de type topologique ne puisse être démontrée que dans les cas où une seconde équation impliquant un ou des paramètres de ``temps'' soit présente, comme cela arrive dans le cas des systèmes définis par des paires de Lax. La question de savoir si de telles équations sur les ``temps'' seront de type ``équations aux différences'' ou `` équations différentielles'' reste ouverte à ce jour. Quoi qu'il en soit, il s'agit là de pistes de recherche prometteuses pour lesquelles l'exemple que j'ai développé dans \cite{DifferenceTopRec} présente beaucoup d'outils ainsi qu'une stratégie générale. En particulier, une compréhension précise du schéma de la figure $7$ présenté ci-dessus semble accessible dans un futur proche au moins dans le cas de nombreux problèmes issus de géométrie énumérative. Une application aux équations aux différences vérifiées par les polynômes de Jones associés à des n\oe uds mériterait d'être tentée comme une étape pour démontrer les conjectures proposées dans \cite{Knot1,Knot0}.

\section{Conclusion du chapitre}	
A partir d'un système différentiel linéaire dépendant d'un paramètre formel $\hbar$, ce chapitre a montré comment définir une courbe spectrale associée ainsi que des ``fonctions de corrélations'' qui possèdent des propriétés analogues à celles rencontrées dans la récurrence topologique. Moyennant certaines hypothèses sur le système, on peut alors identifier ces fonctions de corrélation avec celles produites par la récurrence topologique appliquée à la courbe spectrale. Dans le cas des paires de Lax, il existe une méthode générale utilisant la seconde équation sur le paramètre de temps $t$ qui permet de démontrer les hypothèses nécessaires à l'identification. Cette méthode a ainsi été appliquée avec succès pour le cas des six équations de Painlevé \cite{P2,IwakiMarchal} ainsi que dans le cas des généralisations récentes concernant certaines courbes de genre $0$ \cite{Lie,TTProperty}. Enfin, par analogie avec les modèles de matrices aléatoires où le déterminant peut être réécrit comme une exponentielle de trace, une formule de fonction d'onde a été récemment proposée et devrait satisfaire une équation différentielle ordinaire obtenue par une ``quantification'' de la courbe spectrale initiale. Cette courbe spectrale quantique a été mise en évidence dans beaucoup de cas particuliers et donne lieu aujourd'hui à une activité de recherche intense. Comme nous l'avons vu également, cette quantification semble pouvoir être effectuée également sur des problèmes définis directement sur les groupes de Lie plutôt que sur les algèbres de Lie associées en partant de systèmes d'équations aux différences. Toutes ces activités de recherche devraient prochainement préciser le lien fort existant entre la récurrence topologique et les systèmes intégrables que nous avons développé dans ce chapitre.

\selectlanguage{french}
\chapter{Universalité des matrices aléatoires et systèmes intégrables}
 \label{chap3}
\thispagestyle{fancy}

Le lien entre les modèles de matrices aléatoires Hermitiennes et les systèmes intégrables décrits dans le premier et le second chapitre est antérieur à la découverte de la récurrence topologique puisqu'il était déjà connu depuis plusieurs dizaines d'années avant la formalisation de la récurrence topologique dans \cite{EO}. On peut ainsi en retrouver les grandes lignes dans le livre référence de Mehta \cite{MehtaBook}. Cette étude classique utilise le formalisme des polynômes orthogonaux (et biorthogonaux dans le cas de modèles à deux matrices) et fournit une alternative à la récurrence topologique avec l'obtention de systèmes déterminantaux universels caractérisant les statistiques locales des valeurs propres. Ces systèmes déterminantaux s'avèrent être en plus des systèmes intégrables en lien avec les équations de Painlevé, ce qui au regard des chapitres précédents, n'apparait plus comme une surprise. Cette approche, en plus d'être historique, fournit un complément à la récurrence topologique, qui a également été utilisée pour retrouver ces résultats existants. Le but de ce chapitre est de présenter brièvement le formalisme des polynômes orthogonaux ainsi que les systèmes déterminantaux classiques qui en découlent. Cette partie classique sera complétée par une seconde montrant comment la récurrence topologique permet de retrouver rapidement ces résultats.

\section{Polynômes orthogonaux, systèmes déterminantaux et universalité locale}

Historiquement, l'étude des intégrales Hermitiennes de la forme
\beq \label{ZNChap3} Z_N=\int_{\mathbb{R}^N} d\lambda_1\dots d\lambda_N \Delta(\lambda_1,\dots,\lambda_N)^2 e^{-\frac{N}{T}\underset{i=1}{\overset{N}{\sum}} V(\lambda_i)}\eeq
a été réalisée à l'aide des polynômes orthogonaux $\left(Q_n(x)\right)_{n\geq 0}$ en décomposant:
\bea \label{JJJ}\Delta(\lambda_1,\dots,\lambda_N)^2&=&\det\left(\left(\lambda_i^{j-1}\right)_{1\leq i,j\leq N}\right) ^2\cr
&=&\det\left(\left(P_{i-1}(\lambda_j)\right)_{1\leq i,j\leq N}\right)\det\left(\left(Q_{i-1}(\lambda_j)\right)_{1\leq i,j\leq N}\right)\cr
&=&\sum_{\sigma,\tau\in\mathcal{\sigma}_N} (-1)^{|\sigma|+|\tau|} \left(\prod_{k=1}^N P_{\sigma(k)-1}(\lambda_k)\right)\left(\prod_{k=1}^N Q_{\tau(k)-1}(\lambda_k)\right)
\eea
où les polynômes $P_k$ et $Q_k$ sont des polynômes de degré $k$ moniques arbitraires. En réinjectant la dernière expression de \eqref{JJJ} dans \eqref{ZNChap3} et en choisissant les polynômes $Q_k=P_k$ orthogonaux pour la mesure $d\mu(x)=e^{-\frac{N}{T}\underset{i=1}{\overset{N}{\sum}} V(x)}dx$, c'est-à-dire:
\beq \int_{\mathbb{R}} P_r(x)P_s(x) e^{-\frac{N}{T}\underset{i=1}{\overset{N}{\sum}} V(x)}dx =h_r \delta_{r=s}\eeq
on obtient finalement:
\beq Z_N=N!\,\prod_{i=1}^N h_i\eeq
Cette relation signifie que la fonction de partition peut s'exprimer à l'aide des normes (induites par le produit scalaire décrit ci-dessus) des polynômes orthogonaux:
\beq h_i= \int_{\mathbb{R}} P_i(x)^2 e^{-\frac{N}{T}\underset{i=1}{\overset{N}{\sum}} V(x)}dx\eeq
En s'inspirant de cette méthode, on peut également réécrire les fonctions de corrélations à l'aide des polynômes orthogonaux \cite{MehtaBook} et le problème se résume alors à calculer l'asymptotique des polynômes orthogonaux pour $N\to \infty$. Traditionnellement ce type de problème est étudié par la méthode de ``steepest descent'' et certains problèmes de Riemann-Hilbert. Cette étude permet d'ailleurs de retrouver les résultats précédents \eqref{Coeff} sur la densité limite des valeurs propres et en constitue une preuve rigoureuse. Par ailleurs en utilisant des polynômes biorthogonaux, on peut facilement adapter la méthode précédente au cas $|\Delta(\boldsymbol{\lambda})|^{2\beta}$ avec $\beta\in\{\frac{1}{2},1,2\}$. Au-delà de la densité limite d'équilibre, il est intéressant de s'intéresser aux intégrales marginales partielles:

\begin{definition} Pour $n\geq 1$ on définit:
\beq \label{FonctionsCorrelations}\rho_n(\lambda_1,\dots,\lambda_n)= \frac{N!}{Z_N (N-n)!}\int_\mathbb{R}\dots\int_\mathbb{R} d\lambda_{n+1} \dots d\lambda_N \,\Delta(\boldsymbol{\lambda})^2  e^{-\frac{N}{T}\underset{i=1}{\overset{N}{\sum}} V(\lambda_i)}\eeq
\end{definition} 

Ainsi pour $n=1$ on retrouve (modulo la normalisation adéquate) la densité marginale d'une valeur propre tandis que pour $n\geq 2$, ces fonctions représentent la densité de probabilité associée au problème de trouver simultanément des valeurs propres en $\lambda_1,\dots,\lambda_n$, la position des autres valeurs propres restant libres. Le résultat fondamental concernant ces fonctions $\rho_n$ est connu sous le nom d'universalité décrit dans \cite{MehtaBook}:

\begin{theorem}
Pour les modèles de matrices Hermitiennes, symétriques réelles et quaternioniques auto-duales, les fonctions $\rho_n(x_1,\dots,x_n)$ avec $n>1$ à petite distance (i.e. d'ordre $1/N$) sont indépendantes du potentiel $V(z)$. En particulier, elles peuvent être calculées à l'aide du potentiel gaussien $V(z)=\frac{z^2}{2}$ pour lequel les formules sont explicitement connues (polynômes de Hermite). Enfin, la connaissance de $\rho_2(x_1,x_2)$ est suffisante pour déterminer les autres fonctions $\left(\rho_n(x_1,\dots,x_n)\right)_{n\geq 3}$ à l'aide de formules déterminantales (Cf. \cite{MehtaBook} pour les formules déterminantales spécifiques aux trois ensembles). Dans le c\oe ur de la distribution limite, la fonction $\rho_2(\lambda_1,\lambda_2)\equiv W_2(r=|\lambda_2-\lambda_1|)$ (par définition, $(\lambda_1,\lambda_2)\mapsto \rho_2(\lambda_1,\lambda_2)$ ne peut dépendre que de $|\lambda_2-\lambda_1|$) est donnée en notant $r=N|\lambda_1-\lambda_2|\rho(\lambda_1)$ par:
\beaa \text{Cas Hermitien}: W_2(r)&=&1-\left(\frac{\sin(\pi r)}{\pi r}\right)^2\cr
\text{Cas Réel symétrique}:  W_2(r)&=&1-\left(\frac{\sin(\pi r)}{\pi r}\right)^2-\left(\int_r^\infty\,\frac{\sin(\pi s)}{\pi s} \,ds \right) \frac{d}{dr} \frac{\sin(\pi r)}{\pi r} \cr
\text{Cas Quaternionique}:  W_2(r)&=&1-\left(\frac{\sin(2\pi r)}{2\pi r}\right)^2+\left(\int_0^r\,\frac{\sin(2\pi s)}{2\pi s} \,ds \right) \frac{d}{dr} \frac{\sin(2\pi r)}{2\pi r}\cr
\eeaa
\end{theorem}

\begin{center}
	\includegraphics[height=8cm]{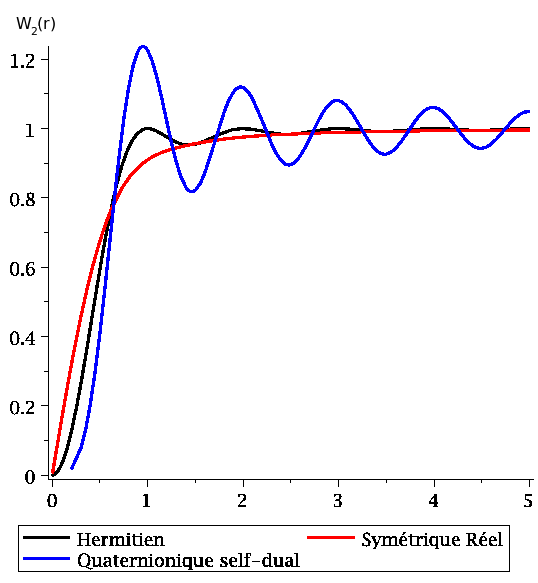}

	Fig. 8: Fonction universelle $r\mapsto W_2(r)$ dans les cas des trois ensembles classiques $\beta\in\left\{1,\frac{1}{2},2\right\}$  
\end{center}

On voit ainsi que la seule donnée de la fonction $\rho_2(x_1,x_2)$ permet de reconstruire toutes les autres lorsque l'on se place au voisinage d'un point intérieur à la distribution limite. A l'aide de ce résultat, il est alors possible de calculer la ``gap probability'', c'est-à-dire la probabilité qu'aucune valeur propre ne se situe dans un intervalle $J=[s,s+\delta]$. Cette probabilité est donnée par un déterminant de Fredholm:
\beq E(J)=\det(\text{Id}-K(J)) \eeq
où $K(J)$ est un opérateur intégral agissant sur l'intervalle $J$. Dans le cas des matrices Hermitiennes, le noyau $K(J)$ tend asymptotiquement dans le c\oe ur de la distribution vers le noyau ``sinus'':
\beq K(J)\mathop{\to}_{n \to \infty} K_{\text{sinus}}(J)=\frac{\text{sin}\,( \pi |J|)}{\pi|J|}\eeq
Notons que dans le cas d'une étude au voisinage d'une extrémité régulière de la distribution limite le noyau $K(J)$ est donné par le noyau d'Airy. 

\medskip

Ce formalisme déterminantal est intéressant mais ne rend pas forcément les calculs plus simples car l'obtention des asymptotiques de déterminants de Fredholm est un problème notoirement difficile. Néanmoins dans un article célèbre \cite{JMMS}, Jimbo, Miwa, Môri et Sato ont obtenu une représentation du noyau sinus à l'aide d'une solution de l'équation de Painlevé $5$:

\begin{theorem}\label{Trou} Représentation du noyau sinus à l'aide de l'équation de Painlevé $5$ \cite{JMMS}:
\beqq \det(\normalfont{\text{Id}}-\lambda K_{\normalfont{\text{sinus}}}([0,s))) = \normalfont{\text{exp}}(\int_{0}^{\pi s} \frac{\sigma(x,\lambda)}{x} dx)\eeqq
où $\sigma(x,\lambda)$ est l'unique solution de l'équation (cas particulier de Painlevé $5$) différentielle:
\beqq (x \sigma''(x,\lambda))^2+4\left(x\sigma'(x,\lambda)-\sigma(x,\lambda)\right) \left(x\sigma'(x,\lambda)-\sigma(x,\lambda)+(\sigma'(x,\lambda))^2 \right)=0\eeqq
avec $\sigma(x,\lambda)\underset{x\to 0}{\to}-\frac{\lambda}{\pi}x -\frac{\lambda^2}{\pi^2}x^2$.
\end{theorem}

De la même façon, Tracy et Widom ont montré que le noyau d'Airy peut être réécrit en termes d'une solution de l'équation de Painlevé $2$:

\begin{theorem} Représentation du noyau d'Airy à l'aide de l'équation de Painlevé $2$ (\cite{TW}):
\beqq 
\det(\normalfont{\text{Id}}-K_{\normalfont{\text{Airy}}}([s,+\infty))=\normalfont{\text{exp}}(-\int_s^\infty (x-s)q(x)dx)\eeqq
où $\sigma(x)$ est l'unique solution de l'équation de Painlevé $2$ (appelée solution d'Hastings-McLeod):
\beqq q''(x)=xq(x)+2q(x)^3 \virg q(x)\underset{x\to \infty}{\sim} Ai(x)\eeqq
\end{theorem}

On voit donc apparaître à nouveau des systèmes intégrables classiques de Painlevé dans l'expression de ces déterminants de Fredholm. Notons que ces résultats sont des résultats locaux au voisinage d'un point de la distribution limite d'équilibre. Les deux cas présentés ci-dessus représentent les cas génériques, c'est-à-dire autour d'un point intérieur à la distribution ou autour d'une extrémité régulière de la distribution (i.e. $\rho_\infty(x)\propto \sqrt{x-a}$). Ils montrent que ces statistiques locales sont universelles et régies par des systèmes intégrables de Painlevé. D'autres résultats concernant ces mêmes statistiques au voisinage de points singuliers ont été étudiés:

\begin{itemize}\item Extrémités singulières de la forme $\rho_\infty(x)\propto (x-a)^{\frac{2m+1}{2}}$ avec $m\geq 0$ dans \cite{2m1}
\item Cas d'un point singulier intérieur de la forme $\rho_\infty(x)\propto (x-a)^{\frac{p}{q}}$ avec $p\wedge q=1$ et $p,q> 0$ dans \cite{BergereEynard,BBE14,TTProperty}. On parle alors de modèles minimaux de type $(p,q)$
\item Cas d'une extrémité de type bord dur (régulière ou non) de type $\rho_\infty(x)\propto (x-a)^{-\frac{p}{q}}$et $p,q> 0$ dans \cite{LaguerreForrester,UniversalityHardEdge}
\end{itemize}

\medskip

On peut par exemple illustrer le cas d'un point singulier intérieur de la forme $(p,q)$ par:

\begin{center}
	\includegraphics[height=8cm]{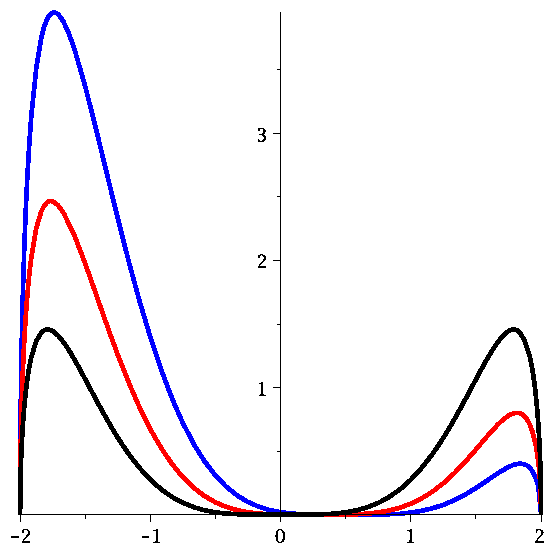}

	Fig. 9: Illustration d'un point critique intérieur suivant $3$ valeurs du couple $(p,q)$   
\end{center}

Dans la prochaine partie nous allons voir comment la récurrence topologique permet de retrouver ces résultats et comment elle fournit une nouvelle vision sur l'universalité et sur le lien avec les systèmes intégrables.

\section{Récurrence topologique et universalité}

La récurrence topologique fournit un formalisme perturbatif pour étudier l'universalité dans les modèles de matrices Hermitiennes. Elle permet également de traiter les comportements au voisinage des points singuliers de la distribution limite d'équilibre de façon analogue aux cas réguliers. Le formalisme peut être illustré par le cas d'un point régulier dans le centre de la distribution limite d'équilibre en relation avec l'équation de Painlevé $5$ et le noyau sinus. Ce travail a été développé de façon précise dans mon article \cite{P5}. La probabilité d'avoir un trou de valeurs propres (gap probability) entre $x_0$ et $x_1$ est donnée par:
\beq p_2(x_0,x_1)=\frac{ \int_{\left(\mathbb{R}\setminus\left[x_0,x_1\right]\right)^N} \Delta(\boldsymbol{\lambda})^2\,\,\underset{i=1}{\overset{N}{\prod}} \text{exp}^{-\frac{N}{T}\,V(\lambda_i)}\,d\lambda_i} {\int_{\mathbb R^N} \Delta(\boldsymbol{\lambda})^2\,\,\underset{i=1}{\overset{N}{\prod}} \text{exp}^{-\frac{N}{T}\,V(\lambda_i)}\,d\lambda_i}\eeq

Le calcul des équations de boucles pour le numérateur a été décrit dans le chapitre \ref{chap1} puisqu'il s'agit d'une intégrale Hermitienne avec bords durs en $x_0$ et $x_1$. Le calcul donne:
\bea \label{spec} &&W_1(x)^2+W_2(x,x)=\frac{N}{T}V'(x)W_1(x)-\frac{N}{T}P_1(x) \cr
&& +\frac{1}{(x-x_0)(x-x_1)}\,\left(N^2-\frac{N}{T}<\Tr (M+x-x_0-x_1)V'(M)>\right) 
\eea
où $P_1(x)=\left< \underset{i=1}{\overset{\infty}{\sum}} \frac{V'(x)-V'(\lambda_{i})}{x-\lambda_{i}}\right>$. En supposant que les conditions d'un développement topologique sont réunies sur le potentiel, on obtient alors la courbe spectrale en posant $y(x)=W_1^{(0)}(x)-\frac{V'(x)}{2}$ par:
\beq y(x)^2=\frac{V'^2(x)}{4} -P_1^{(0)}(x) + \frac{a_0}{x-x_0}+\frac{a_1}{x-x_1}\eeq
Il est maintenant important de se rappeler que l'universalité n'a lieu que localement au voisinage d'un point, c'est-à-dire dans un voisinage dont la taille diminue en $O\left(\frac{1}{N}\right)$. Or pour l'instant $x_0$ et $x_1$ sont fixés et l'écart $|x_1-x_0|$ ne dépend pas de $N$. Pour se localiser au voisinage de $x_0$, on procède donc à une ``double limite d'échelle'', c'est-à-dire que l'on définit:
\bea \label{scaling limit} x(\xi)&=&x_0+\xi (x_1-x_0)\cr
x_1-x_0&=&\frac{s}{N}\cr
&N\to \infty \, ,\, x_0 \text{ fixé}&
\eea
Ainsi l'écart $|x_1-x_0|$ est bien d'ordre $\frac{1}{N}$ et la variable $\xi$ décrit localement l'espace entre $x_0$ et $x_1$. L'utilisation de cette double limite d'échelle dans la courbe spectrale globale \eqref{spec} donne la courbe spectrale locale:
\beq \label{SpectralCurveLocal}y^2(\xi)=s^2+\frac{b_0(s)}{\xi}+\frac{b_1(s)}{\xi-1} \eeq 
Si l'on suppose maintenant que $x_0$ (qui est donné par $s=0$) est un point régulier intérieur à la distribution limite alors il apparait que la densité limite locale ne peut être constituée que d'un seul intervalle. Cela contraint la courbe spectrale locale à être de genre $0$ c'est-à-dire à n'avoir qu'un seul ou deux points de branchements. A l'aide de cette information, on obtient finalement:
\beq \label{FinalSpecCurve}y^2(\xi)=\frac{s^2\left(\xi-\frac{1}{2}\right)^2}{\xi(\xi-1)} \eeq 
Notons que cette courbe spectrale locale est universelle dans le sens où elle ne dépend plus du tout du potentiel de départ ni de la position du point $x_0$ (tant qu'il reste un point intérieur régulier). En utilisant la récurrence topologique on peut alors calculer aisément les premiers ordres des invariants symplectiques. On trouve:
\bea \label{FgP5reduit} F^{(0)}=-\frac{s^2}{8} \,\,\,\,\,&,&\,\,\,\,\, F^{(1)}=\frac{1}{4}\ln\, s\cr
F^{(2)}=\frac{1}{8s^2} \,\,\,\,\,&,&\,\,\,\,\,F^{(3)}=-\frac{5}{8s^4} \rule[0pt]{0pt}{20pt} \cr
F^{(4)}=\frac{131}{12s^6} \,\,\,\,\,&,&\,\,\,\,\,F^{(5)}=-\frac{6575}{16 s^8}\rule[0pt]{0pt}{20pt}
\eea
A partir desquels on peut reconstruire une fonction $\tau$ formelle de la forme:
\beq \label{logtauTopRec}\ln\, \tau_{\text{top. rec.}}^{\text{P5}}(s)\overset{\text{déf}}{=}\sum_{g=0}^\infty F^{(g)}\hbar^{2g-2}=-\frac{s^2}{8\hbar^2}+\frac{1}{4}\ln\, s+\frac{\hbar^2}{8s^2}-\frac{5\hbar^4}{8s^4}+\frac{131\hbar^6}{12s^6}-\frac{6575\hbar^8}{16 s^8}+O(\hbar^{10})\eeq
On peut alors comparer ces résultats avec ceux de la paire de Lax correspondante à l'équation de Painlevé $5$ fournie en \eqref{P5}. Plus précisément, le cas présent est un cas particulier de l'équation de Painlevé $5$ dans lequel toutes les monodromies sont nulles: $\theta_0=\theta_1=\theta_\infty=0$. On trouve alors une paire de Lax simplifiée par rapport au cas général de \eqref{P5}:
\bea\label{P5Reduite} \DD(\xi,t)&=&\begin{pmatrix}\frac{t}{2}+\frac{pq}{\xi}-\frac{pq}{\xi-1}& -\frac{pq}{\xi}+\frac{p}{2q(\xi-1)}\\
\frac{pq}{\xi}-\frac{pq^2}{\xi-1}&-\left(\frac{t}{2}+\frac{pq}{\xi}-\frac{pq}{\xi-1}\right)\end{pmatrix}\cr
\RR(\xi,t) &=& \begin{pmatrix}\frac{\xi}{2}-\frac{p(q-1)^2}{2t} & -\frac{p(q-1)}{t}\\
-\frac{pq(q-1)}{t}& -\left(\frac{\xi}{2}-\frac{p(q-1)^2}{2t}\right)\end{pmatrix}
\eea
La fonction $\sigma(t)$ d'Okamoto (en lien avec la fonction $\tau$ de Jimbo-Miwa et la structure Hamiltonienne) est donnée par $\dot{\sigma}(t)=-pq$ et satisfait l'équation différentielle:
\beq -(\hbar t \ddot{\sigma})^2+(t\dot{\sigma}-\sigma)(t\dot{\sigma}-\sigma-4\dot{\sigma}^2)=0\eeq
A l'aide du changement de variables:
\beq \label{chgtvr} s=\frac{i t}{2\pi} \,\,,\,\, \hat{\sigma}(s)=\sigma\left(-2\pi i s\right)\,\,,\,\, \hbar=1\eeq
On obtient que $\hat{\sigma}(s)$ satisfait l'équation différentielle:
\beq \label{EqDiffsigma}\frac{1}{\pi^2}(s\ddot{\hat{\sigma}})^2+4(s\dot{\hat{\sigma}}-\hat{\sigma})(s\dot{\hat{\sigma}}-\hat{\sigma}+\frac{1}{\pi^2}(\dot{\hat{\sigma}})^2)=0\eeq
qui correspond exactement à la probabilité de trou donnée dans le théorème \ref{Trou}:
\beq \label{E2} E_2(0,s)=\text{exp}\left(\int_0^s \frac{\hat{\sigma}(u)}{u}du\right)\eeq
On peut alors utiliser les formules déterminantales décrites dans le chapitre \ref{chap2} pour les comparer avec les quantités obtenues du côté de la double limite d'échelle issue des matrices aléatoires. La courbe spectrale associée à la paire de Lax \eqref{P5Reduite} est donnée par:
\beq y(\xi)^2 =  s^2 +\frac{\hat{\sigma}^{(0)}(s)}{\pi^2\xi}-\frac{\hat{\sigma}^{(0)}(s)}{\pi^2(\xi-1)}\eeq
c'est-à-dire de la même forme que \eqref{SpectralCurveLocal}. Le calcul direct des premiers ordres de $\hat{\sigma}(s)$ à l'aide de \eqref{EqDiffsigma} permet d'identifier la courbe précédente avec \eqref{FinalSpecCurve}. La fonction $\tau$ de Jimbo Miwa est également directement reliée à la fonction $\hat{\sigma}$ par:
\beq \hat{\sigma}(s)=\hbar^2\pi^2 s\frac{d}{ds}\ln\, \tau_{\text{JM}}(s)\eeq
Le calcul des premiers ordres en $\hbar^k$ à l'aide de \eqref{EqDiffsigma} donne:
\beq \ln \,\tau_{\text{JM}}(s)=-\frac{s^2}{8\hbar^2}+\frac{1}{4}\,\ln\, s+\frac{\hbar^2}{8s^2}-\frac{5\hbar^4}{8s^4}+\frac{131\hbar^6}{12s^6}-\frac{6375\hbar^8}{16s^8}+O\left(\hbar^{10}\right)\eeq
qui s'identifie donc exactement avec \eqref{logtauTopRec}.

\medskip

Une fois les questions de paramétrisation réglées \eqref{chgtvr}, les courbes spectrales identifiées l'une à l'autre et les premiers ordres calculés et identifiés de part et d'autre, il reste à fournir une démonstration rigoureuse de l'identité suivante:
\beq \label{tautau} \ln \,\tau_{\text{JM}}(s)\overset{\text{?}}{=} \ln\, \tau_{\text{top. rec.}}^{\text{P5}}(s)\eeq
ainsi qu'une démonstration permettant d'identifier les différentielles de la récurrence topologique $\omega_n^{(g)}$ (calculées sur la courbe \eqref{FinalSpecCurve}) avec celles produites par les formules déterminantales $W_n^{(g)}$ du côté de la paire de Lax \eqref{P5Reduite}. Pour ce faire, on peut parfaitement appliquer la méthode décrite dans le chapitre précédent consistant à vérifier les conditions de type topologique du côté du système intégrable. Cela a été fait dans mon article \cite{P5} et complété pour la partie de l'ordre dominant dans mon autre article \cite{IwakiMarchal}. On arrive ainsi à montrer l'égalité \eqref{tautau} à tout ordre en $\hbar$ ce qui permet de justifier que les statistiques locales autour d'un point régulier intérieur peuvent être reconstruites (comme la probabilité de trou \eqref{E2}) de façon formelle par la récurrence topologique appliquée à une courbe spectrale provenant de la paire de Lax de Painlevé $5$:

\begin{theorem}[Identification à tout ordre] On a l'identification formelle (i.e. terme à terme pour chaque puissance de $\hbar$) à tout ordre \cite{IwakiMarchal}:
\beqq \ln \,\tau_{\text{JM}}= \ln\, \tau_{\text{top. rec.}}^{\text{P5}}\eeqq
\end{theorem}

On note que cette méthode possède la faiblesse de n'être qu'une méthode perturbative en $\hbar$ et qu'elle ne garantit pas la convergence des différentes séries impliquées. En revanche, son avantage majeur est qu'elle permet très facilement l'étude d'autres types de points (extrémités régulières ou singulières, points singuliers intérieurs) puisque du côté des matrices aléatoires et de la double limite d'échelle, le formalisme et les résultats peuvent être facilement adaptés à d'autres types de points de la distribution limite des valeurs propres. Relativisons néanmoins en mentionnant que du côté des paires de Lax, il est nécessaire de deviner par avance quel sera le système intégrable (i.e. la paire de Lax) qui devra être utilisé (mais le calcul des premiers ordres du côté des matrices aléatoires peut permettre de s'en faire une idée). Ce type d'approche a été réalisée pour des extrémités régulières \cite{BE09}, des points singuliers de type $\rho_\infty(x)=(x-a)^{\frac{2m+1}{2}}$ dans mon article \cite{P2} et de manière partielle pour les points de type $\rho_\infty(x)=(x-a)^{\frac{p}{q}}$ avec $p,q>0$ dans \cite{BBE14}. La démonstration rigoureuse et complète de ces derniers cas a été récemment réalisée dans mon article \cite{TTProperty}. 

\section{Conclusion du chapitre}

Ce chapitre montre que la récurrence topologique et les méthodes de double limite d'échelle permettent de démontrer de façon alternative le caractère universel des statistiques locales dans les modèles de matrices Hermitiennes. En particulier, nous avons montré que l'on retrouve de façon naturelle les systèmes de Lax de certaines équations de Painlevé et que l'on peut, moyennant l'utilisation des formules déterminantales et la preuve des conditions de type topologique (présentées dans le chapitre précédent), identifier formellement, à tout ordre, les quantités produites par la récurrence topologique avec leurs homologues du côté intégrable. Cette approche perturbative (formelle) est ainsi particulièrement efficace pour traiter les différents types de points singuliers qui peuvent apparaître du côté des matrices aléatoires Hermitiennes. Cette technique pourrait ainsi être envisagée dans le cas de bords durs singuliers ou sur d'autres types de singularités qui apparaîtraient un jour dans la littérature. 
\selectlanguage{french}
\chapter{Conclusion}
\thispagestyle{fancy}

Bien que la récurrence topologique ne soit apparue que depuis seulement une dizaine d'années, elle a connu un essor considérable dans beaucoup de domaines des mathématiques et de la physique théorique. Dans cette habilitation à diriger des recherches, j'ai en particulier mis l'accent sur l'utilisation de la récurrence topologique pour le calcul de certaines probabilités ou de certaines fonctions de partition issues de modèles d'intégrales de matrices Hermitiennes ou unitaires. Cet aspect probabiliste historique reste encore d'actualité et pourrait trouver une généralisation aux cas d'interactions de type ensembles $\beta$ dans les prochaines années. En parallèle de ces travaux, j'ai également montré que la récurrence topologique peut être utilisée pour l'étude des systèmes intégrables. Elle permet tout d'abord de retrouver de façon assez directe les résultats d'universalité et les relations avec les équations de Painlevé historiquement abordés par la méthode des polynômes orthogonaux. De façon plus générale, elle permet aussi, sous certaines conditions, de reconstruire le développement perturbatif et WKB de quantités issues directement d'une paire de Lax. De nombreux travaux de recherche sont en cours sur ce sujet avec, en ligne de mire, la recherche et la compréhension de courbes spectrales ``quantiques''. Il est presque certain que ce domaine est amené à évoluer rapidement dans les prochaines années. Enfin la généralisation des formules déterminantales à des groupes de Lie généraux est également en bonne voie et devrait ouvrir de nouvelles perspectives en physique théorique et dans la théorie des cordes topologiques.
Ainsi, de part son caractère polymorphe et ses applications variées, la récurrence topologique est sans aucun doute promise à un bel avenir mathématique.

\bibliographystyle{unsrt}
\thispagestyle{fancy}
\bibliography{BiblioExt}  
\thispagestyle{fancy}

\bibliographystyleA{unsrt}
\thispagestyle{fancy}
\bibliographyA{MyBiblio} 
\thispagestyle{fancy}

\end{document}